\pdfoutput=1
\documentclass[vecphys]{svmult}

\usepackage[pdftex]{graphicx}
\usepackage{epstopdf}
\usepackage{mathptmx,helvet,courier}
\usepackage{multicol}
\usepackage[bottom]{footmisc}
\usepackage{cite}
\usepackage{amsmath}
\usepackage{amssymb}
\usepackage{color}
\usepackage{blindtext}

\begin{document}

\title*{Nonlinear Beam Propagation in a Class of Complex Non-$\mathcal{PT}$-Symmetric Potentials}
%\titlerunning{}
\author{
%Jes\'us
J. Cuevas-Maraver
  \and
%Panayotis
P. G. Kevrekidis
\and
%Dimitrios
D. J. Frantzeskakis
\and
%Yannis
Y. Kominis
}

\institute{
Jes\'us Cuevas-Maraver
\at Grupo de F\'{\i}sica No Lineal, Universidad de
Sevilla, Departamento de F\'{i}sica Aplicada I,
Escuela Polit\'ecnica Superior.
C/ Virgen de \'{A}frica, 7, 41011-Sevilla, Spain \\
Instituto de Matem\'{a}ticas de la Universidad de Sevilla (IMUS). Edificio Celestino Mutis. Avda. Reina Mercedes s/n, 41012-Sevilla, Spain
\email{jcuevas@us.es}
\and
Panayotis G. Kevrekidis
\at Department of Mathematics and Statistics, University of Massachusetts,
Amherst, MA 01003-4515, USA
\and
Dimitrios J. Frantzeskakis
\at
Department of Physics, National and Kapodistrian University of Athens, Panepistimiopolis, Zografos,
Athens 15784, Greece
\and
Yannis Kominis
\at School of Applied Mathematical and Physical Science, National Technical University of Athens, Zographou GR-15773, Greece
}

\maketitle

\abstract{The subject of $\mathcal{PT}$-symmetry and its areas
  of application have been blossoming over the past decade. Here, we
  consider a nonlinear Schr{\"o}dinger model with a complex potential that
  can be tuned controllably away from being $\mathcal{PT}$-symmetric, as it
  might be the case in realistic applications.
  We utilize two parameters: the first one breaks $\mathcal{PT}$-symmetry but
  retains a proportionality between the imaginary and the derivative of the
  real part of the potential; the second one, detunes from this latter
  proportionality.
  It is shown that the departure of the potential from the $\mathcal{PT}$-symmetric form
  does not allow for the numerical identification of exact stationary solutions.
  Nevertheless, it is of crucial importance to consider the dynamical evolution
  of initial beam profiles.
  In that light, we define a suitable notion of optimization and
  find that even for non $\mathcal{PT}$-symmetric cases,
  the beam dynamics, both in 1D and 2D --although prone to weak growth or decay-- suggests that
  the optimized profiles do not change significantly under propagation for specific
  parameter regimes.
}


\section{Introduction}

The original suggestion of Bender and collaborators~\cite{Bender1,Bender2}
of a new class of systems that respect parity and time-reversal
(so-called $\mathcal{PT}$-symmetric systems) was motivated by the
consideration of the foundations of quantum mechanics and the
examination of the need of Hermitianity within them. The argument
of Bender and collaborators was that such systems, even if non-Hermitian
and featuring gain and loss, could give rise to real spectra,
thus presenting a candidacy for being associated with measurable
quantities.

This proposal found a fertile ground for its development in areas,
arguably, different than where it was originally proposed. In particular,
the work of Christodoulides and co-workers in nonlinear
optics a decade later, spearheaded an array of experimental realizations
of such media (capitalizing on the ubiquitous in optics loss and
on controllable gain)~\cite{Ruter,Peng2014,peng2014b,ncomms2015,RevPT,Konotop}.
Other experiments swiftly followed in areas ranging from
electronic circuits~\cite{Schindler1,Schindler2,Factor} to
mechanical systems~\cite{Bender3}, bringing about not only
experimental accessibility, but also an intense theoretical
focus on this theme. These threads of research have now been summarized
in two rather comprehensive recent reviews~\cite{RevPT,Konotop}.

While $\mathcal{PT}$-symmetric variants of
other nonlinear wave models have more recently been proposed, including
the $\mathcal{PT}$-symmetric variants of the Dirac equations~\cite{ptnlde}
and of the Klein-Gordon equation~\cite{ptkg}, the main focus of
associated interest has been on models of the nonlinear Schr{\"o}dinger (NLS)
type. This is natural given the relevance at the paraxial
approximation level of such a model in applications stemming from
nonlinear optics and related themes~\cite{RevPT,Konotop}. In this
important case, the $\mathcal{PT}$-invariance is consonant with
complex external potentials $\tilde{V}$, of the form $\tilde{V}=V+i W$,
subject to the constraint that $\tilde{V}^{\ast }(x)=\tilde{V}(-x)$.
This implies that the real part, $V$, of the potential needs
to be even, while the imaginary part, $W$, of the potential
needs to be odd to ensure $\mathcal{PT}$-symmetry.
The expectation, thus, has been that typically Hamiltonian
and $\mathcal{PT}$-symmetric systems featuring gain and loss
will possess continuous families of soliton solutions;
otherwise, the models  will possess solutions for
isolated values within the parameter space.

However, more recent investigations have started to challenge
this belief. On the one hand, work on complex, asymmetric
so-called Wadati potentials has produced mono-parametric
continuous families of stationary solutions~\cite{abdu,kono}.
On the other hand, the notion of partial $\mathcal{PT}$-symmetry
has been explored, e.g., with models that possess the symmetry
in one of the directions but not in another~\cite{jian,jenn}.
In fact, in the recent work of~\cite{Kom15,kom15b} that motivated
the present study, it was shown that to identify critical points
one can localize a soliton
\footnote{Below, we use the term ``soliton'' in a loose sense, without
implying complete integrability \cite{mja}.}
in a way such that its intensity has
a vanishing total overlap with the imaginary part of the potential,
assuming that the real part of the potential is {proportional to} the anti-derivative
of the imaginary part (but without making any assumptions on the
parity of either).

In the present work, we revisit these considerations.
In particular, we discuss the results of the important
contribution of~\cite{ny16}. This work suggests (and indeed
conjectures) that the {\it only} complex potentials that could
feature continuous families of stationary solutions
although non-${\mathcal PT}$-symmetric are the ones of the Wadati
type. In our case, we have considered potentials that depart
from this form and either satisfy --or controllably depart from --
the simpler mass and momentum balance conditions of~\cite{Kom15,kom15b}.
We observe that in such settings, waveforms ``optimizing'' the vector
field (which we define as bringing it {\it very} --but
nor arbitrarily-- close to vanishing) may exist,
but still are not true solutions, in line with the above conjecture.
We develop diagnostics that explore how these optimized
beams behave dynamically, and identify their slow growth or
decay.
%correlating their proximity to true solutions with the
%rate of departure from the configurations.
We do this for two
different broad multi-parametric families
of potentials to showcase the generality of our conclusions.
We then extend relevant considerations also to 2D settings,
showing how symmetry breaking bifurcation scenarios can be
traced via our optimized beam approach.

Our presentation will be structured as follows. In section~2,
we introduce the model, connect our considerations to
those of~\cite{ny16} and justify the selection of the
complex potential. In section~3,
we explore the optimized beams and the associated dynamics of the relevant
waveforms numerically. Then, in section~4,
we generalize these notions in a two-dimensional setting.
In section~5, we proceed to summarize our findings and
propose a number of directions for future study. Finally,
in the Appendix, details of the numerical
method used to optimize the dynamical beams are presented.

\section{{The one-dimensional potential}}

As explained in the previous section, motivated by
the development in the analysis of NLS models with
complex potentials, we consider the rather broad
setting of the form:
\begin{equation}\label{eq:dyn}
    i\psi_t=-\psi_{xx}+[V(x)+i W(x)]\psi-|\psi|^2\psi,
\end{equation}
with subscripts denoting partial derivatives. In the context
of optics, $\psi(x,t)$ represents the complex electric field
envelope, $t$ is the propagation distance, $x$ corresponds to
the transverse direction, while the variation of the dielectric
permittivity plays the role of the external potential, with
$V(x)$ and $W(x)$ being its real and imaginary parts, respectively
\cite{RevPT,Konotop}. In the recent analysis of~\cite{Kom15,kom15b},
assuming the existence of bright solitons (as is natural in the
focusing nonlinearity setup under consideration),
dynamical evolution equations were obtained
for the soliton mass and velocity.
Here, we use as our motivating point for constructing standing
wave structures of Eq.~(\ref{eq:dyn})
the stationary form of
these equations, which read
(cf. Eqs.~(5)-(6) of Ref.~\cite{kom15b}):
\begin{eqnarray}
  \int_{-\infty}^{\infty} |\psi(x)|^2 W(x+x_0) dx=
  \int_{-\infty}^{\infty} |\psi(x)|^2 V'(x+x_0) dx=0,
  \label{eqn0}
  \end{eqnarray}
where $x_0$ denotes the soliton center. The first one among
these equations,
corresponds to a ``power-balance'' (or mass balance)
condition, implying that the soliton
has a transverse profile such that it experiences gain and loss in an
overall balanced fashion across its spatial extent.
The second equation corresponds to a ``momentum-balance'' condition,
i.e., the total force exerted on the solitary wave vanishes,
hence the coherent structure is at an equilibrium.

This pair of stationarity conditions in Eq.~(\ref{eqn0}) reduces to a single
one, provided that $V'=-C W$, with $C$ being a constant. In that
context, the resulting
condition posits the following:
if a soliton can be placed relative to the
gain/loss profile so that its intensity has an overall vanishing
overlap with the imaginary part of the potential, then the
existence of a fixed point (and thus a stationary soliton
solution) may be expected.

However, it should be kept in mind that these conditions are
{\it necessary but not sufficient} for the existence of
a stationary configuration. In particular, a recent ingenious
calculation shed some light on this problem for a general potential
in the work of~\cite{ny16}. Using a standing wave deomposition
$$\psi=r(x)e^{i \int^x \theta(x') dx'} e^{i \mu t}$$ in Eq.~(\ref{eq:dyn}), the following
ordinary differential equations were derived:
\begin{eqnarray}
  &&r_{xx} - \mu r - V r + r^3 - \theta^2 r = 0,
\label{extra1}
  \\
  &&(r^2 \theta)_x = W r^2.
  \label{extra2}
\end{eqnarray}
It was then realized that, in the absence of external potential,
two quantities, namely $J_1=r^2 \theta$ (the ``angular momentum''
in the classical mechanical analogy of the problem) and
$J_2=r_x^2 - \mu r^2 + r^4/2 + r^2 \theta^2$ (the ``first integral''
or energy in the classical analogue) are conserved, i.e., % in that
$d J_i/dx=0$. For $J_1$, Eq.~(\ref{extra2}) yields its evolution
in the presence of the potential while for $J_2$, direct calculation
shows:
\begin{eqnarray}
  \frac{dJ_2}{dx}=V (r^2)_x + 2 W r^2 \theta = S_x - r^2 V_x-2 (r^2 \theta)_x
  \int W dx,
  \label{extra3}
\end{eqnarray}
with $S=V r^2 +2 r^2 \theta \int W dx$. Combining the last terms,
upon substitution of $(r^2 \theta)_x$ from Eq.~(\ref{extra2})
allows us to infer that this pair of terms will vanish if the coefficient
multiplying $r^2$, namely $V_x -2 W \int W dx$, vanishes; this occurs
if the potential has the form:
$$V+i W=- [g^2 + i g'(x)] + c,$$
where $c$ is a constant.
A shooting argument presented in~\cite{ny16} suggests that there
are 3 real constants (2 complex ones, yet one of them can be
considered as real due to the %overal
phase invariance) in order
to ``glue'' two complex quantities, namely $\psi$ and $\psi_x$
at some point within the domain. This can only be done when
a conserved quantity exists, which requires the type of
potential suggested above, in the form $-[g^2 + i g'(x)]$. However,
if additional symmetry exists, such as ${\mathcal PT}$-symmetry,
the symmetry alone may impose conditions such as
Im$(\psi(0))$=Re$(\psi_x(0))=0$, which in turn allows for the
shooting to go through (and thus solutions to exist) for a
continuous range of $\mu$'s.

Nevertheless, a natural question is: suppose that the potential
is not of this rather non-generic form, yet it deviates from
the ${\mathcal PT}$-symmetric limit, possibly in ways respecting
the above mass and/or momentum balance conditions of Eq.~(\ref{eqn0});
then what is the fate of the system~? Do stationary states perhaps
exist or do they not, and what are the dynamical implications
of such conditions~? It is this class of questions that we will
aim to make some progress towards in what follows.

To test relevant ideas, we will use two different potentials
 $\tilde{V}_j(x)=V_j(x)+iW_j(x)$, with $j=1,2$. In the first one, $W$ is of the form:
\begin{equation}\label{eq:potential1i}
    W_1(x)=A_1k_1\mathrm{sech}(x-x_d-\delta_1)\tanh(x),
\end{equation}
where $A_1$, $k_1$, $x_d$ and $\delta_1$ are constants, with the
latter two controlling the breaking of the $\mathcal{PT}$-symmetry.
We then use a real potential $V_1$ given by the form:
\begin{equation}\label{eq:potential1r}
V_1(x)=-2A_1\left[\arctan\left(\tanh\left(\frac{x_d-x}{2}\right)\right)\coth(x_d)-
\arctan\left(\tanh\left(\frac{x}{2}\right)\right)\mathrm{csch}(x_d)\right],
\end{equation}
which, in the limit $x_d\rightarrow0$, transforms into $V_1(x)=-A_1\mathrm{sech}(x)$.
The motivation behind this selection is that
if $\delta_1=0$ in Eq.~(\ref{eq:potential1i}) then
$V_1$ is {proportional to} the anti-derivative of $W_1$
(hence ensures that the pair of conditions of Eq.~(\ref{eqn0})
degenerate to a single one). In addition, for $\delta_1=0$ and in the limit $x_d\rightarrow0$, the potential is $\mathcal{PT}$-symmetric. In short,
the two parameters $x_d$ and $\delta_1$ both control the departure from
$\mathcal{PT}$-symmetry, while the latter affects the departure from
proportionality of $V'_1$ and $W_1$.
This selection and these parameters
thus allow us to tailor the properties of the potential,
controlling its departure from the $\mathcal{PT}$-symmetric limit,
but also from the possible degeneracy point of the conditions (\ref{eqn0}).

The second potential is given by

\begin{equation}\label{eq:potential2i}
    W_2(x)=A_2k_2x\mathrm{sech}^2(x-\delta_2-1),
\end{equation}
and
\begin{equation}\label{eq:potential2r}
    V_2(x)=-A_2(\log[\cosh(1-x)]+x\tanh(1-x)),
\end{equation}
where $A_2$, $k_2$ and $\delta_2$ are constants. Contrary to the $\tilde{V}_1$ case, this potential does not possess a $\mathcal{PT}$-symmetric limit.

For both $\tilde{V}_1(x)$ and $\tilde{V}_2(x)$ potentials, if $\delta=0$ then $V'_j(x)=-C_jW_j(x)$ and, as shown in Ref.~\cite{Kom15},
rendering a topic of interest the exploration
of the potential existence of stationary solutions in the vicinity of the
interface between the lossy and amplifying parts when Eq.~(\ref{eqn0}) applies.
In our particular case, the proportionality factor
$C_j$ is $C_j=1/k_j$.

\section{Numerical results}

\subsection{Stationary states}

We start by seeking
stationary localized solutions, of the form $\psi(x,t)=\mathrm{e}^{i \mu t}u(x)$ with $u(x)\in\mathbb{C}$,
which will thus satisfy:
\begin{equation}\label{eq:stat}
    F[u]\equiv\mu u-u_{xx}+[V(x)+i W(x)]u-|u|^2u=0.
\end{equation}
In what follows, we fix $A_1=0.1$ and $A_2=1$, and consider stationary solutions of frequency $\mu=1$. {We will make use of periodic boundary conditions}.

Notice that the potentials of~\cite{ny16}
$\tilde{V}(x)=-[g^2(x)+ig'(x)]$ would, in the present notation,
necessitate:
\begin{equation}\label{eq:Yang}
    [V'(x)]^2=-4V(x)W^2(x).
\end{equation}
It is important to note that the potentials studied in our chapter do not
fulfill this relation
\emph{for any set of parameters} $(A_1, k_1, x_d, \delta_1)$ or $(A_2, k_2, \delta_2)$ ---except for the ``trivial'' $\mathcal{PT}$-limit--- as it can be easily demonstrated.
As a result then, presumably because of the above calculation, the standard
fixed point methods that we have utilized
fail to converge away from the $\mathcal{PT}$-symmetric limit. For this reason, we make use of minimization algorithms in order to obtain optimized profiles of localized waveforms. With these methods, one can seek for local minima of the norm of $F[u]$ instead of zeros of that function. In our problem, we have made use of the Levenberg--Marquardt algorithm {(see
  %Subsection~\ref{subsec:LMA}
Appendix A
  for more details)}, which has been successfully used for computing solitary gravity-capillary water waves \cite{dcd16}, and established a tolerance of $||F[u]||<10^{-3}$ with $||F[u]||$ being the $L^2$-norm of $F[u]$:

\begin{equation}\label{eq:L2}
    ||F[u]||=\sqrt{\int |F[u(x)]|^2\mathrm{d}x}.
\end{equation}

In the particular case of potential $\tilde{V}_1(x)$, we have studied the stability of solitons in the $\mathcal{PT}$-symmetric limit $x_d=\delta_1=0$ as a function of $k_1$, observing that solitons are stable whenever $k_1<k_c$, with $k_c=8.28$. At this point, the soliton experiences a Hopf bifurcation. In order to avoid
any connection of the findings below with the effect of such instability, we have fixed in what follows a value of $k_1$ far enough from $k_c$.
Moreover, since the minimal value attained for
$||F[u]||$ increases with $k_1$, we have restricted
consideration to relatively small values of $k_1$
and more specifically will report results in what follows for    $k_1=1/2$.

\begin{figure}
\begin{center}
\begin{tabular}{cc}
\includegraphics[width=.45\textwidth]{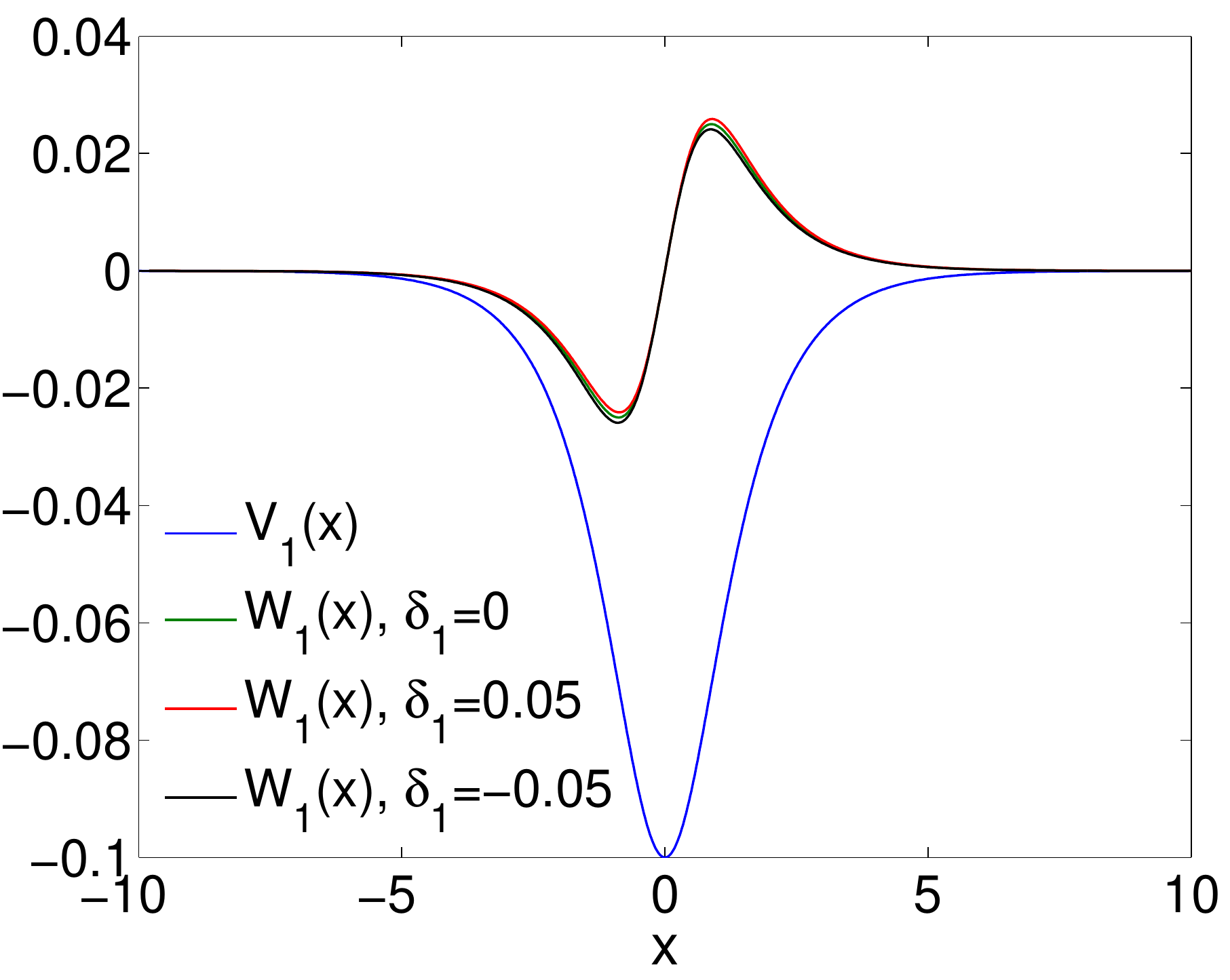} &
\includegraphics[width=.45\textwidth]{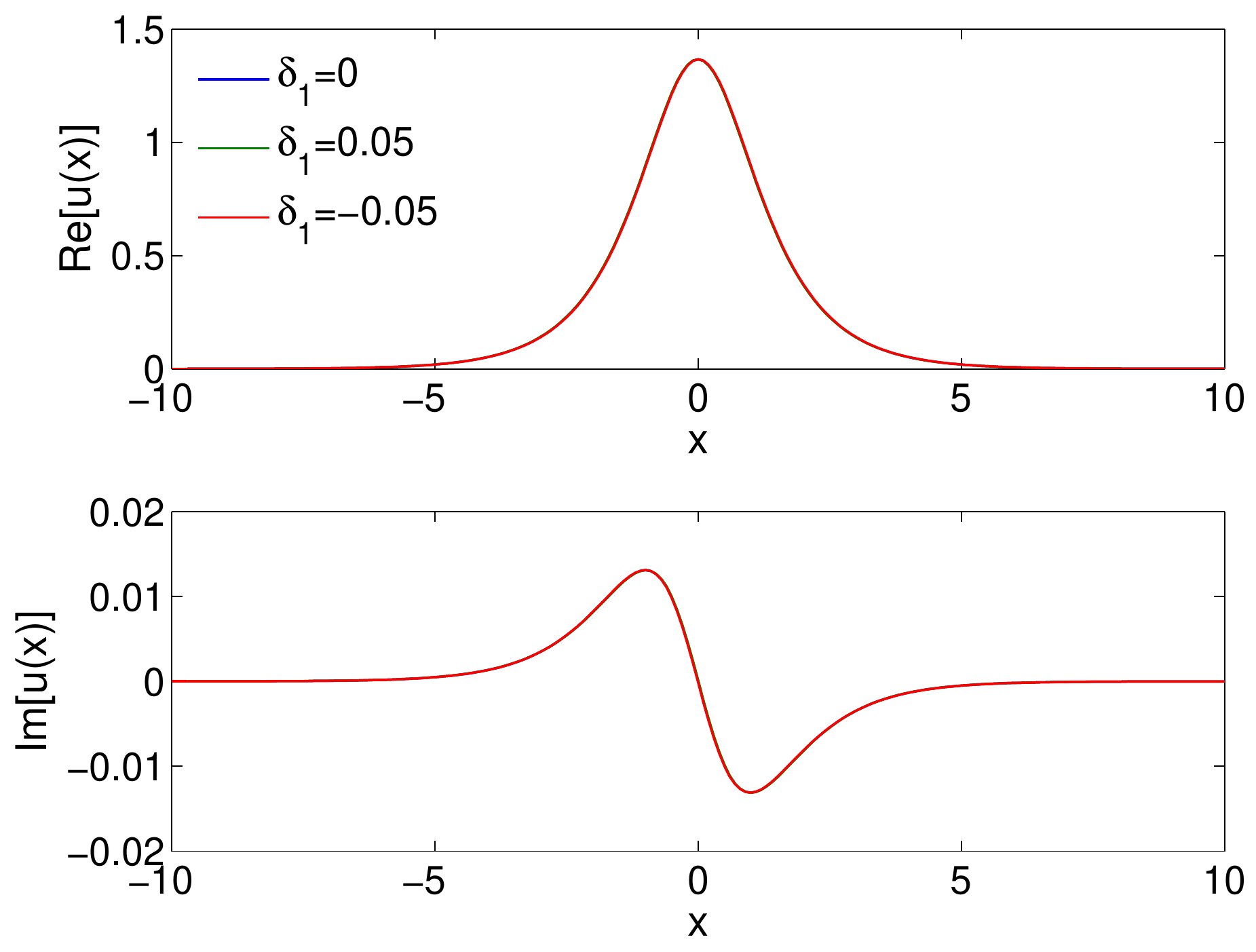} \\
\includegraphics[width=.45\textwidth]{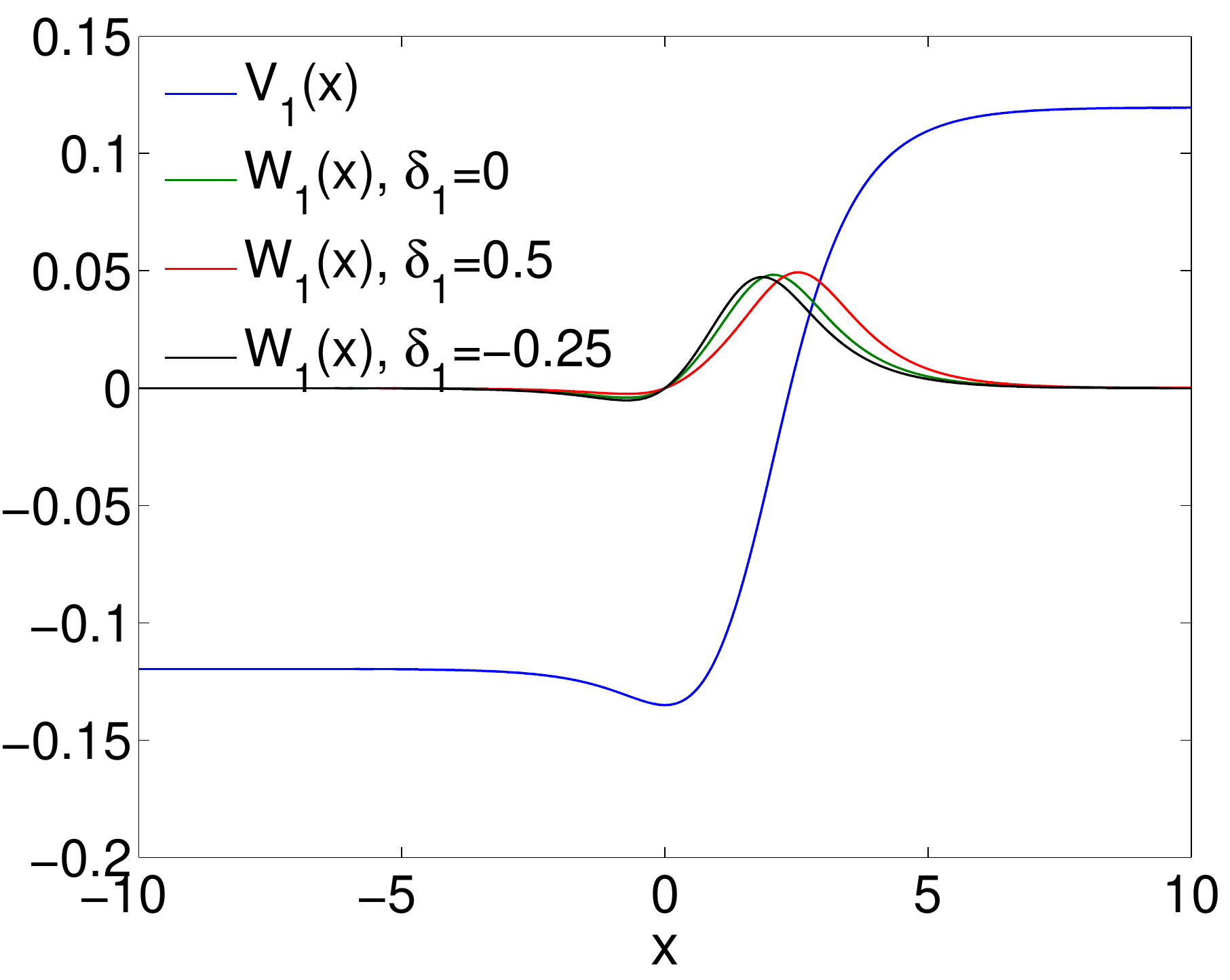} &
\includegraphics[width=.45\textwidth]{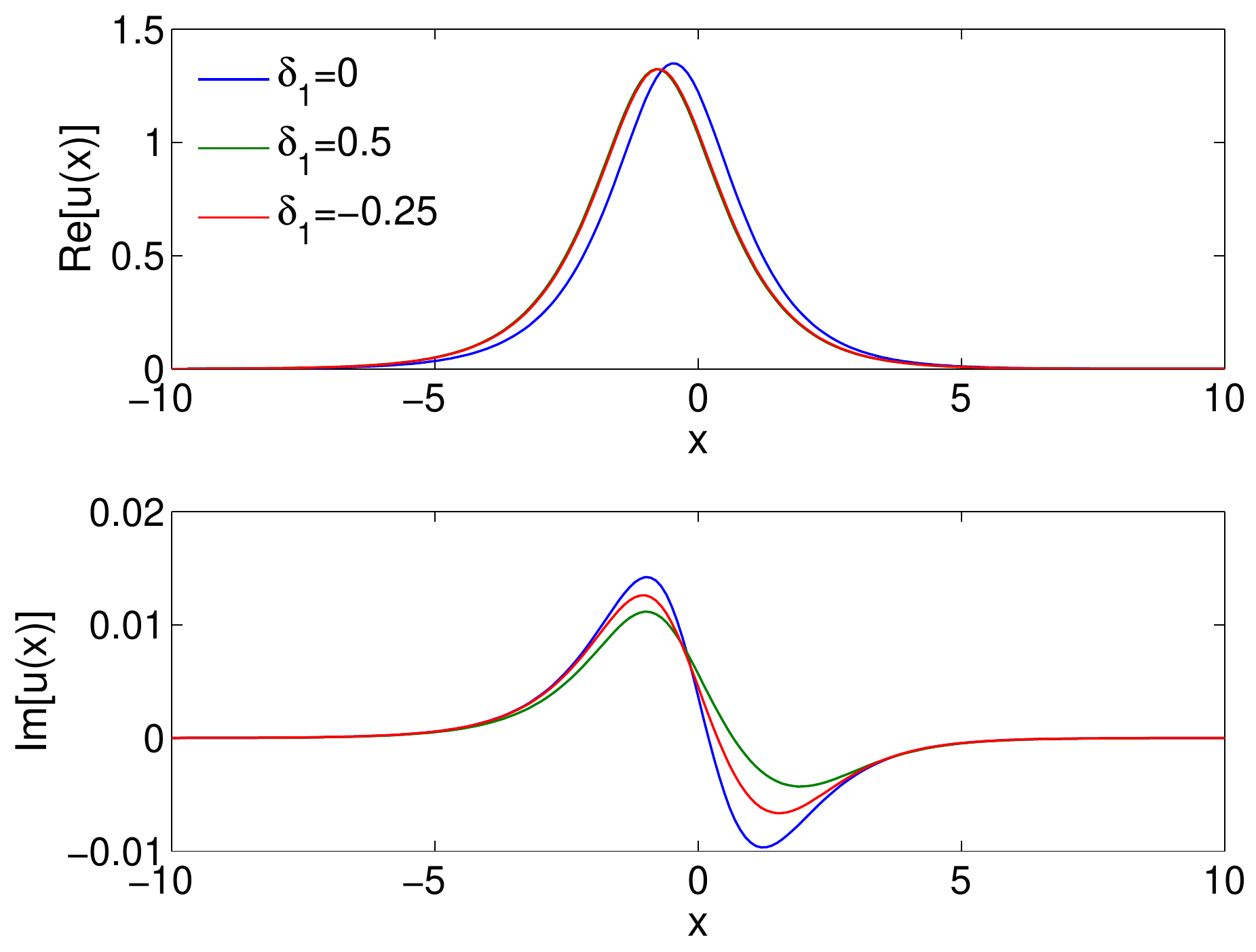} \\
\end{tabular}
\end{center}
\caption{Left panels: Real and imaginary part of the potential $\tilde{V}_1(x)$ for $A_1=0.1$, $k_1=1/2$ and $x_d=0$ (top) and $x_d=1$ (bottom);
green line corresponds to the imaginary part for $\delta_1=0$, whereas red (black) line corresponds to the imaginary part for $\delta_1=0.05$ ($\delta_1=-0.05$).
Right panels:
Beam profiles minimizing $||F[u]||$
(real and imaginary parts) for $A_1=0.1$, $k_1=1/2$ and $x_d=0$ (top) and $x_d=1$ (bottom); the blue line corresponds to $\delta_1=0$,
and the green (red) line corresponds to $\delta_1=0.5$ ($\delta_1=-0.25$).}
\label{fig:potential1}
\end{figure}

\begin{figure}
\begin{center}
\begin{tabular}{cc}
\includegraphics[width=.45\textwidth]{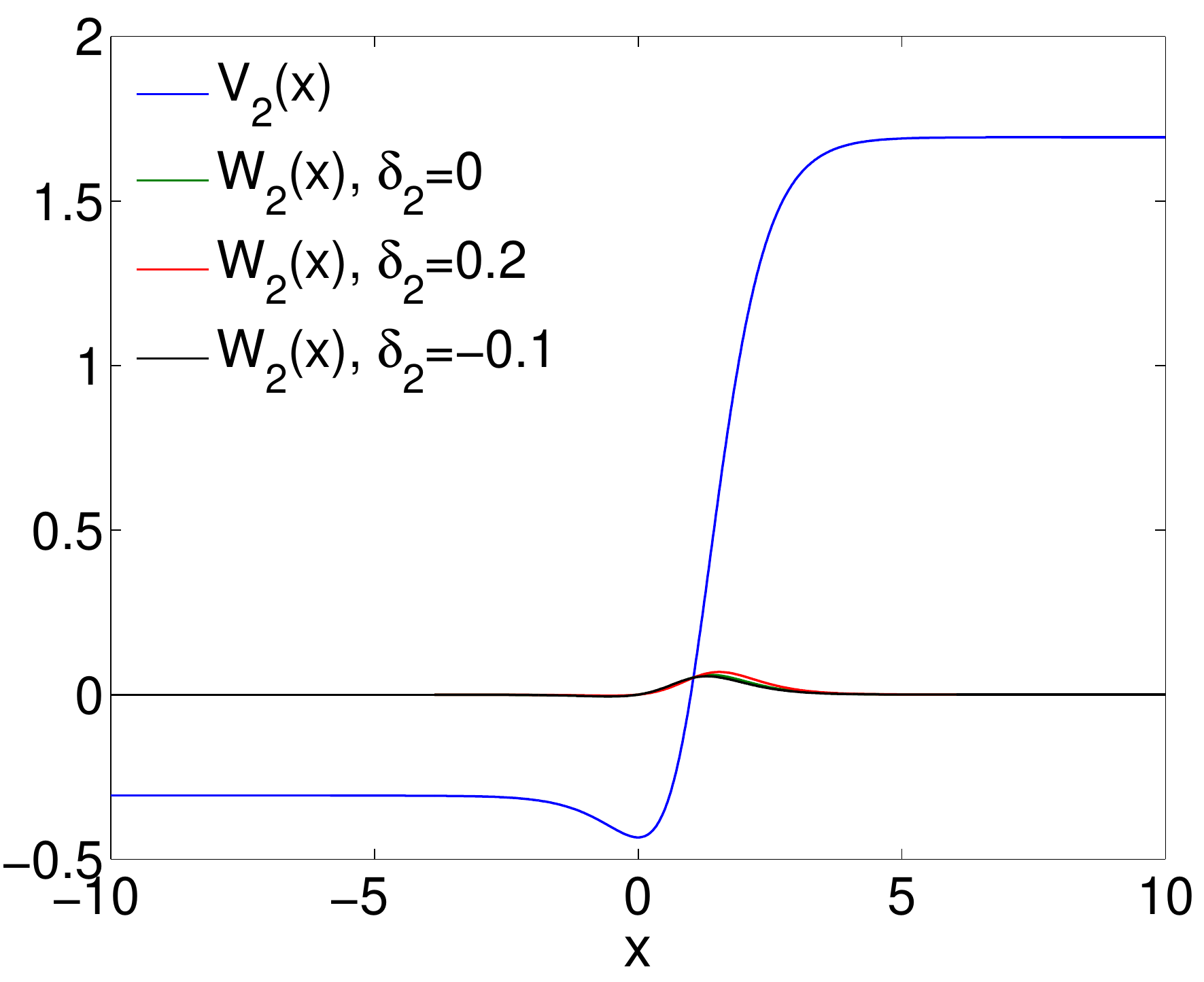} &
\includegraphics[width=.45\textwidth]{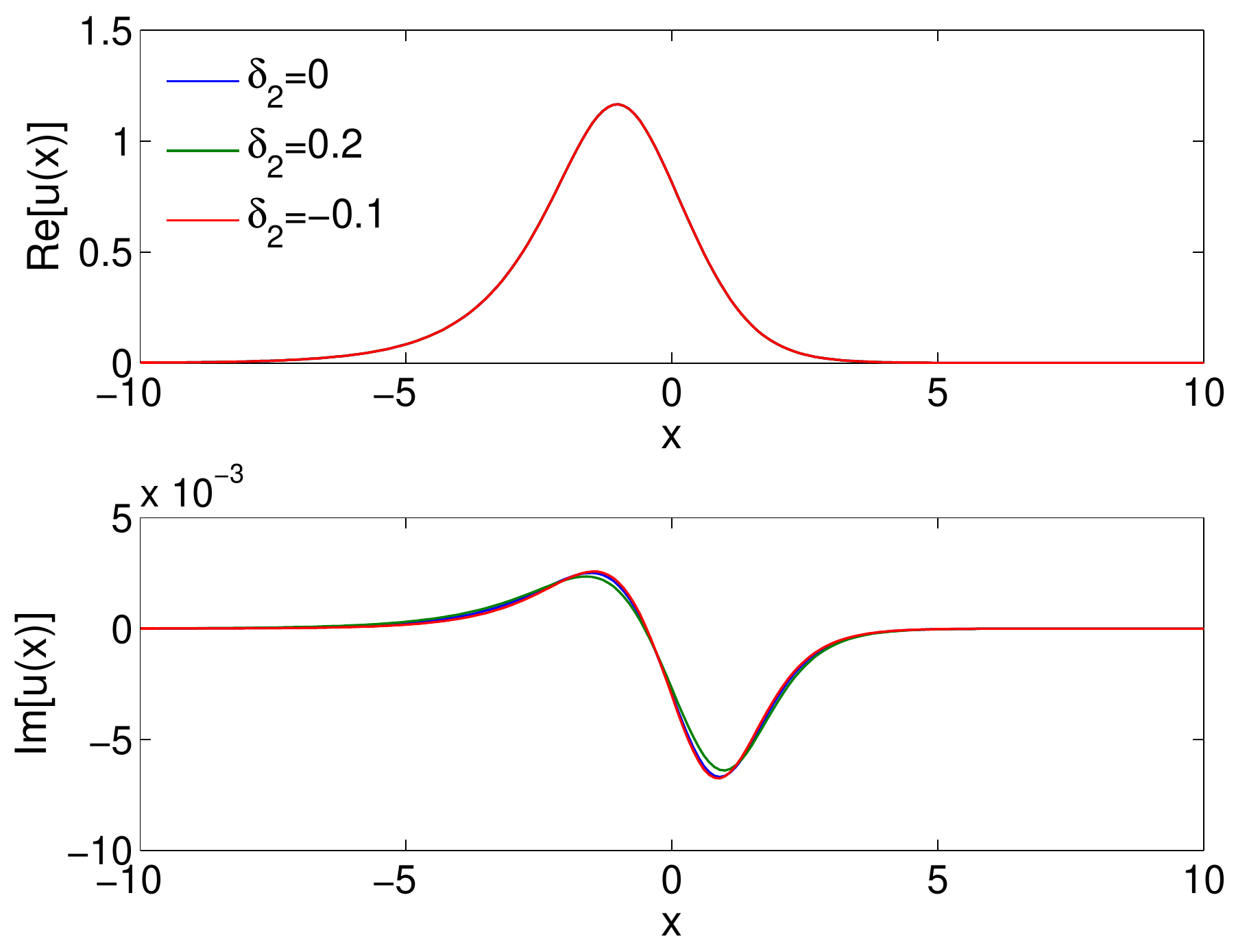} \\
\end{tabular}
\end{center}
\caption{Left panel: Real and imaginary part of the potential $\tilde{V}_2(x)$ for $A_2=1$, and $k_2=0.05$;
  the green line corresponds to the imaginary part for $\delta_2=0$, whereas
  the red (black) line corresponds to the imaginary part for
  $\delta_2=0.2$ ($\delta_2=-0.1$).
Right panels:
optimized beam profiles (real and imaginary parts) for $A_2=1$, and $k_2=0.01$;
the blue line corresponds to $\delta_2=0$,
and the green (red) line corresponds to $\delta_2=0.2$ ($\delta_2=-0.1$).}
\label{fig:potential2}
\end{figure}

Figures~\ref{fig:potential1} and \ref{fig:potential2} show the potential profile for two different ($x_d$,$\delta_1$) and ($k_2$,$\delta_2$)
parameter sets. These figures also show the profile of the
waveforms minimizing  $||F[u]||$
for such potentials, which
will be considered further in what follows.
These beam profiles will be referred to as ``optimized'' in the
sense of the above minimization.
In particular, their real
part is nodeless, while their imaginary part features a zero crossing.
Naturally, the profiles are asymmetric mirroring the lack of
definite parity of the potentials' real and imaginary part.
It is interesting to see that, despite the breaking of both the $\mathcal{PT}$-symmetry
and the violation of conditions such as the one in Eq.~(\ref{eq:Yang}),
there still exist spatially asymmetric structures
almost satisfying the equations of motion.
This naturally poses the question of the dynamical implications of
such profiles in the evolution problem of Eq.~(\ref{eq:dyn}), as we will see below.

\subsection{Dynamics}

\begin{figure}
\begin{center}
\begin{tabular}{cc}
\includegraphics[width=.45\textwidth]{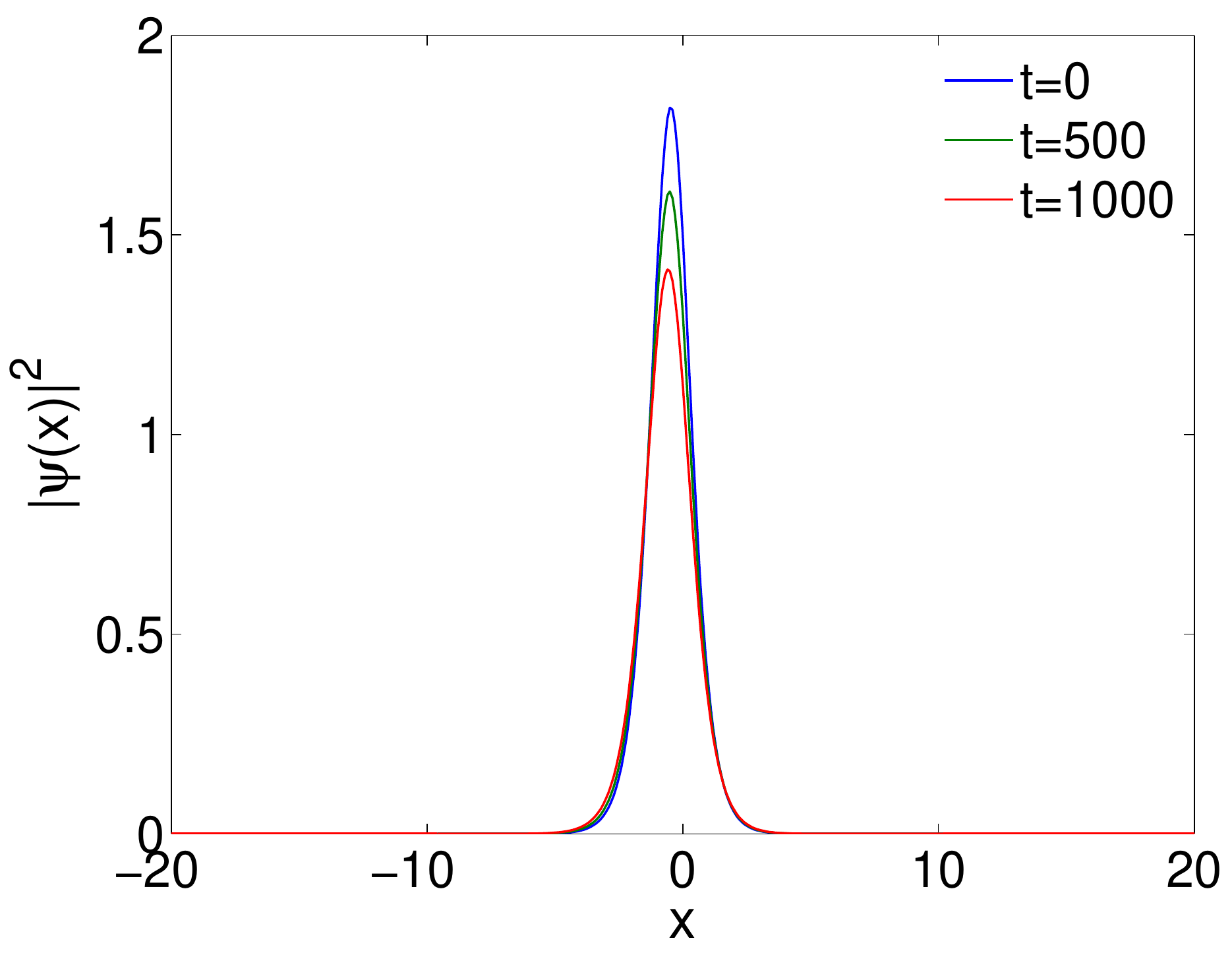} &
\includegraphics[width=.45\textwidth]{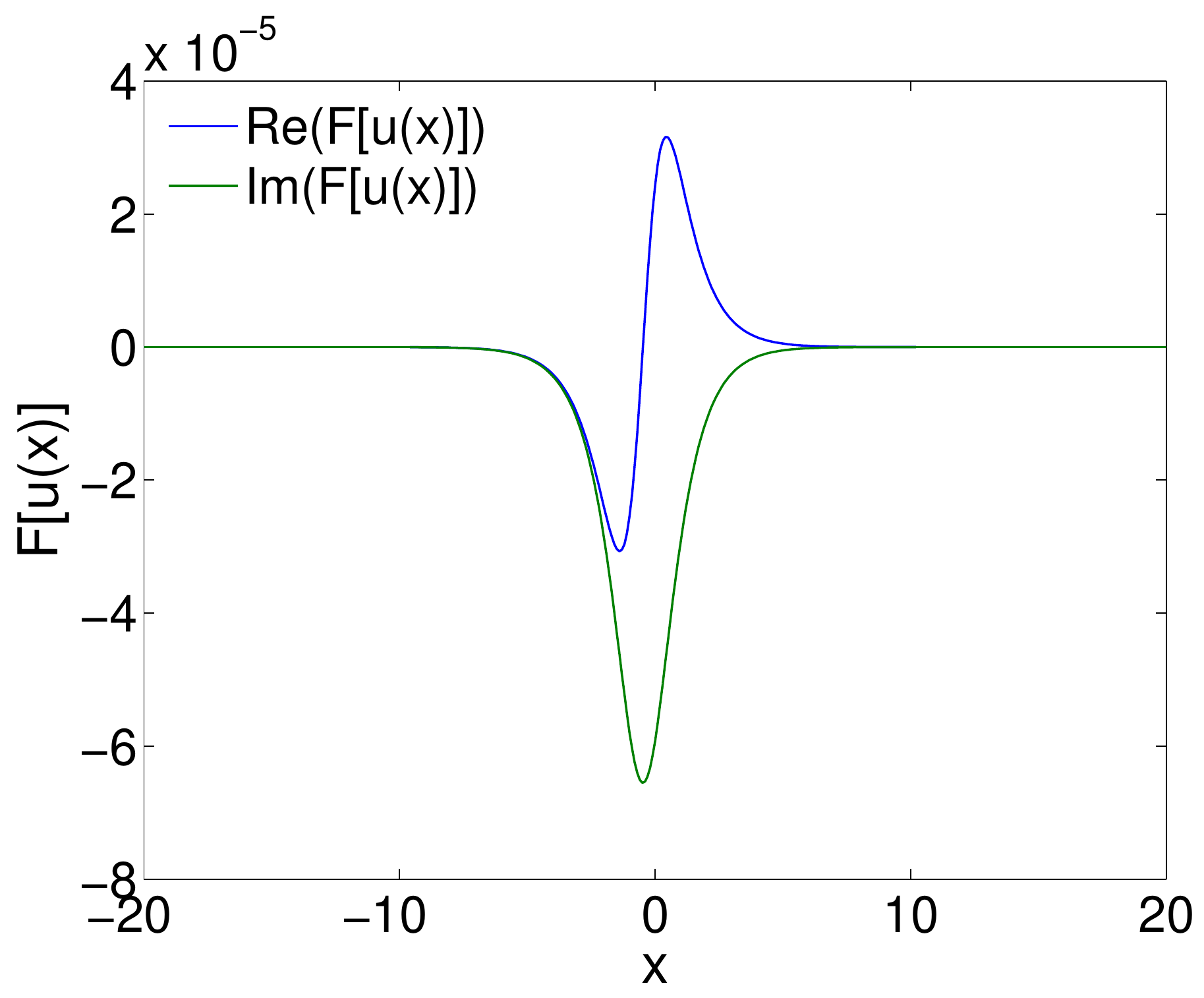} \\
\includegraphics[width=.45\textwidth]{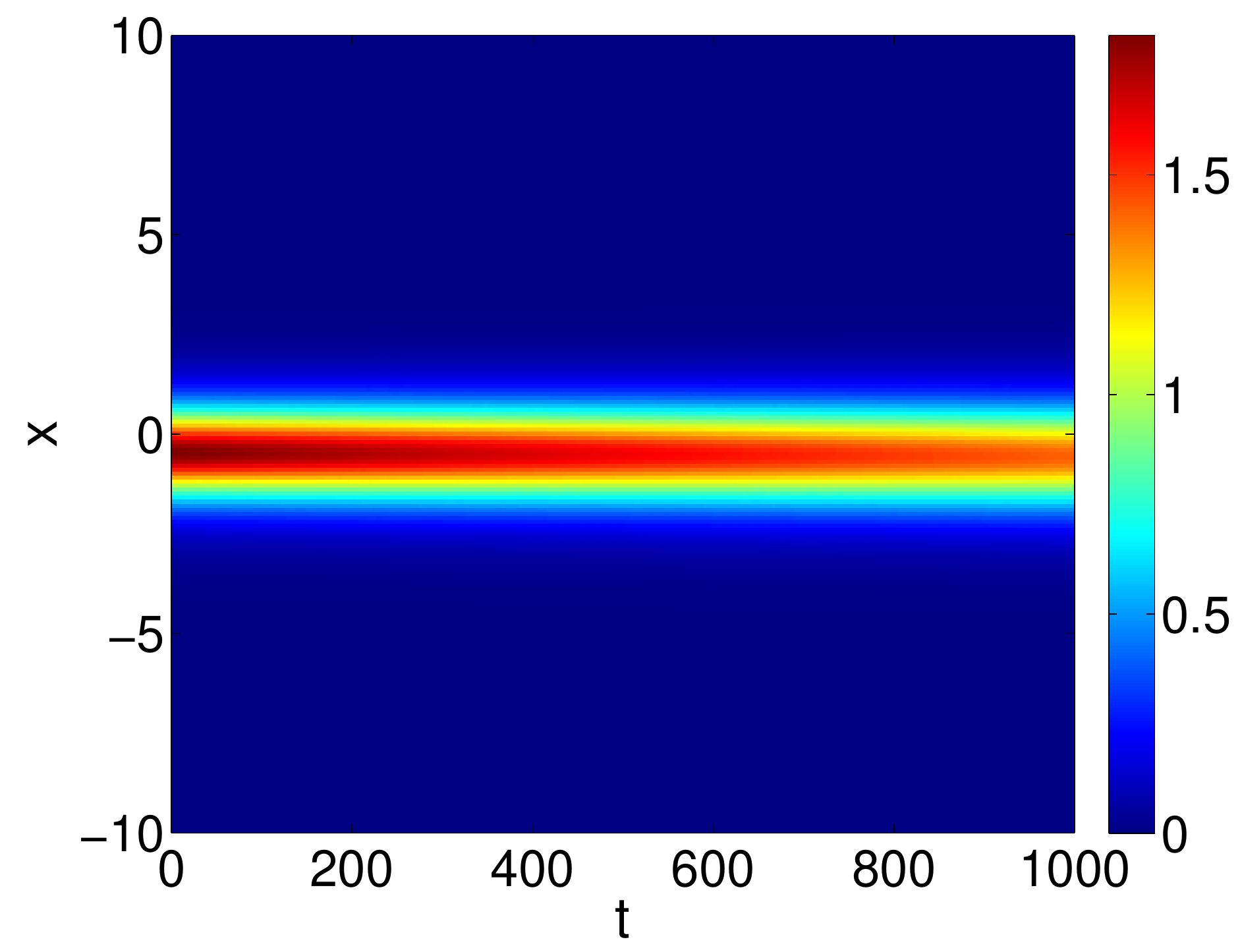} &
\includegraphics[width=.45\textwidth]{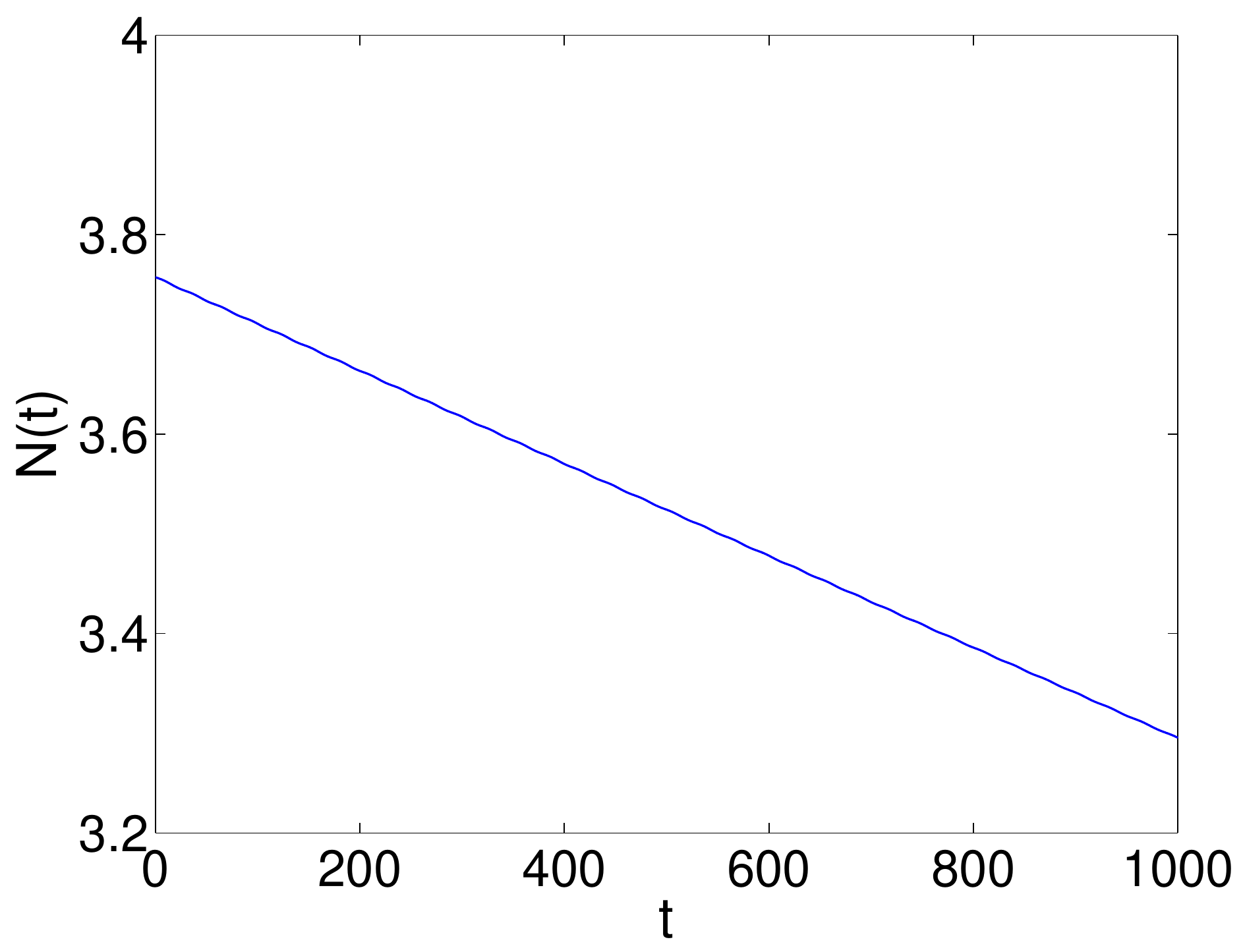} \\
\end{tabular}
\end{center}
\caption{
Optimized beam dynamics in the potential $\tilde{V}_1(x)$ for $A_1=0.1$, $k_1=1/2$, $x_d=1$ and $\delta_1=0$. The top left panel shows the density profile at $t=0$, $t=500$ and $t=1000$,
while the top left panel shows the real and imaginary part of $F[u]$.
The bottom left panel shows the space-time contour plot of the density evolution,
and the bottom right panel shows the evolution of the norm $N(t)$.
The values of diagnostic quantities are $\lambda=-3.63\times10^{-3}$ and $\sigma=-1.07\times10^{-4}$.}
\label{fig:dynamics1}
\end{figure}

\begin{figure}
\begin{center}
\begin{tabular}{cc}
\includegraphics[width=.45\textwidth]{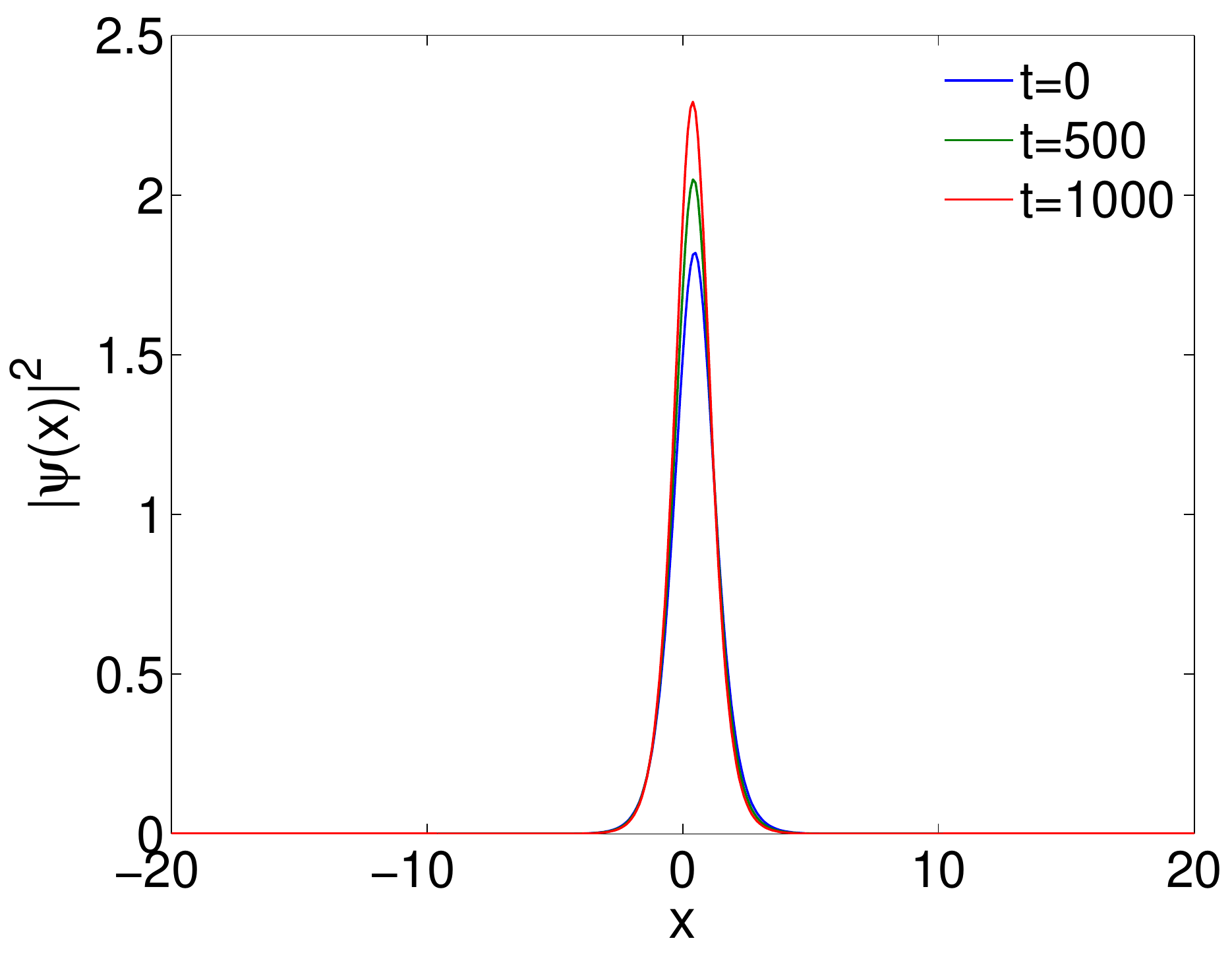} &
\includegraphics[width=.45\textwidth]{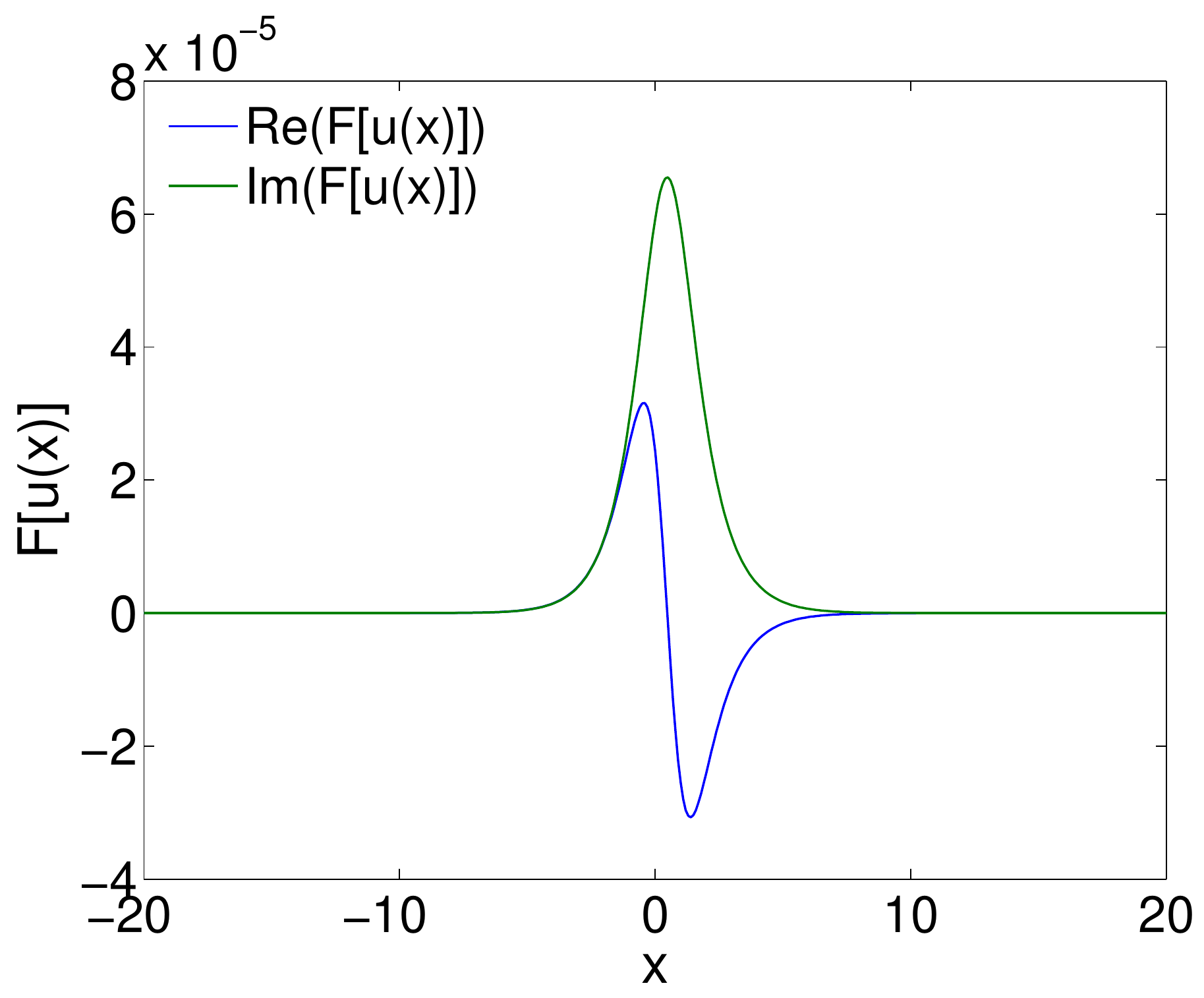} \\
\includegraphics[width=.45\textwidth]{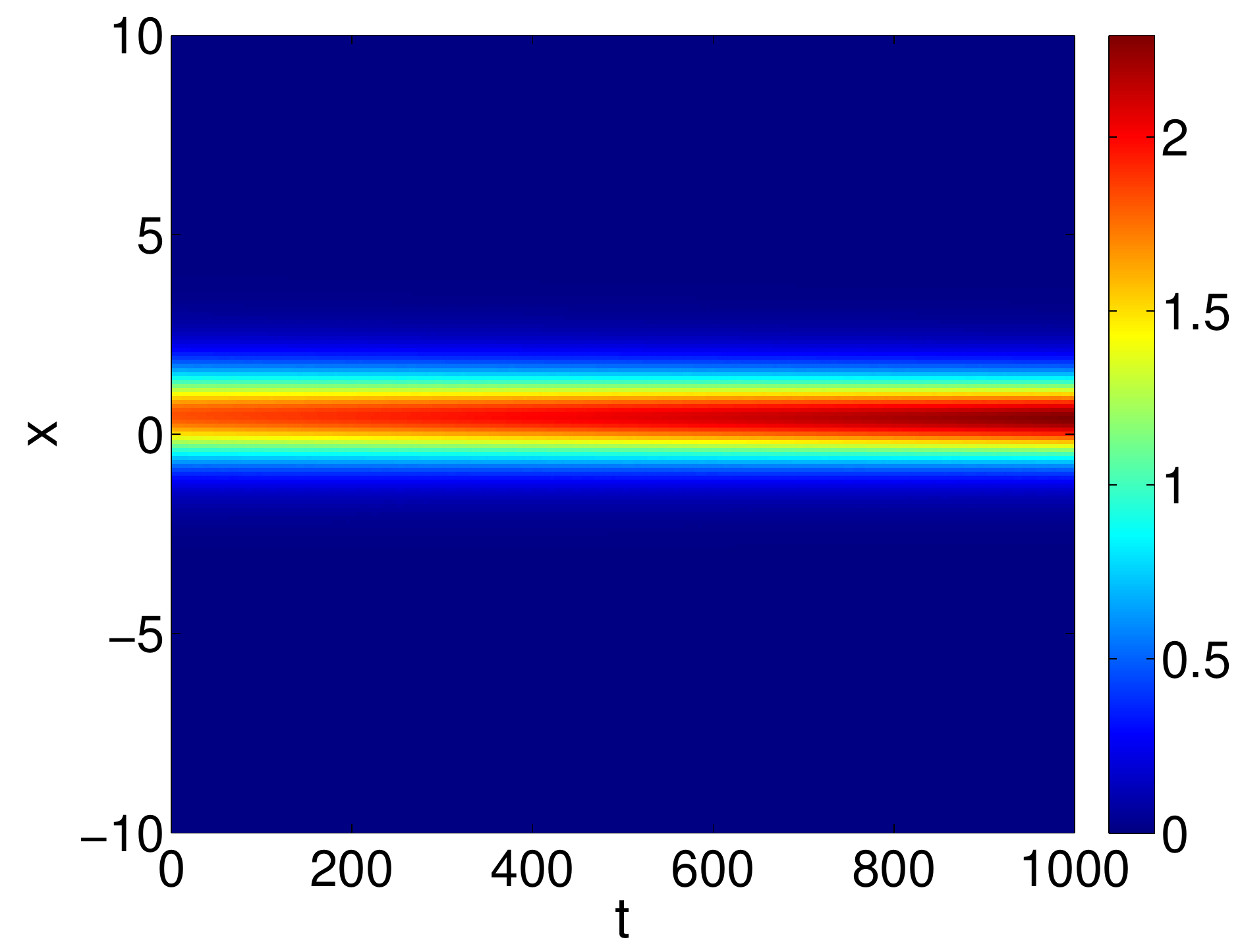} &
\includegraphics[width=.45\textwidth]{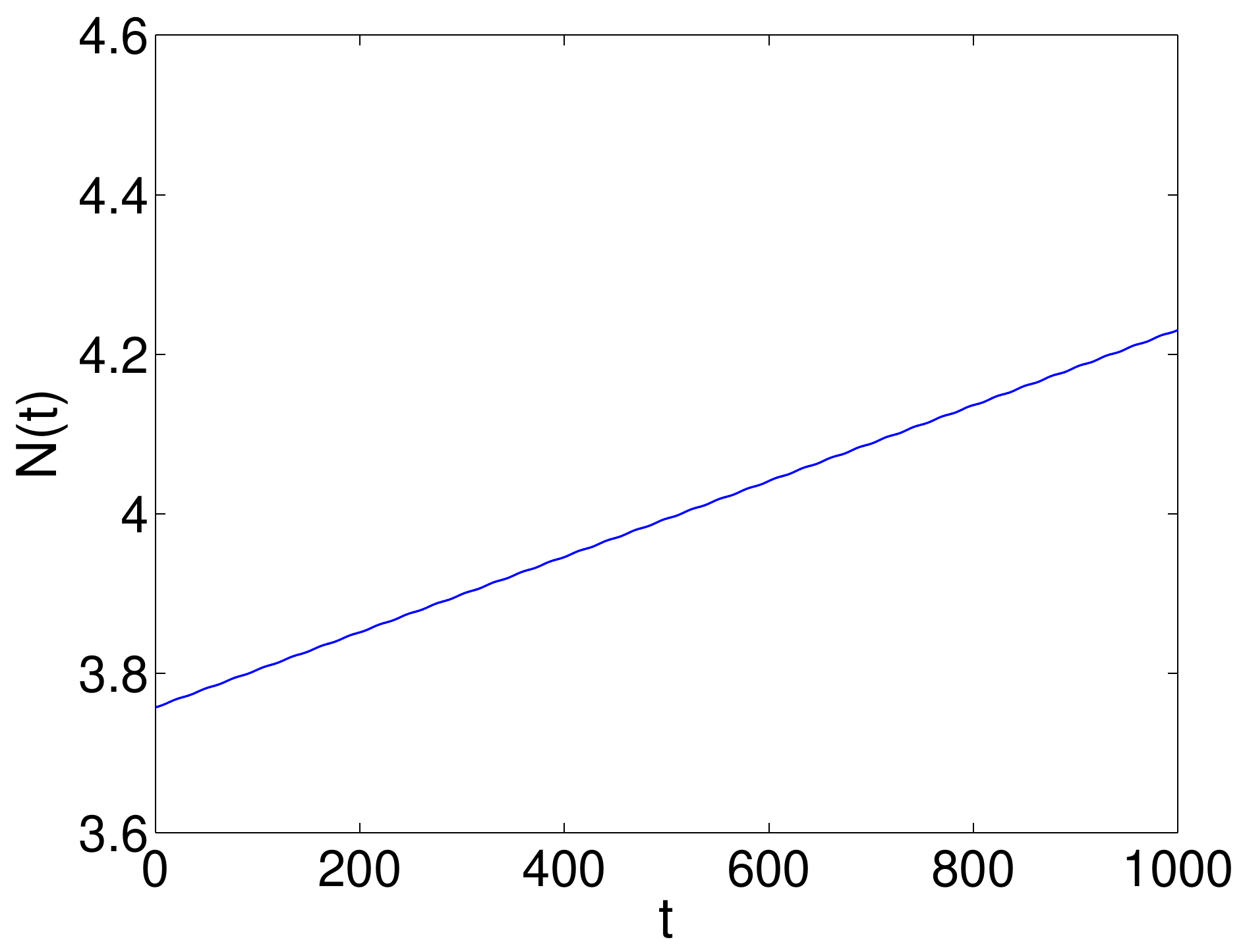} \\
\end{tabular}
\end{center}
\caption{
Same as Fig.~\ref{fig:dynamics1}, but for $x_d=-1$. The values of diagnostic quantities are $\lambda=3.63\times10^{-3}$ and $\sigma=1.07\times10^{-4}$.}
\label{fig:dynamics2}
\end{figure}

\begin{figure}
\begin{center}
\begin{tabular}{cc}
\includegraphics[width=.45\textwidth]{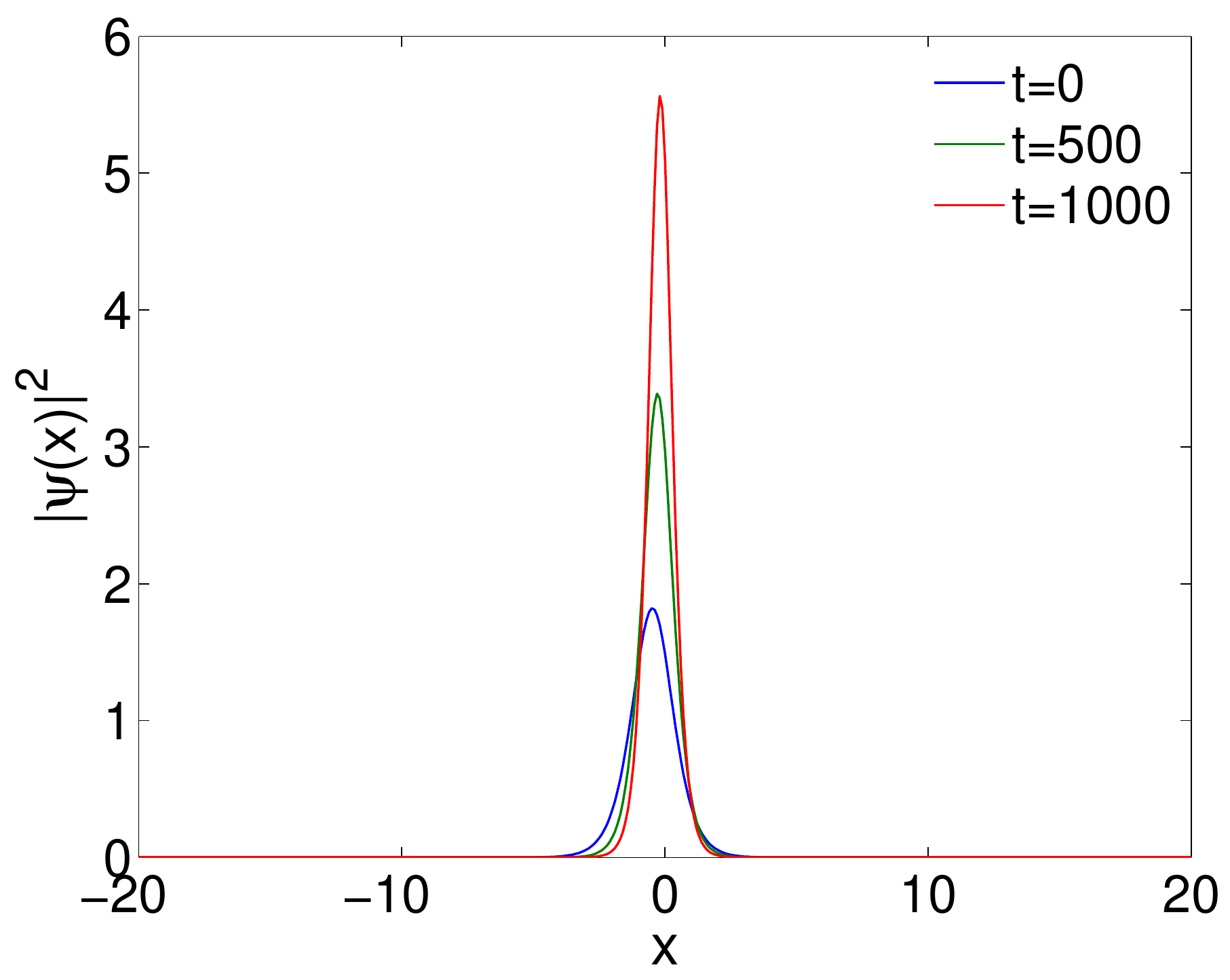} &
\includegraphics[width=.45\textwidth]{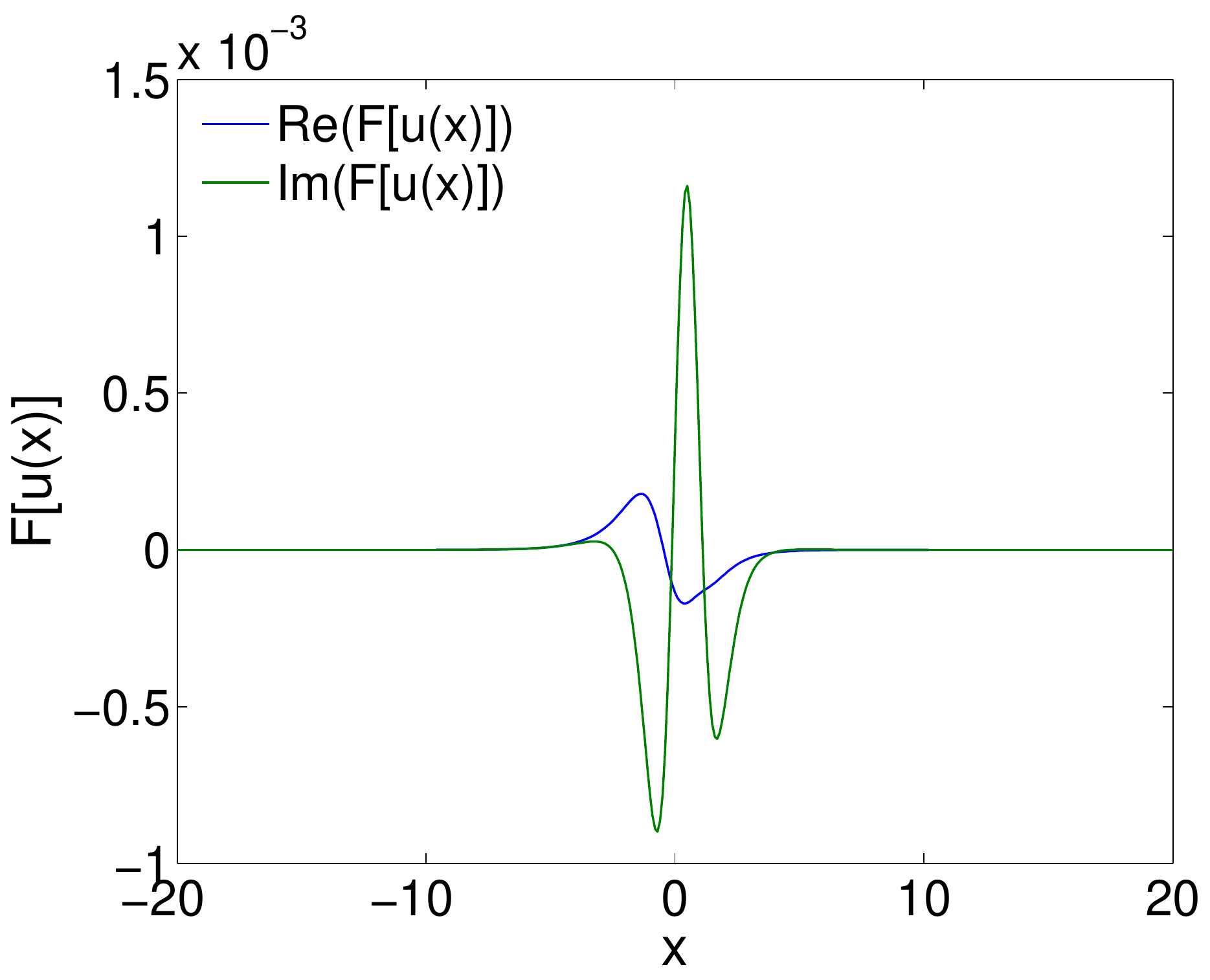} \\
\includegraphics[width=.45\textwidth]{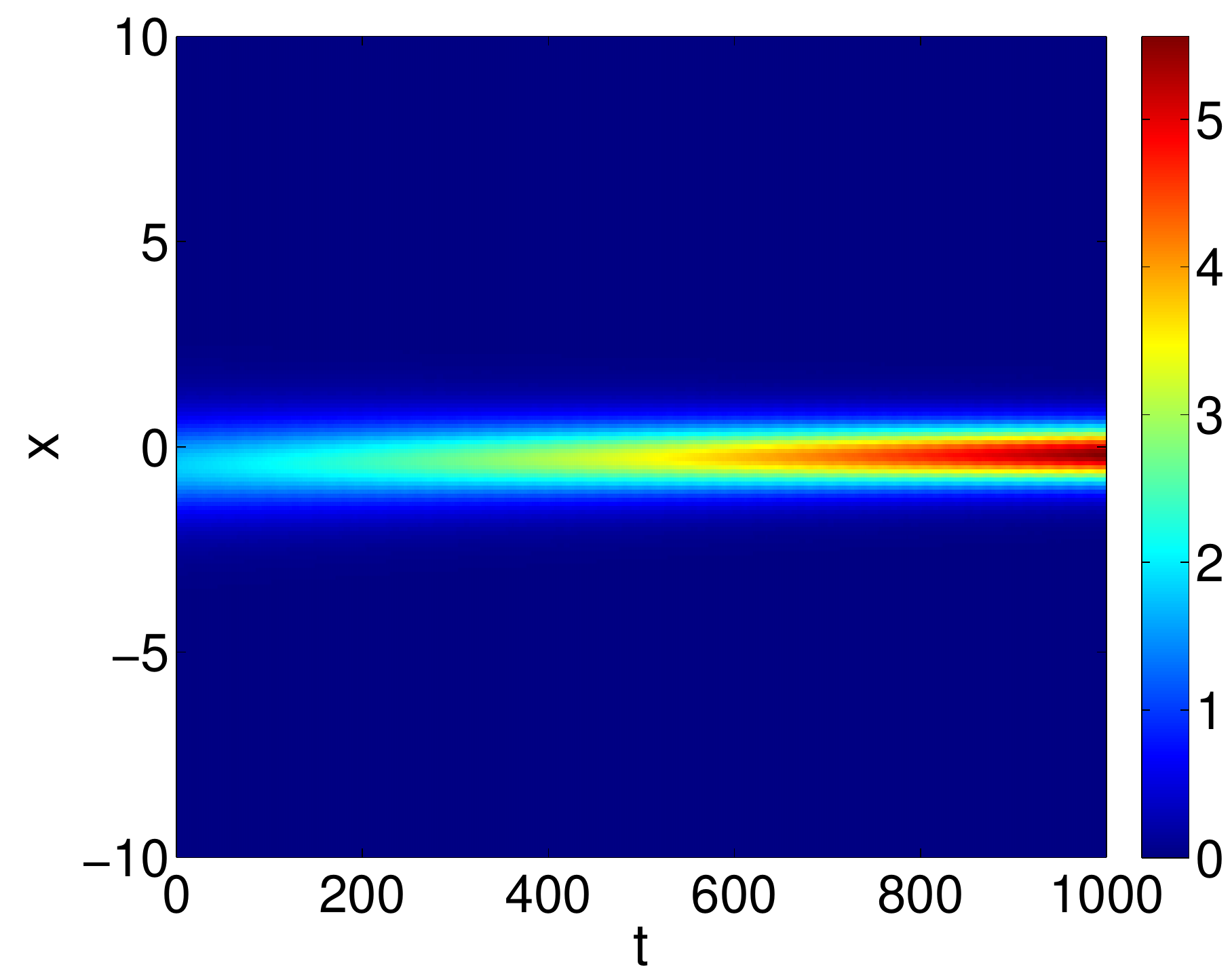} &
\includegraphics[width=.45\textwidth]{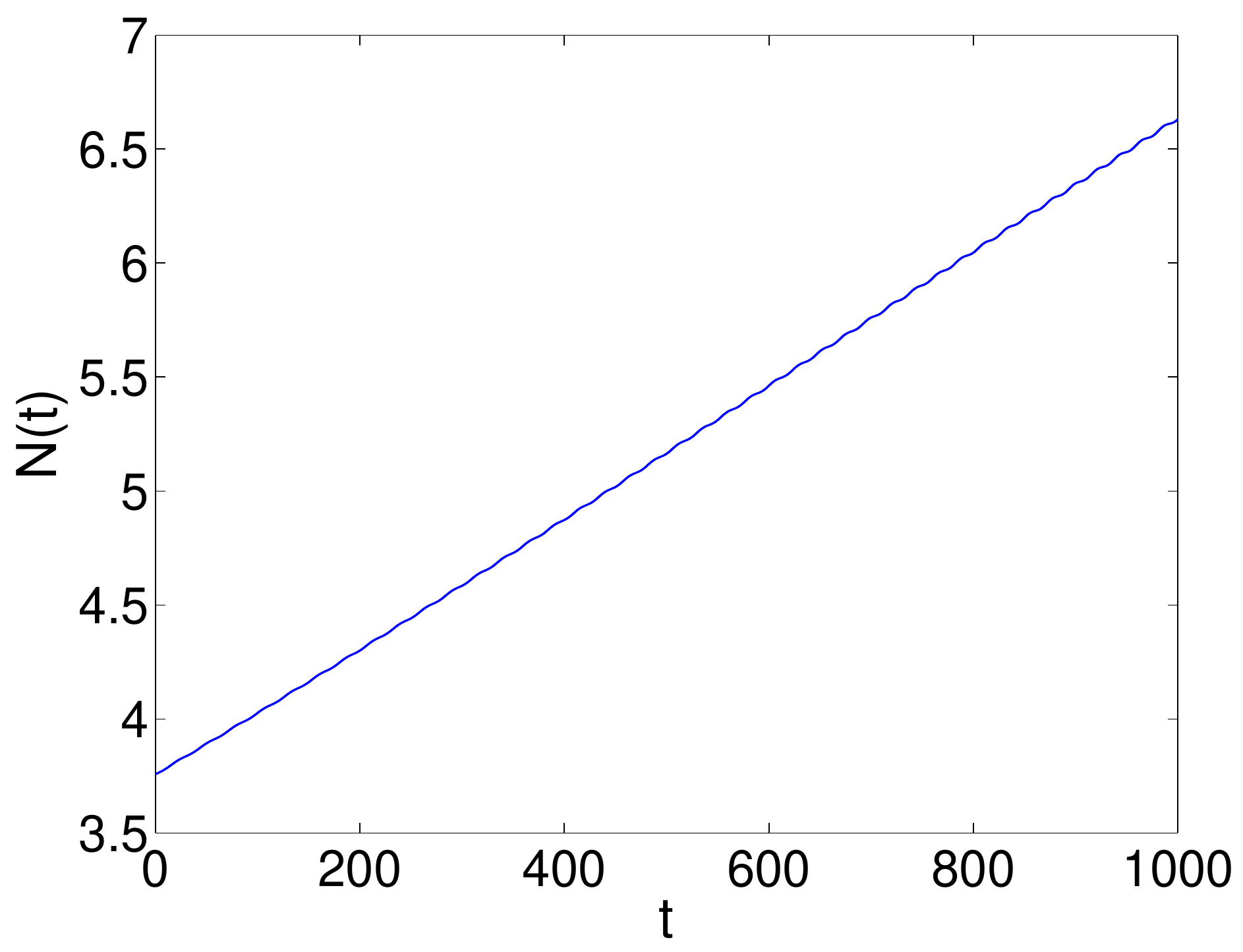} \\
\end{tabular}
\end{center}
\caption{
Same as Fig.~\ref{fig:dynamics1}, but for $\delta_1=0.1$. The values of diagnostic quantities are $\lambda=2.07\times10^{-2}$ and $\sigma=6.07\times10^{-4}$.}
\label{fig:dynamics3}
\end{figure}

\begin{figure}
\begin{center}
\begin{tabular}{cc}
\includegraphics[width=.45\textwidth]{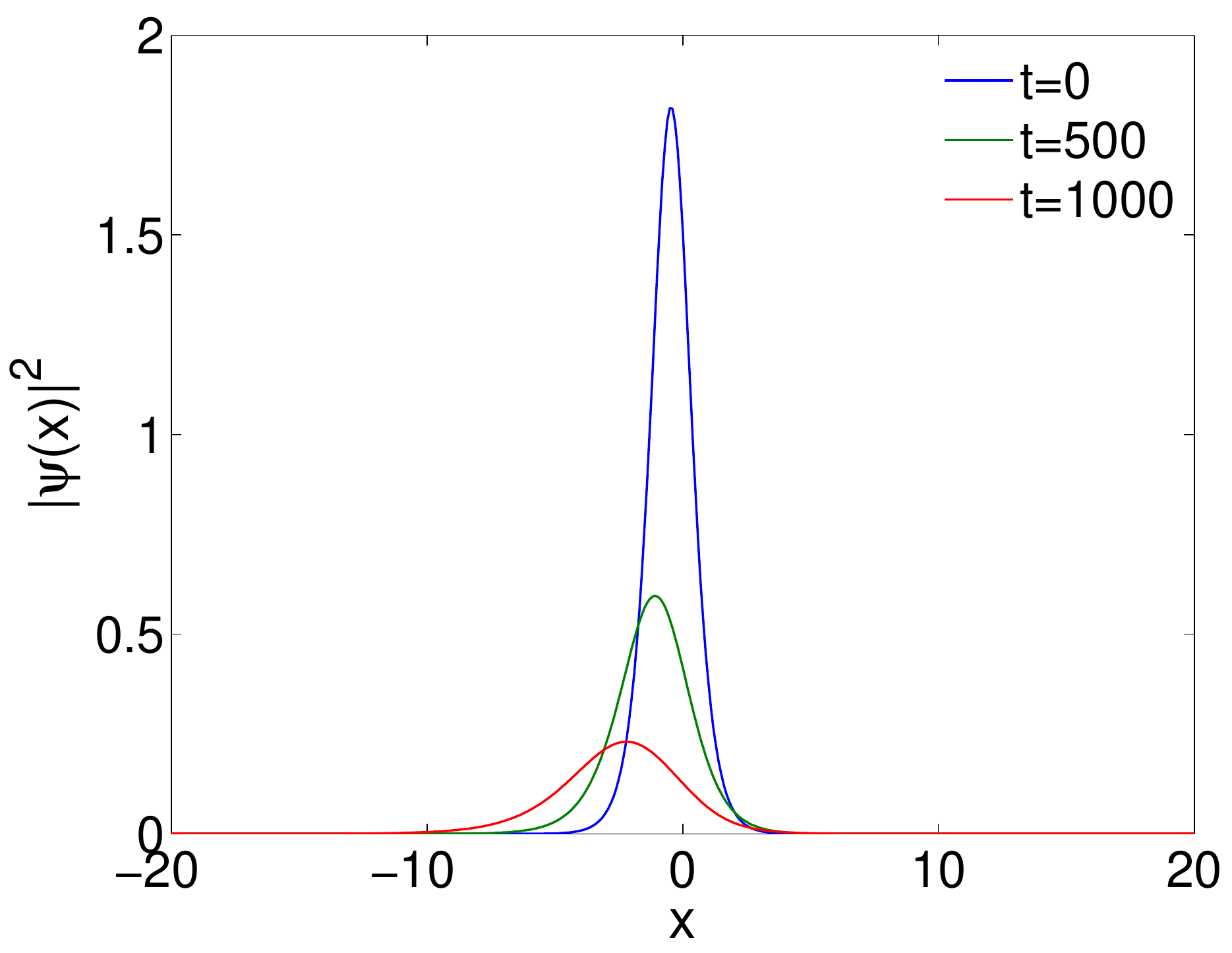} &
\includegraphics[width=.45\textwidth]{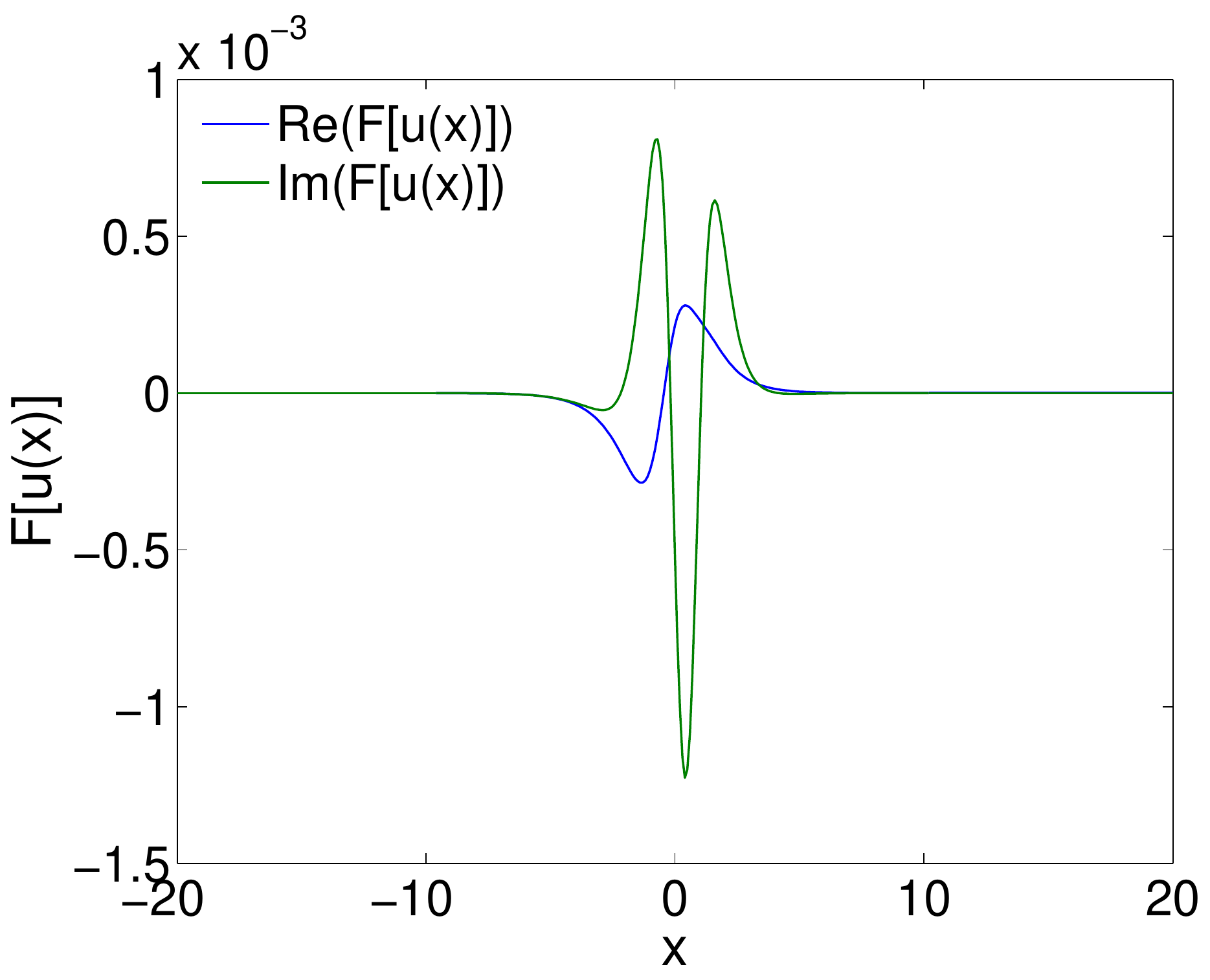} \\
\includegraphics[width=.45\textwidth]{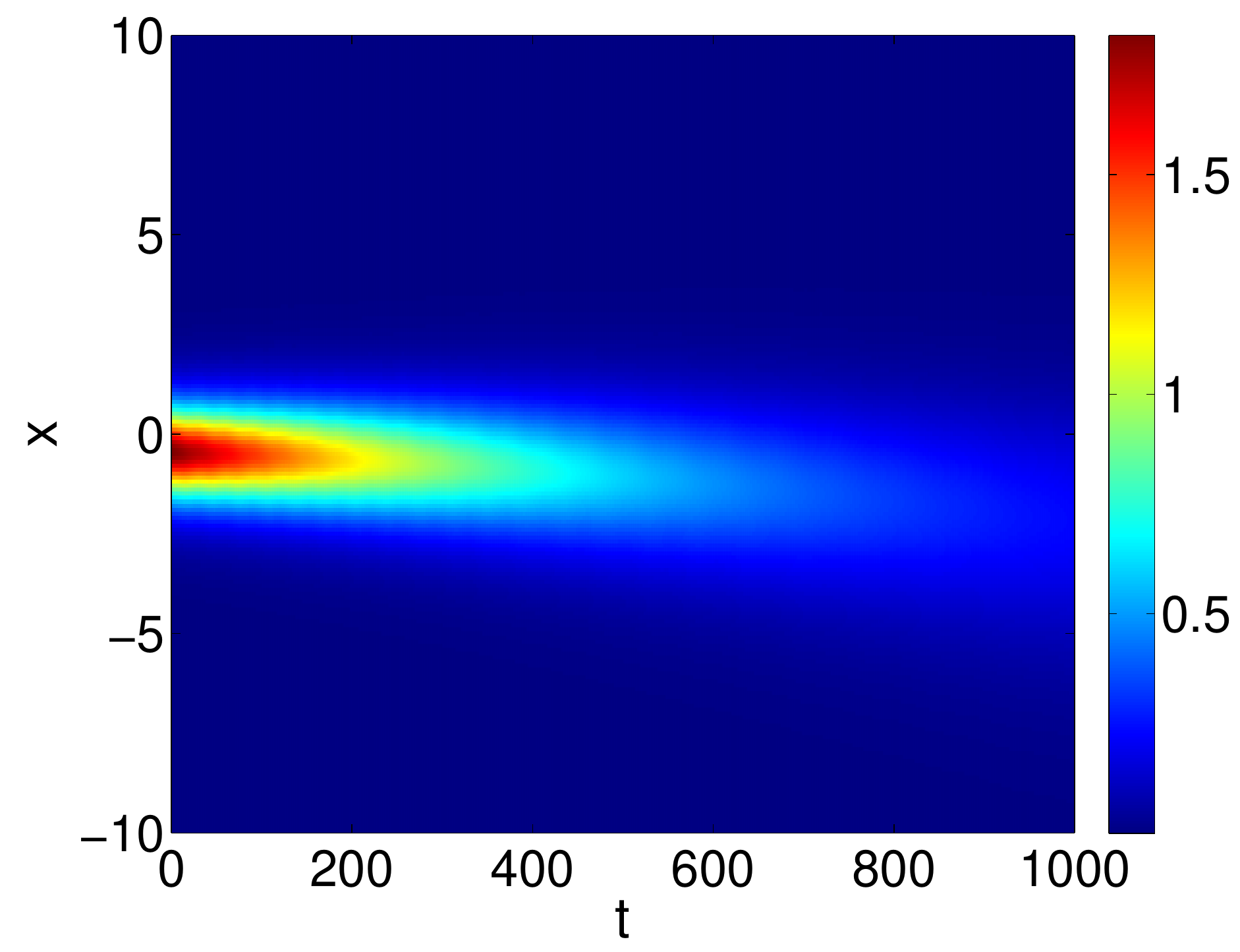} &
\includegraphics[width=.45\textwidth]{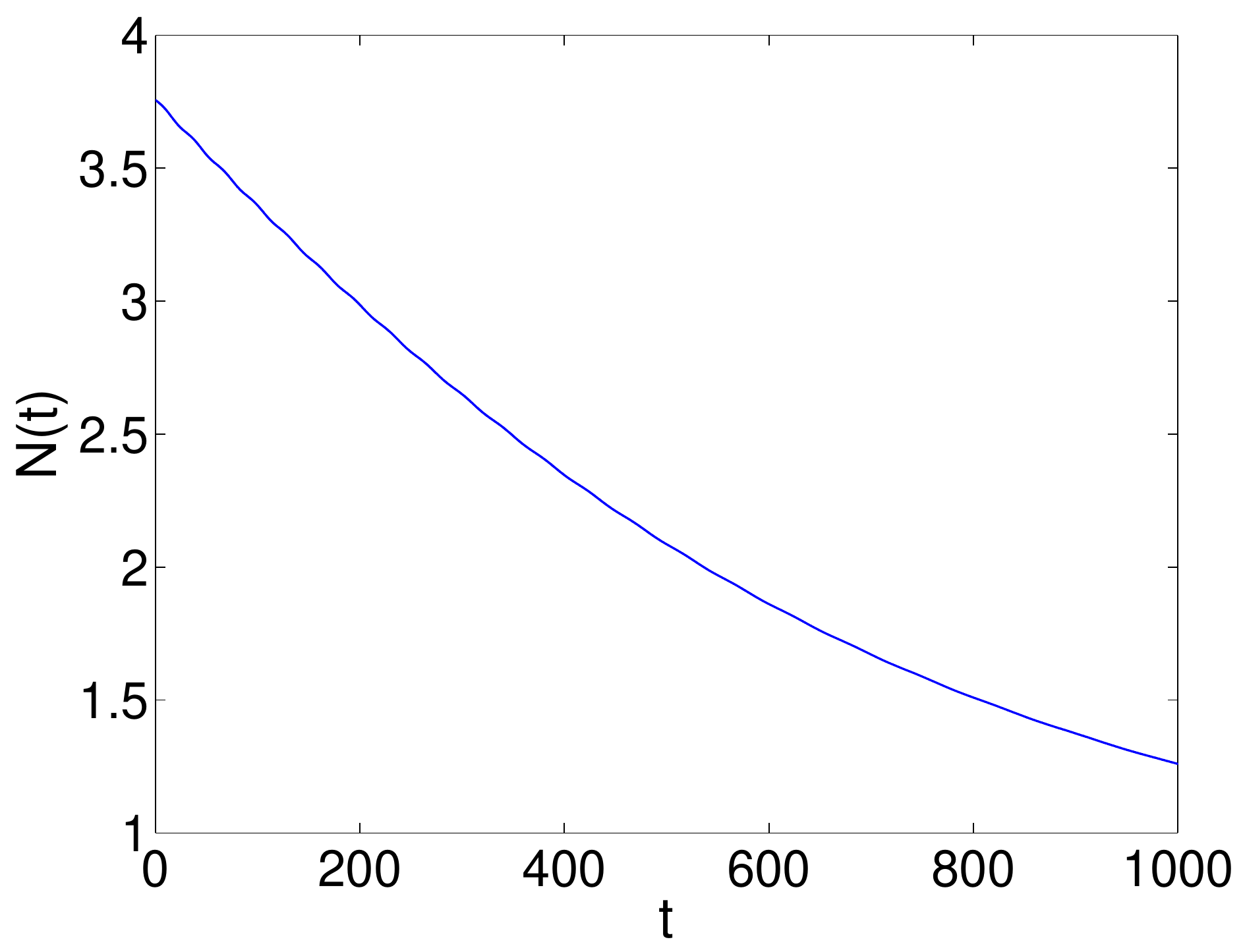} \\
\end{tabular}
\end{center}
\caption{
Same as Fig. \ref{fig:dynamics1} but for $\delta_1=-0.1$. The values of diagnostic
quantities are $\lambda=-3.15\times10^{-2}$ and $\sigma=-9.40\times10^{-4}$.}
\label{fig:dynamics4}
\end{figure}

We now analyze the dynamics of several case examples for the NLS equation with potential $\tilde{V}_1(x)$, using as initial condition the optimized beam profiles found by the Levenberg-Marquardt algorithm. Figures \ref{fig:dynamics1} and \ref{fig:dynamics2} show the outcome
of the simulations for $x_d=1$ and $x_d=-1$, respectively, when $\delta_1=0$ is fixed;
on the other hand, Figs.~\ref{fig:dynamics3} and \ref{fig:dynamics4} correspond, respectively,
to $\delta_1=-0.1$ and $\delta_1=0.1$, when $x_d=1$ is fixed.
In these figures, we show the density $|\psi(x)|^2$ at different
time instants (top left), the real and imaginary part of $F[u]$ (top right),
a space-time contour plot of the evolution of the localized beam density $|\psi(x,t)|^2$
(bottom left), and the (squared) $L^2$-norm (power/mass in optics/atomic physics), $N(t)$ (bottom right),
defined as

\begin{equation}
    N(t)=\int |\psi(x,t)|^2 \mathrm{d}x.
\end{equation}

%One can observe a clear correlation between the qualitative shape of $F[u]$ and the growing/decaying character of the dynamics. In other words, in the growing case, the real part is similar to $-\mathrm{sech}(x)\tanh(x)$ whereas the imaginary part is similar to $\mathrm{sech}(x)$; for the decaying case, the signs are reversed.
%This correlation  (of the real and imaginary parts of $F[u]$ with those
%of $\tilde{V}_1$) is one that we have observed in numerous other case examples.

{One can observe a clear correlation between the qualitative shape of $\mathrm{Im}\{F[u]\}$ and the growing/decaying character of the dynamics. In other words, in the growing case, this quantity is predominantly positive, whereas for the decaying case, it is predominantly negative.}

\begin{figure}
\begin{center}
\begin{tabular}{cc}
\includegraphics[width=.45\textwidth]{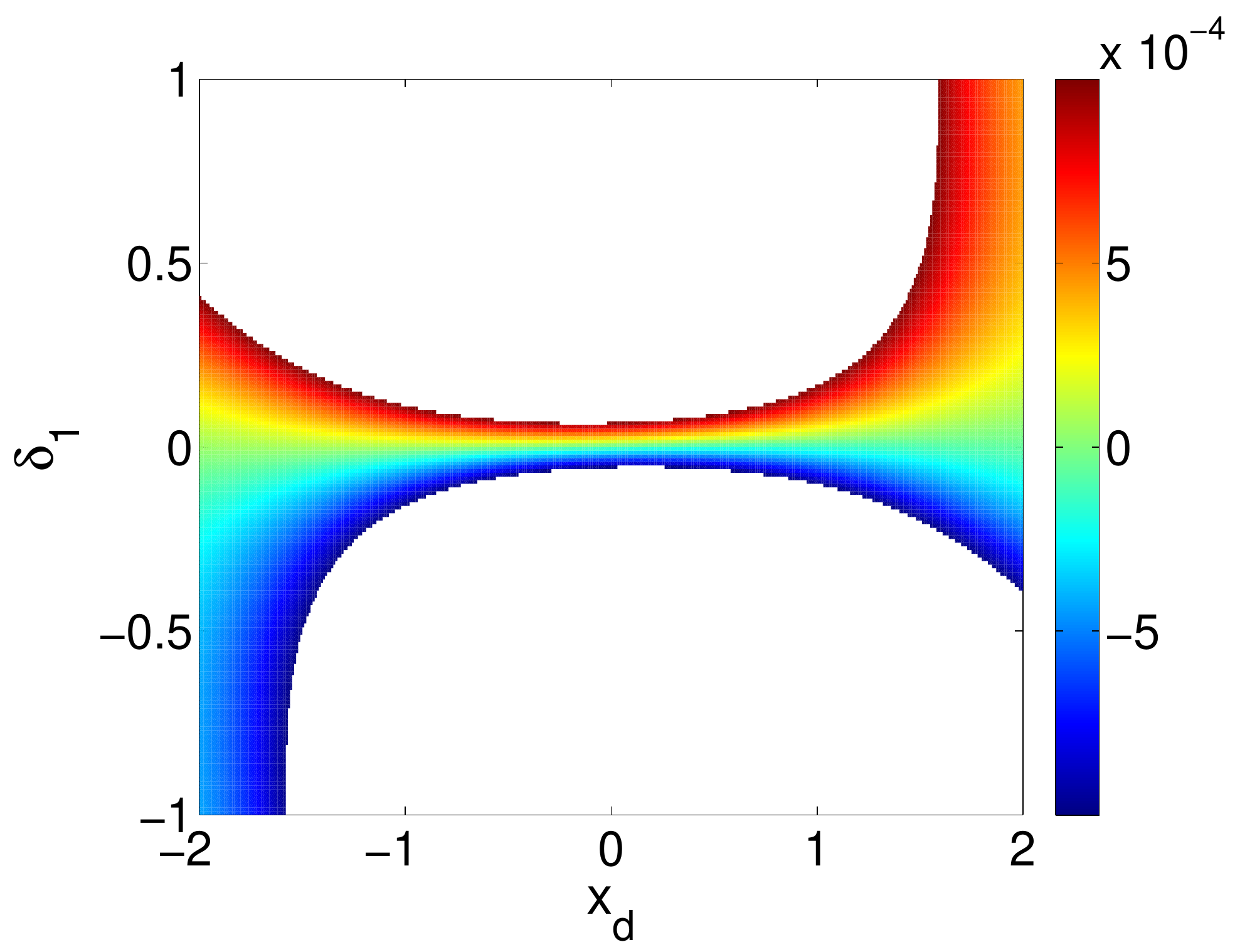} &
\includegraphics[width=.45\textwidth]{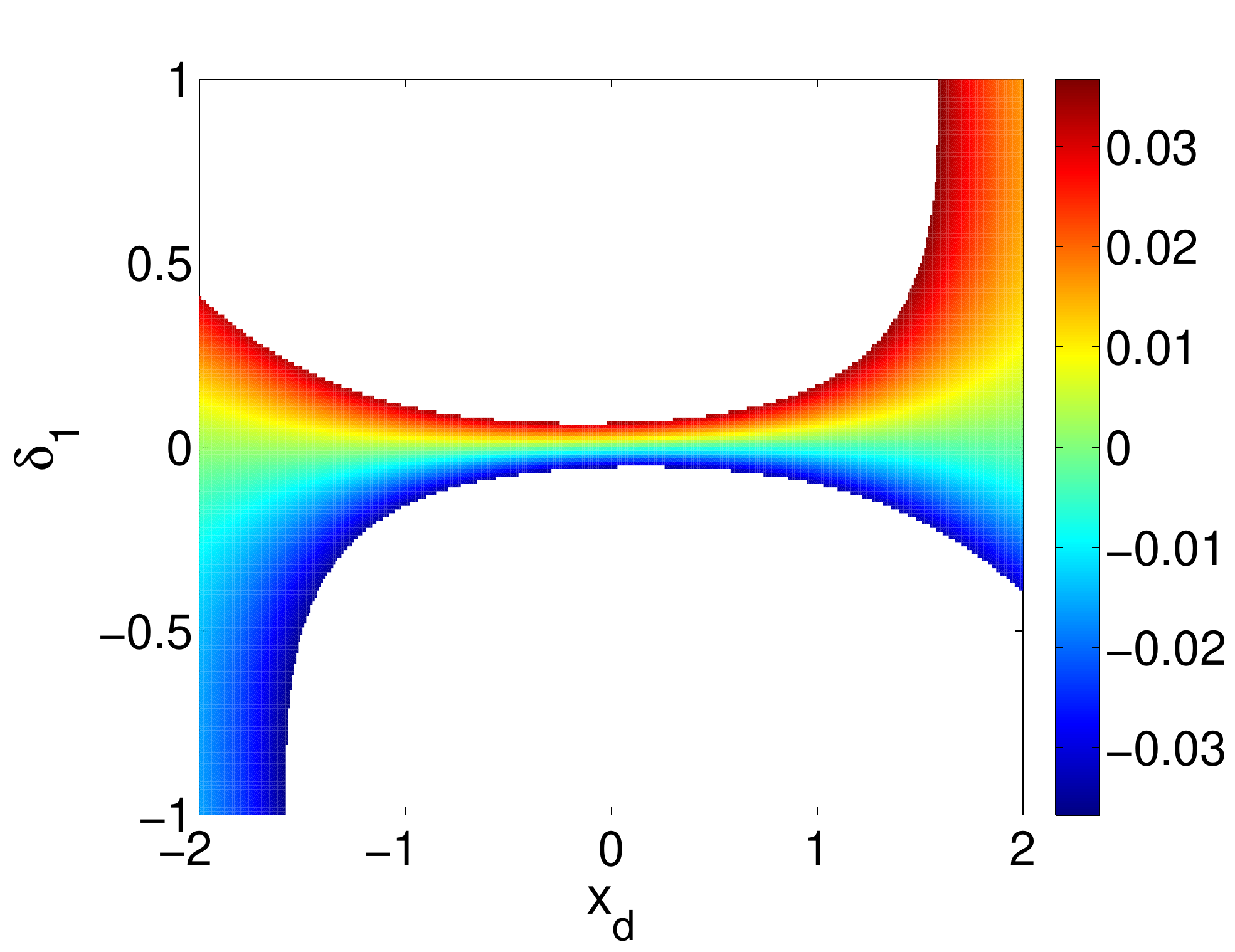} \\
\multicolumn{2}{c}{\includegraphics[width=.45\textwidth]{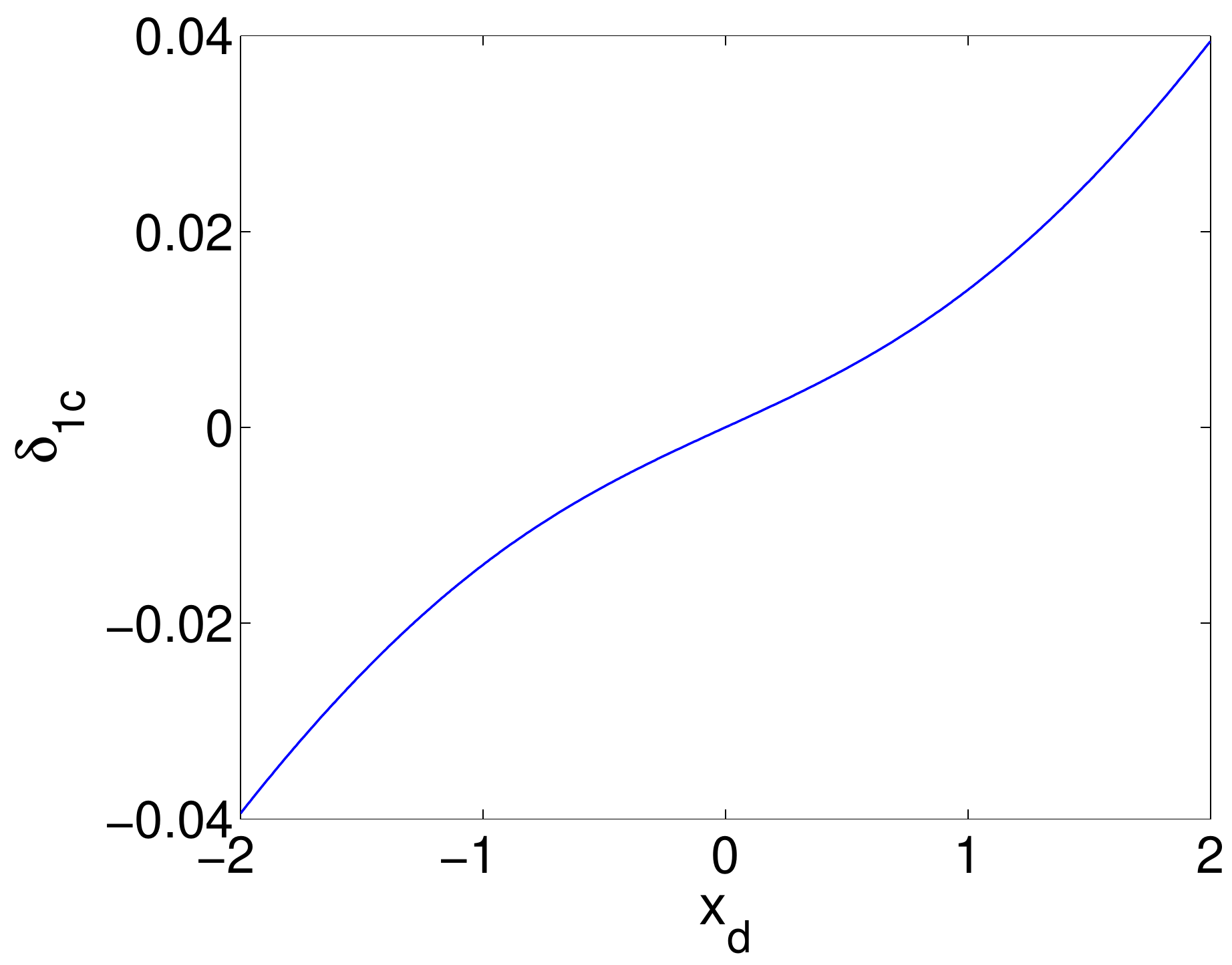}}\\
\end{tabular}
\end{center}
\caption{Top panels: Dependence of diagnostic
  quantities $\sigma$ (left) and $\lambda$ (right) as a function of $x_d$ and $\delta_1$ for the potential $\tilde{V}_1(x)$ with $A_1=0.1$ and $k_1=1/2$.
  The bottom panel depicts the curve $\delta_{1c}(x_d)$ at which both $\sigma$ and $\lambda$ vanish.
}
\label{fig:plane1}
\end{figure}

Moreover, it seems that a larger growth rate (i.e., a faster
increase or decrease of $N$) is associated to a larger $||F[u]||$. In order to
showcase this fact, we have depicted in Fig.~\ref{fig:plane1} the dependence of
diagnostic quantities $\lambda$ and $\sigma$, that we have
accordingly defined as
\begin{equation}\label{eq:lambda}
    \lambda=\frac{\mathrm{d}N}{\mathrm{d}t}\bigg{|}_{t=0}.
\end{equation}
\begin{equation}\label{eq:sigma}
    \sigma=S||F[u]||,
\end{equation}
with
\begin{equation*}
    S=\mathrm{sgn}\left\{\int {\rm Im}\{F[u(x)]\}\mathrm{d}x\right\}
\end{equation*}
{The quantity $\sigma$  takes into account both the (minimized) norm of $||F[u]||$ and the form of $\mathrm{Im}\{F[u(x)]\}$ through $S$ --that is, if the imaginary part of $F[u(x)]$ is chiefly positive or negative.}
%The quantity $\sigma$  takes into account both the (minimized) norm of $||F[u]||$ and the form of {$F[u(x)]$}
%%the profile
%through $\mathrm{sgn}(\mathrm{Im}(F[u(x^*)]))$ --that is, if the imaginary part of {$F[u(x)]$}%the profile
%is like a $\mathrm{sech}(x)$ or a $-\mathrm{sech}(x)$.
Notice that the blank regions
    correspond to solutions for which $||F[u]||$ is higher than the prescribed tolerance of $10^{-3}$. On the other hand, $\lambda$ characterizes the
    rate of ``departure'' from the optimized beam profile obtained from
    this minimization procedure.

\begin{figure}
\begin{center}
\begin{tabular}{cc}
\includegraphics[width=.45\textwidth]{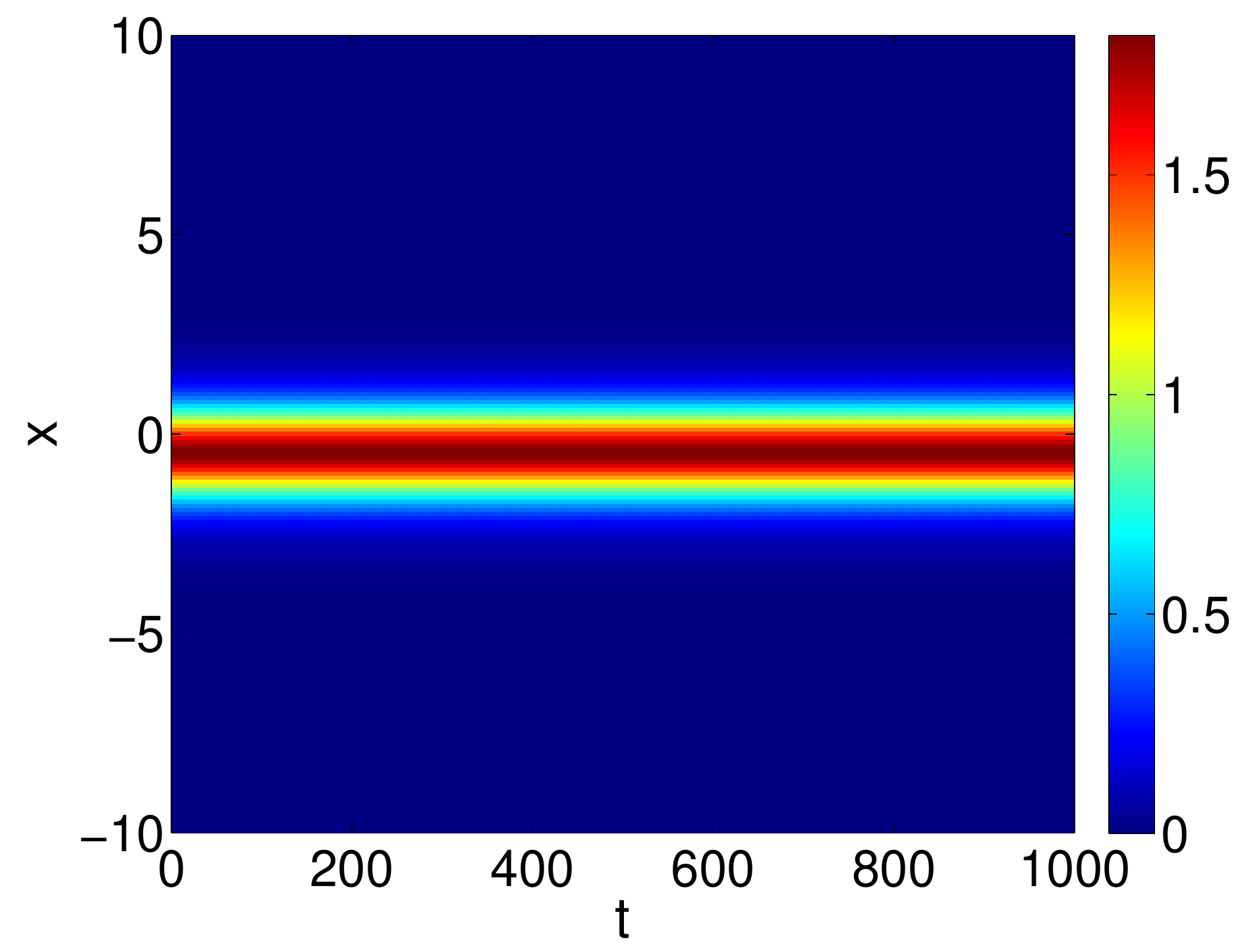} &
\includegraphics[width=.45\textwidth]{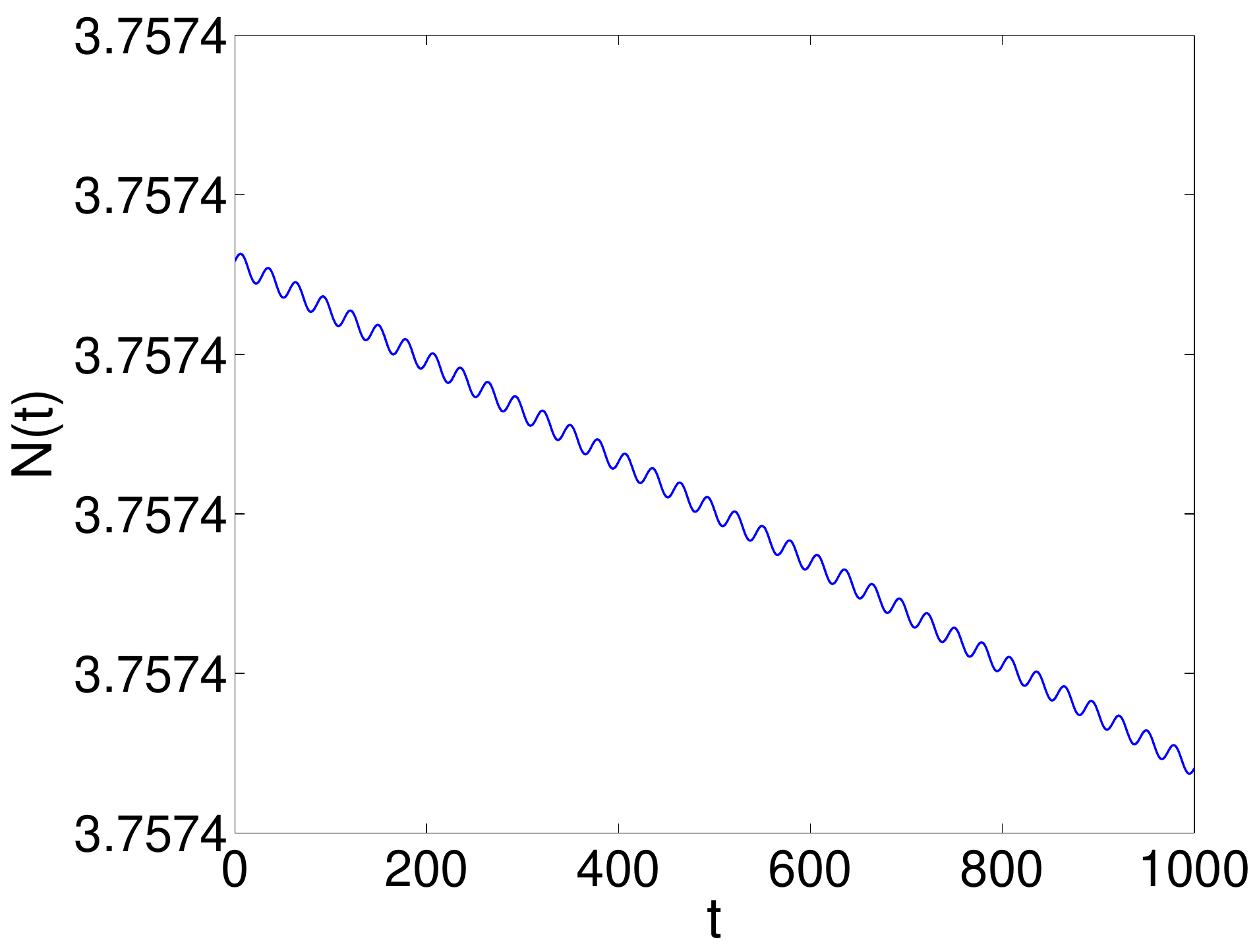} \\
\includegraphics[width=.45\textwidth]{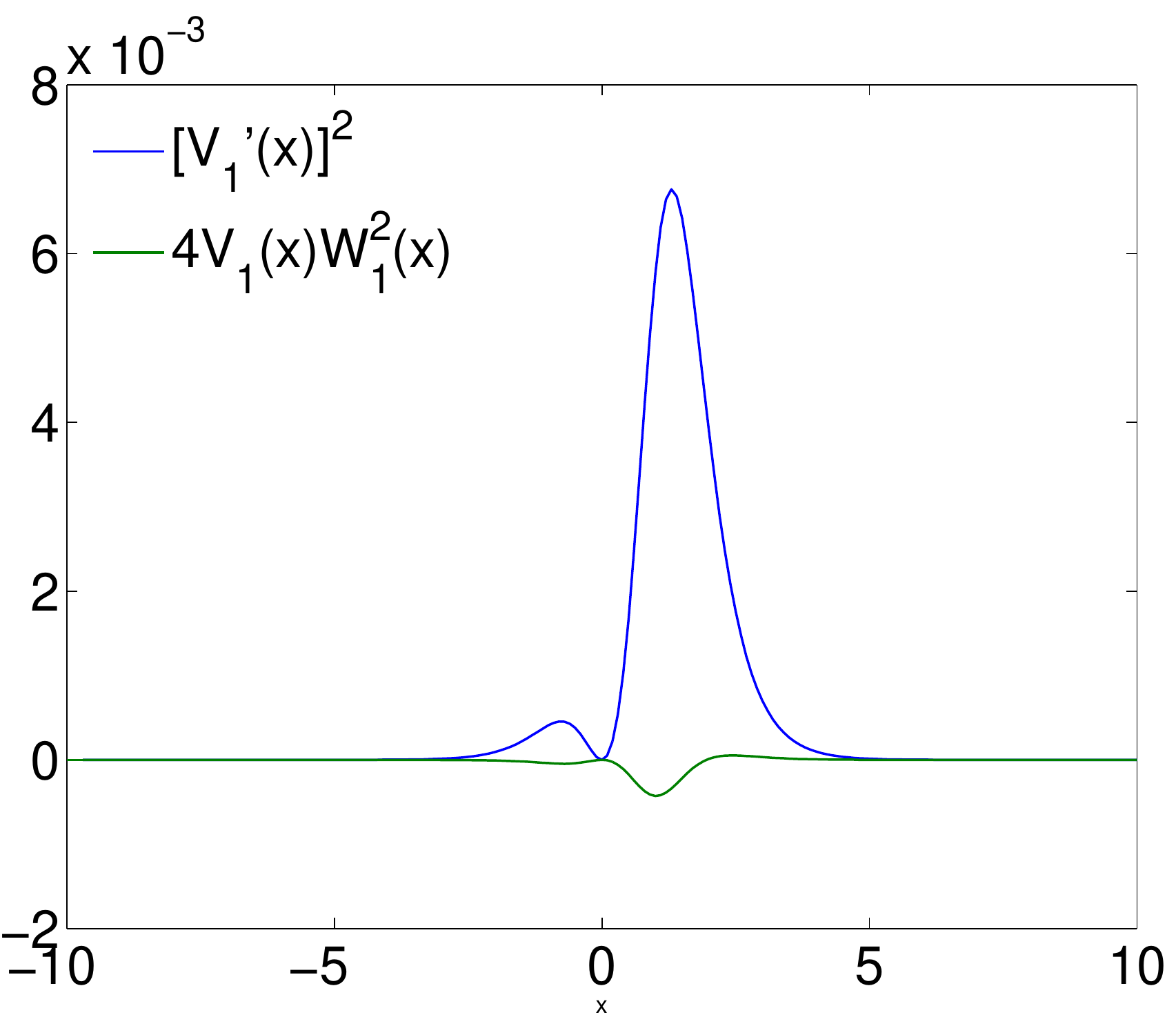} &
\includegraphics[width=.45\textwidth]{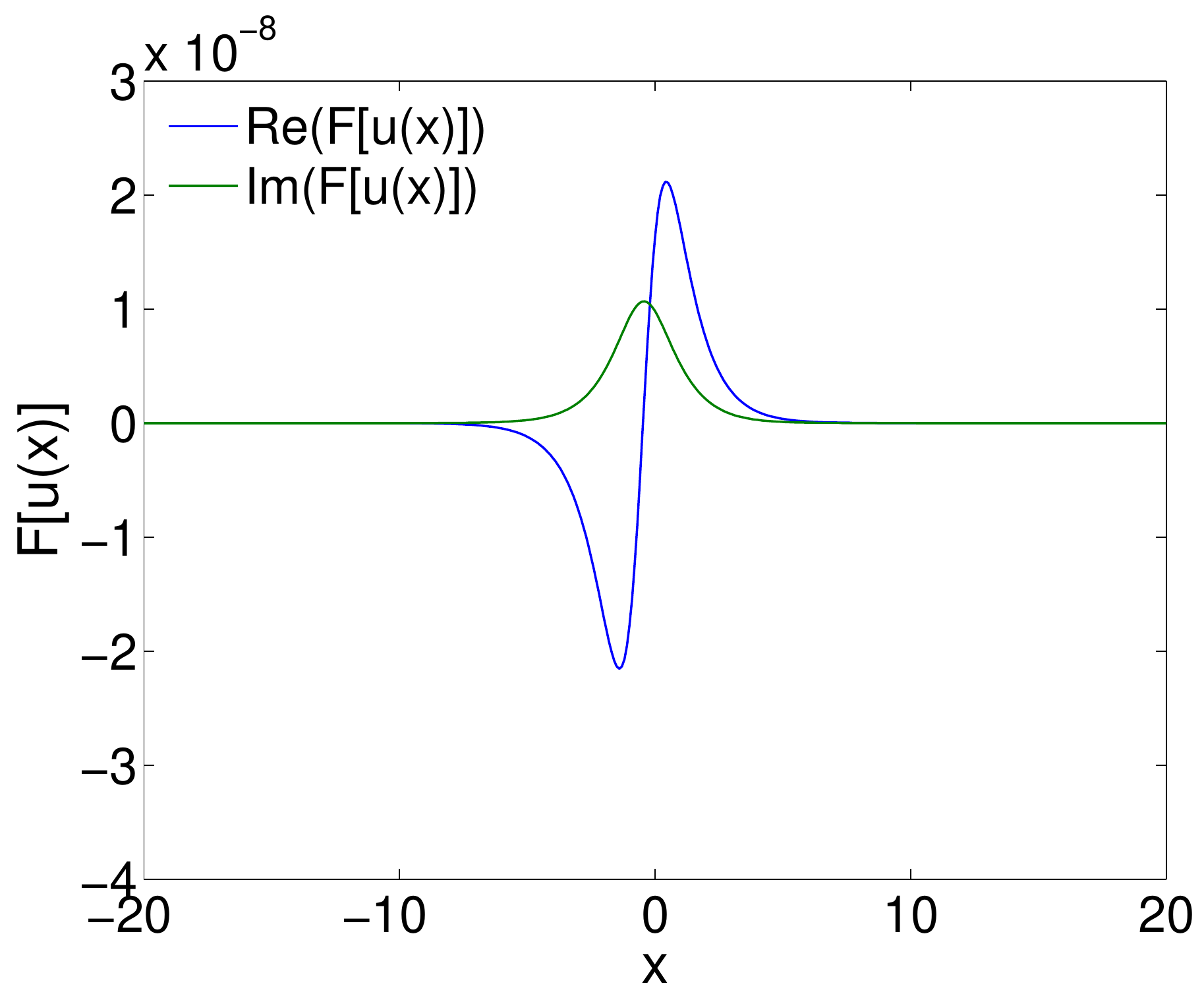} \\
\end{tabular}
\end{center}
\caption{
Optimized beam dynamics with almost zero $||F[u]||$ in the potential $\tilde{V}_1(x)$
for $A_1=0.1$, $k_1=1/2$, $x_d=1$ and $\delta_1=0.014038$. The top left panel shows
the space-time contour plot of the density evolution, while the top right panel shows
the evolution of the norm $N(t)$. The bottom left panel compares $[V_1'(x)]^2$ and $4V(x)W^2(x)$,
%in order to
showing that Eq. (\ref{eq:Yang}) does not hold. The bottom right panel depicts %shows
the real and imaginary part of $F[u]$. The values of the diagnostic quantities are $\lambda=-6.09\times10^{-8}$ and $\sigma=-3.87\times10^{-8}$.}
\label{fig:dynamics5}
\end{figure}

Figure~\ref{fig:plane1} shows a clear correlation between $\sigma$ and $\lambda$.
Notice the symmetry between the outcomes when the transformation
$(x_d,\delta_1)\rightarrow(-x_d,-\delta_1)$ is applied, which is also manifested
in the values of $\lambda$ and $\sigma$ displayed in the captions of
Figs.~\ref{fig:dynamics1} and \ref{fig:dynamics2}. From this
figure it is also clear that, roughly speaking,
when $x_d \delta_1<0$, $N(t)$ grows with time, whereas the opposite takes place
when $x_d \delta_1>0$. This is not always true, as there is a critical value
$\delta_{1c}$ (close to zero) separating the growing ($\lambda>0$) and decaying ($\lambda<0$) dynamics, which is tantamount to the separation of the regions
with $\sigma>0$ and $\sigma<0$. The dependence of $\delta_{1c}$ versus $x_d$
is also depicted in Fig.~\ref{fig:plane1}; having in mind the continuous dependence of $\sigma$ and $\lambda$ with $x_d$ and $\delta_1$, it is clear that $\sigma=0$ just at the curve $\delta_{1c}(x_d)=0$, so one can find stationary soliton solutions.
This is manifested in Fig.~\ref{fig:dynamics5}, where, for a set of parameters
very close to the curve $\delta_{1c}(x_d)=0$ (in particular, $x_d=1$ and $\delta_1=0.014038$),
the decay is very slow (with $\lambda\lesssim10^{-7}$), but not identically zero,
as $||F[u]||\sim 10^{-8}$. Interestingly, as shown in the bottom left panel of the figure,
the relation~(\ref{eq:Yang}) is not fulfilled. Consequently, there is a range of
parameter values for which states with a very small value of $||F[u]||$
can be obtained \emph{even if} the
potential is not of the form $-(g^2+ig')$.%}

\begin{figure}
\begin{center}
\begin{tabular}{cc}
\includegraphics[width=.45\textwidth]{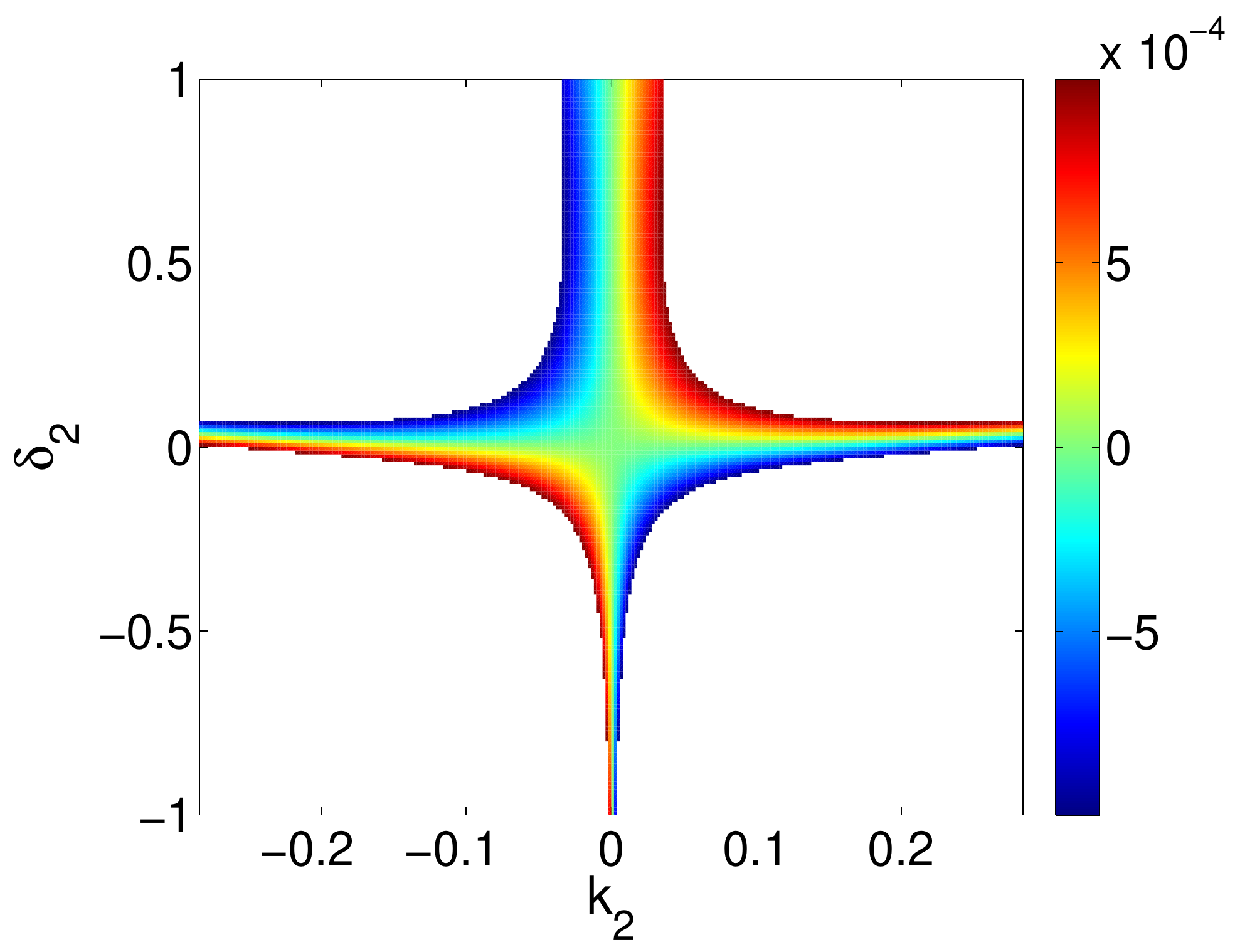} &
\includegraphics[width=.45\textwidth]{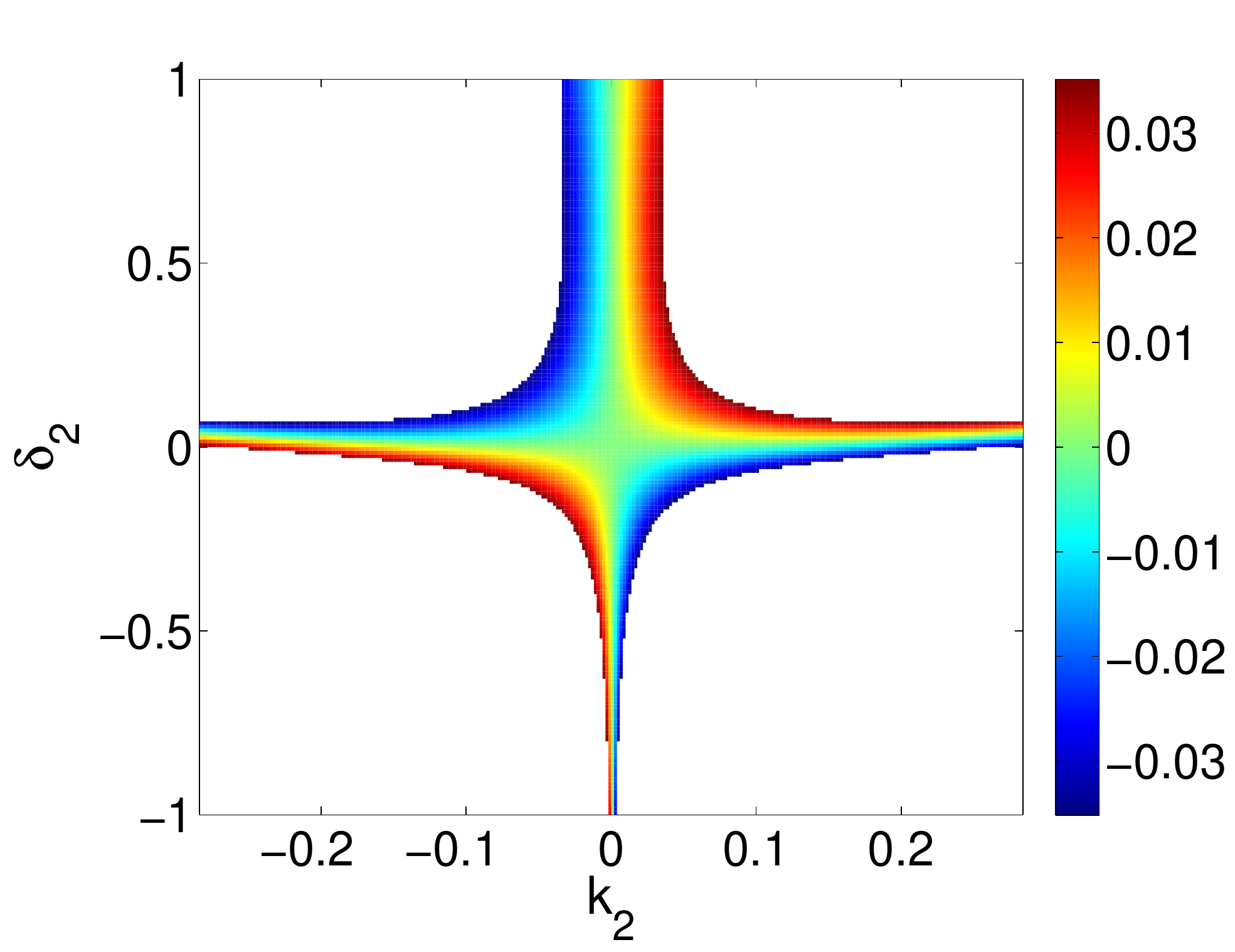} \\
\multicolumn{2}{c}{\includegraphics[width=.45\textwidth]{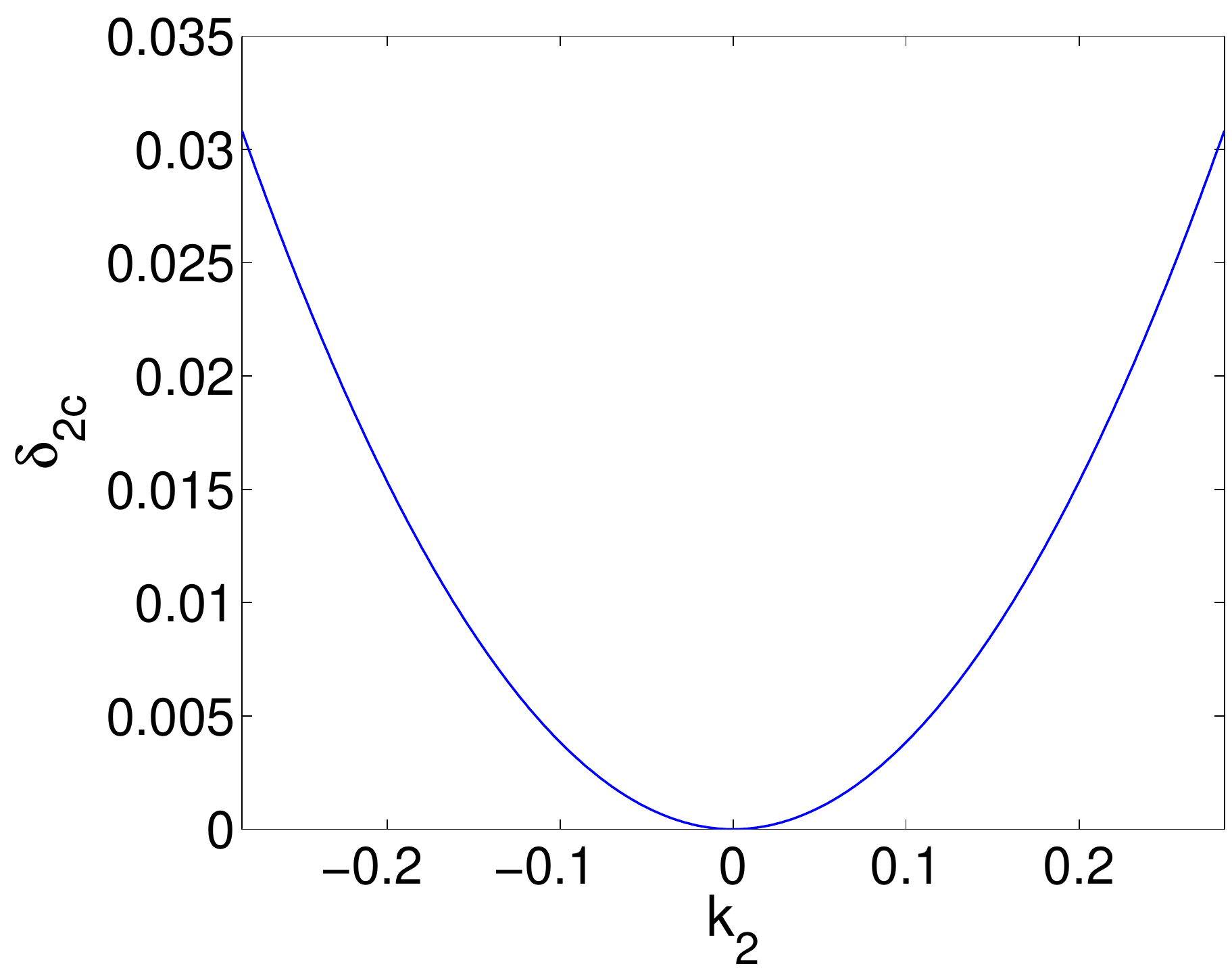}}\\
\end{tabular}
\end{center}
\caption{Top panels: Dependence of diagnostic quantities $\sigma$ (left) and $\lambda$ (right) as a function of $k_2$ and $\delta_2$ for the potential $\tilde{V}_2(x)$ with $A_2=1$. Bottom panel depicts the curve $\delta_{2c}(x_d)$ at which both $\sigma$ and $\lambda$ vanish.
}
\label{fig:plane2}
\end{figure}

In the case of the NLS equation with potential $\tilde{V}_2(x)$, we only focus
on the dependence of $\lambda$ and $\sigma$ with respect to parameters $k_2$ and $\delta_2$,
as the outcome of simulations is essentially the same as in the previous case.
Namely, for non-vanishing values of $\lambda$ and $\sigma$,
a growth or decay of the solutions is identified for typical values
of $\delta_2$, as shown in Fig.~\ref{fig:plane2}.
However, this growth or decay is quite slow, as achieved by the
optimization of the beam via the Levenberg--Marquardt algorithm.
Notice there is an anti-symmetry in the outcome when
the
transformation $k_2\rightarrow-k_2$
is applied. In addition, both $\sigma$ and $\lambda$ are equal to
zero at $k_2=0$ as at that point the potential is real and the solutions are stationary.
Once again, the nearly parabolic curve in the
$(\delta_2,k_2)$ plane where $\lambda=\sigma=0$
enables us to identify parameter values in the vicinity of which
states with particularly small $||F[u]||$ appear to exist.

\section{{Symmetry breaking in two-dimensional potentials}}

It is of particular interest to extend the above one-dimensional
considerations
towards the emergence of asymmetric optimized beam families
in the 2D version of Eq. (\ref{eq:dyn}) that reads:
\begin{equation}\label{eq:dyn2D}
    i\psi_t=-(\psi_{xx}+\psi_{yy})+[V(x,y)+i W(x,y)]\psi-|\psi|^2\psi.
\end{equation}
In this case, stationary solutions, $\psi(x,y,t)=\mathrm{e}^{i \mu t}u(x,y)$ with $u(x,y)\in\mathbb{C}$,
will satisfy:
\begin{equation}\label{eq:stat2D}
    F[u]\equiv\mu u-(u_{xx}+u_{yy})+[V(x,y)+i W(x,y)]u-|u|^2u=0.
\end{equation}

In Ref. \cite{yang2d}, it is shown that not only symmetric solitons exist but also symmetry breaking is possible if the potential $\tilde{V}(x,y)=V(x,y)+i W(x,y)$ is of the form

\begin{equation}\label{eq:cond2D1}
    \tilde{V}(x,y)=-[g^2(x)+\alpha g(x)+ig'(x)+h(y)]
\end{equation}
with $g(x)$ being a spatially even real function, $h(y)$ being a real function and $\alpha$ a real constant. Notice that this potential is partially-$\mathcal{PT}$-symmetric (denoted also as P$\mathcal{PT}$-symmetric), i.e.,
\begin{equation}
    \tilde{V}^*(x,y)=\tilde{V}(-x,y)
\end{equation}
The linear spectrum of this potential can be purely real. In this case, a family of $\mathcal{PT}$-symmetric solitons can emerge from the edge of the continuous spectrum; two degenerate branches of asymmetric solitons, which do not respect the P$\mathcal{PT}$ symmetry, bifurcate from the symmetric soliton branch through a pitchfork bifurcation.

The symmetry breaking bifurcation can also be observed either if the potential possesses double P$\mathcal{PT}$ symmetry
\begin{equation}\label{eq:cond2D2}
    \tilde{V}^*(x,y)=\tilde{V}(-x,y)\qquad\textrm{and }\tilde{V}^*(x,y)=\tilde{V}(x,-y)
\end{equation}
or $\mathcal{PT}$- and one P$\mathcal{PT}$-symmetry simultaneously
\begin{equation}\label{eq:cond2D3}
    \tilde{V}^*(x,y)=\tilde{V}(-x,-y)\qquad\textrm{and}\qquad
    \tilde{V}^*(x,y)=\tilde{V}(-x,y)\textrm{ or }\tilde{V}^*(x,y)=\tilde{V}(x,-y).
\end{equation}
In such cases of double symmetries, there is no need for the potential to have
a special form as in Eq.~(\ref{eq:cond2D1}). In addition, the soliton branch that emerges from the spectrum edge possesses both symmetries whereas the bifurcating branch loses one of the symmetries although it retains the other.

A later work \cite{ch16} reports the existence of the same branching behaviour in a $\mathcal{PT}$-symmetric potential which
also features a partial $\mathcal{PT}$-symmetry along the x-direction.
%does not feature any P$\mathcal{PT}$ symmetry. In other words, this potential does not meet any of the conditions (\ref{eq:cond2D1}), (\ref{eq:cond2D2}), (\ref{eq:cond2D3}) proposed in \cite{yang2d} and, consequently, the asymmetric solitons branches should not exist. We have checked, as shown below, that such branches correspond actually to optimized beams.
More specifically, the potential used in \cite{ch16} is given by

\begin{equation}\label{eq:potential2D}
    V_3(x,y)=-[G^2(x,y)+G(x,y)], \quad W_3(x,y)=k_3\partial_x G(x,y)
\end{equation}
with
\begin{equation}
    G(x,y)=A_3\mathrm{e}^{-y^2}(\mathrm{e}^{-(x-1)^2}+\mathrm{e}^{-(x+1)^2}).
\end{equation}
Notice that the symmetries mentioned above are applicable as
a result of the even nature of the $G(x,y)$.

\begin{figure}
\begin{center}
\begin{tabular}{cc}
\includegraphics[width=.45\textwidth]{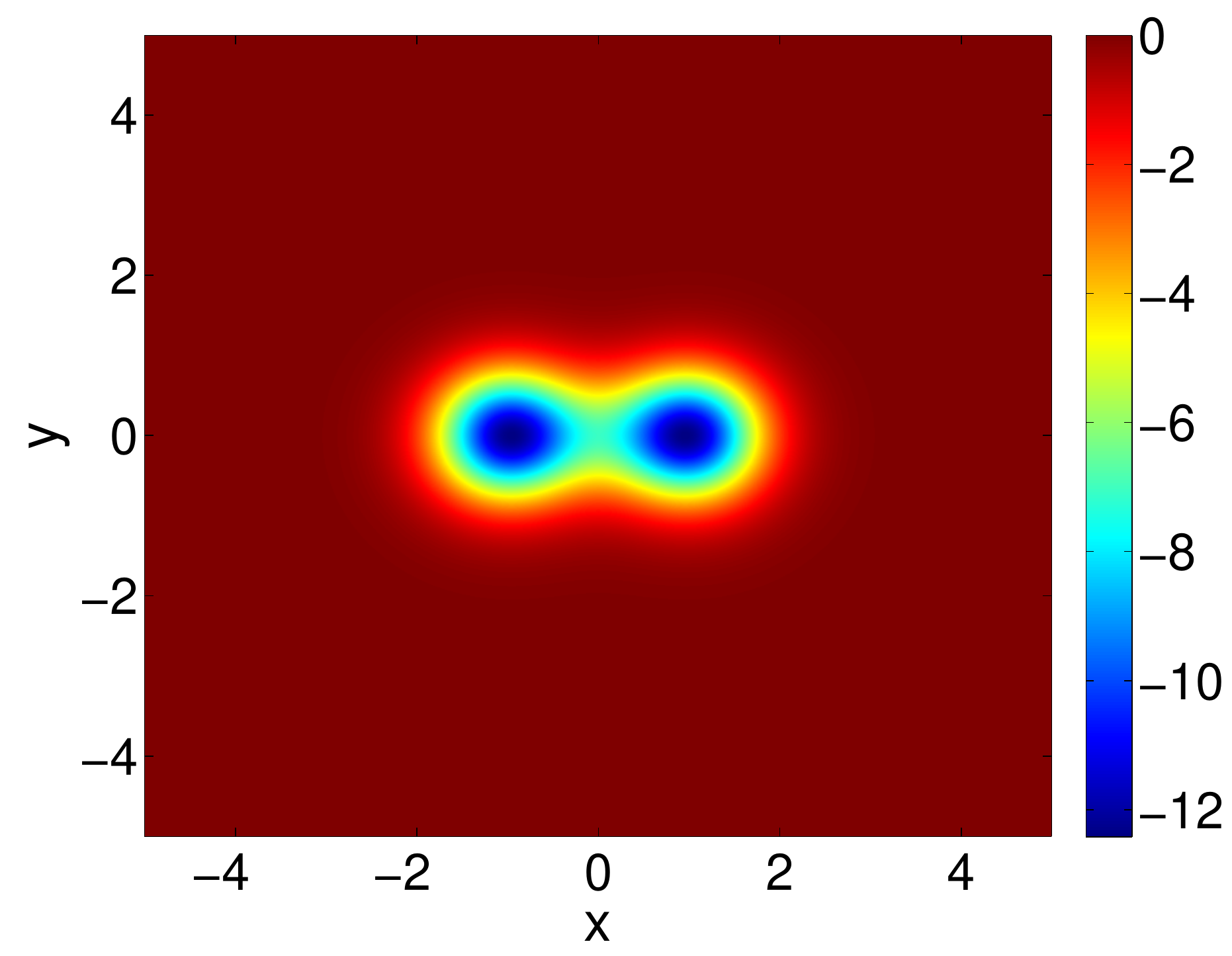} &
\includegraphics[width=.45\textwidth]{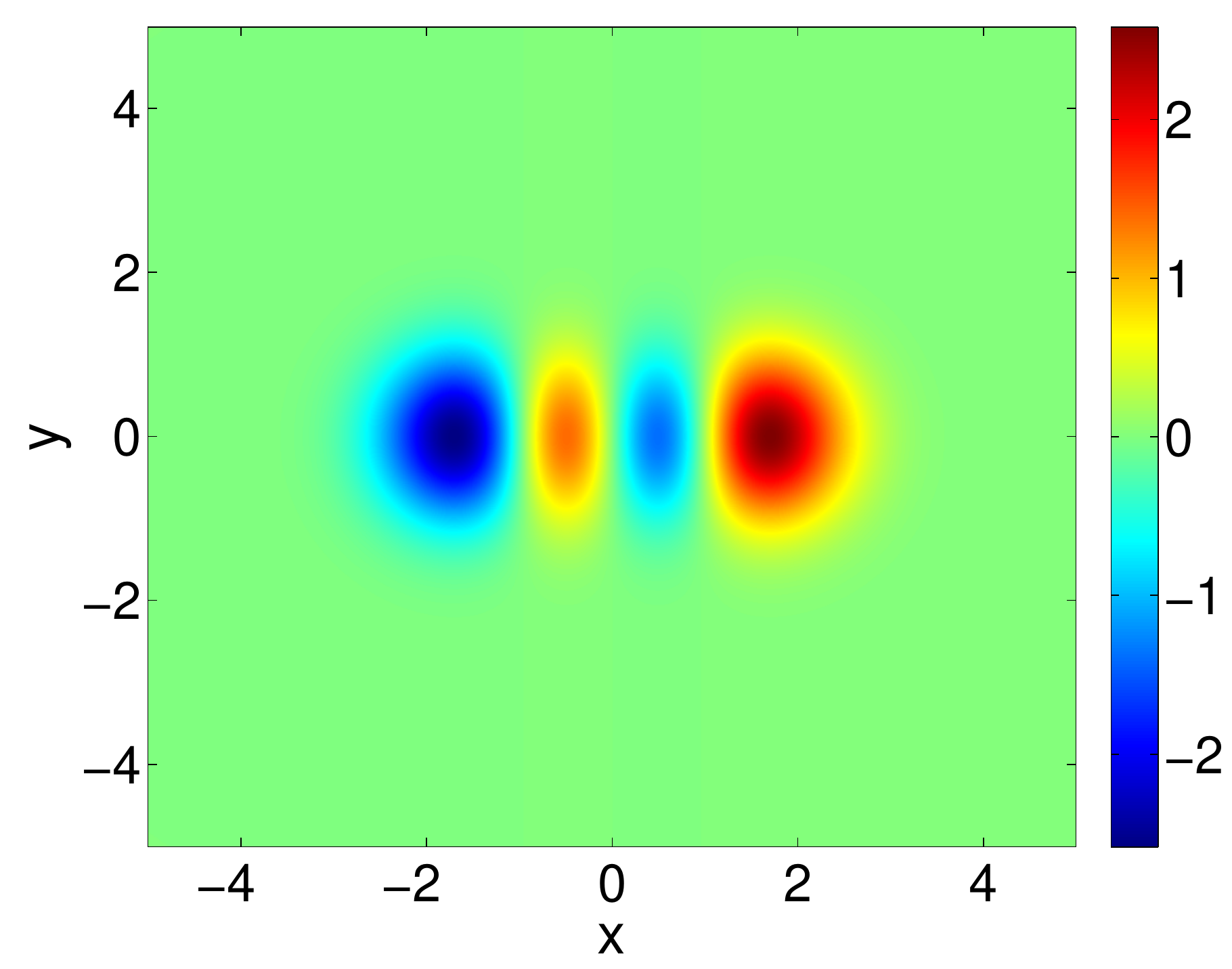} \\
\end{tabular}
\end{center}
\caption{Real (left) and imaginary (right) part of the 2D potential $\tilde{V}_3(x,y)$ for $A_3=3$ and $k_3=1$.}
\label{fig:potential2D}
\end{figure}

To give an associated example
of the resulting symmetry breaking, we use, as in \cite{ch16}, $A_3=3$ and $k_3=1$. The resulting profile of the potential is shown in Fig.~\ref{fig:potential2D}. $\mathcal{PT}$-symmetric solitons are calculated by means of
the Newton--Raphson method and the corresponding branch emerges from $\mu=5.810$; asymmetric solitons (actually, optimized beams) are attained by using the Levenberg-Marquardt algorithm, with a tolerance of $||F[u]||<10^{-2}$. Now, the $L^2$-norm is defined as

\begin{equation}\label{eq:L22D}
    ||F[u]||=\sqrt{\int |F[u(x)]|^2\mathrm{d}x\mathrm{d}y}.
\end{equation}

\begin{figure}
\begin{center}
\begin{tabular}{cc}
\multicolumn{2}{c}{\includegraphics[width=.45\textwidth]{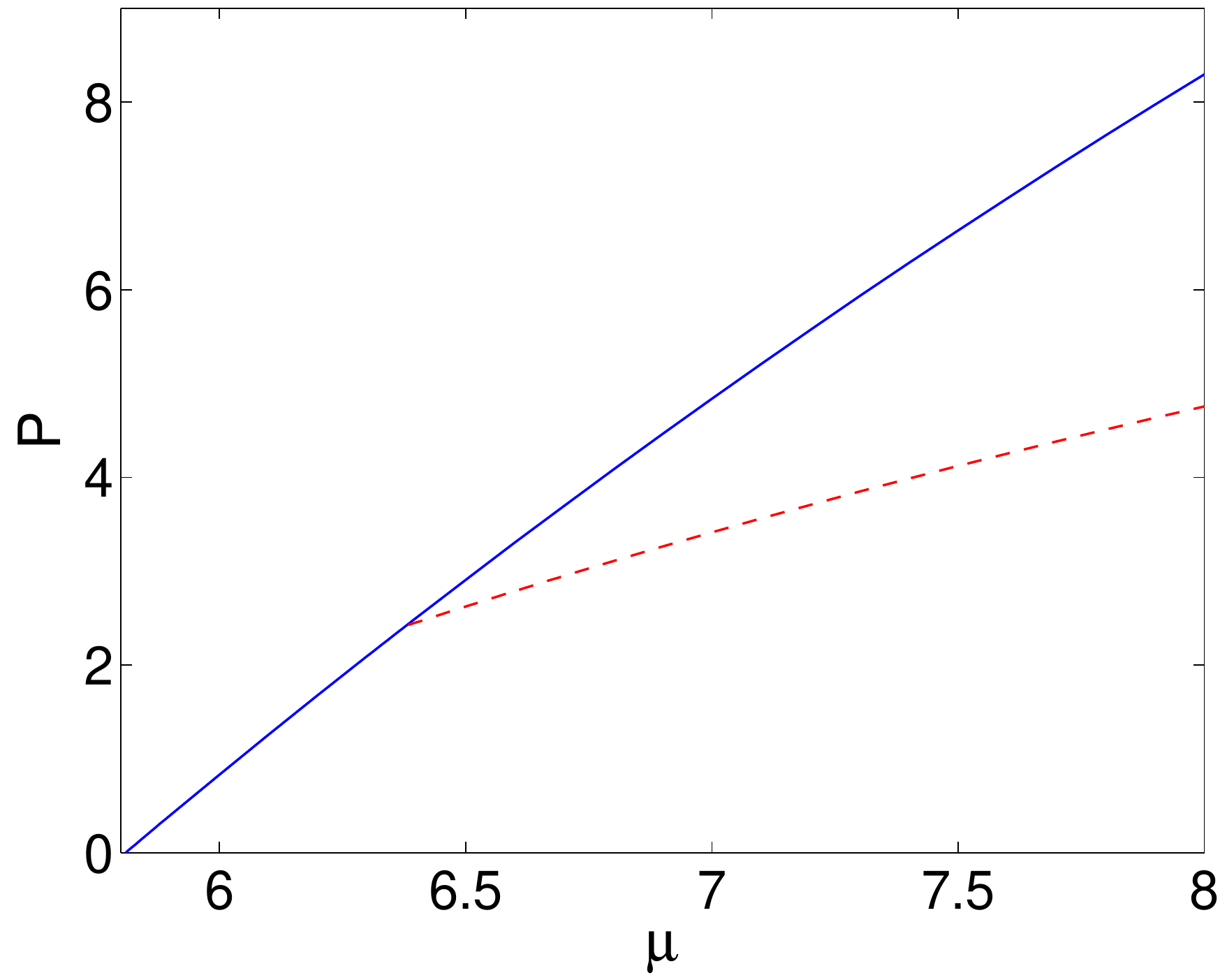}} \\
\includegraphics[width=.45\textwidth]{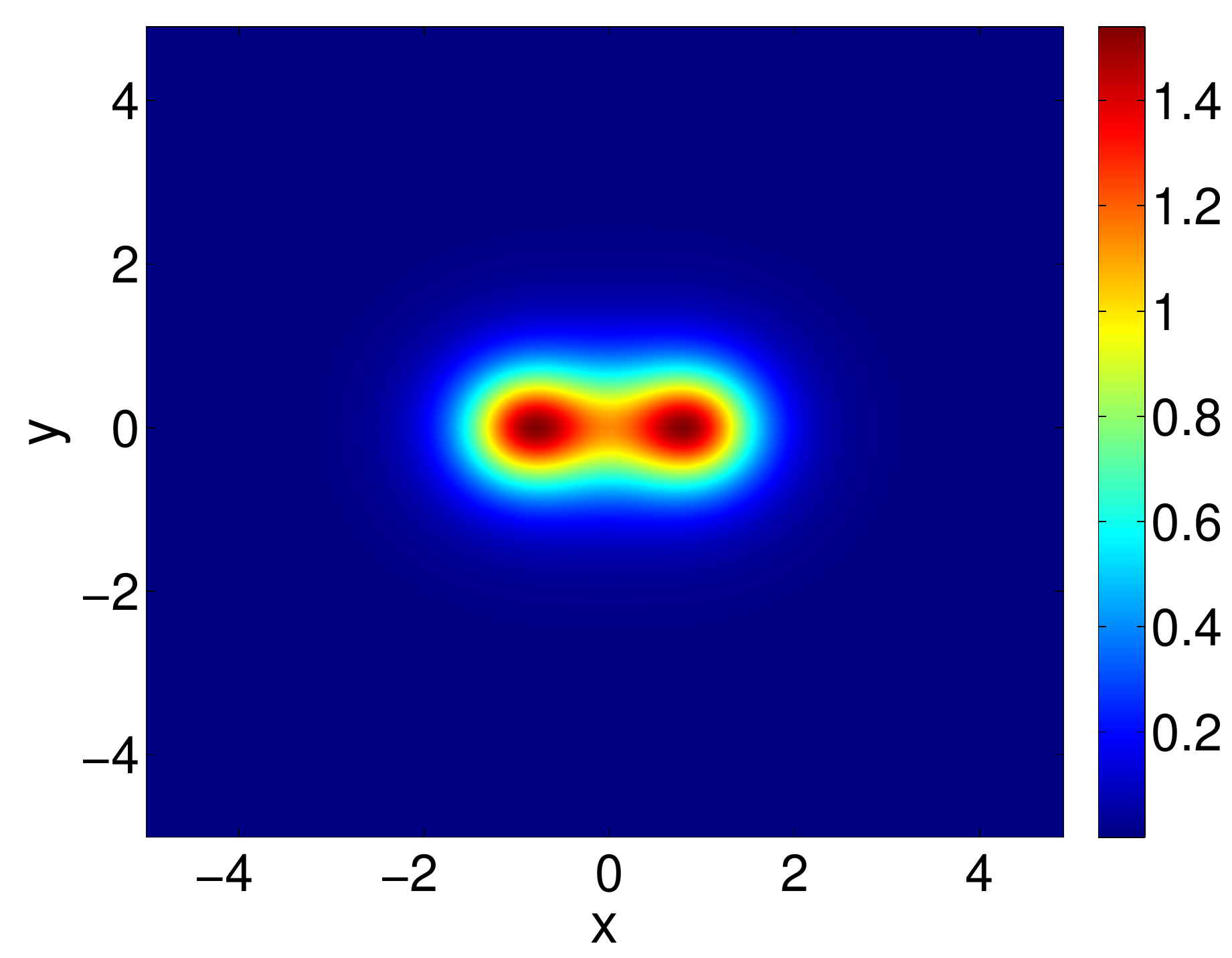} &
\includegraphics[width=.45\textwidth]{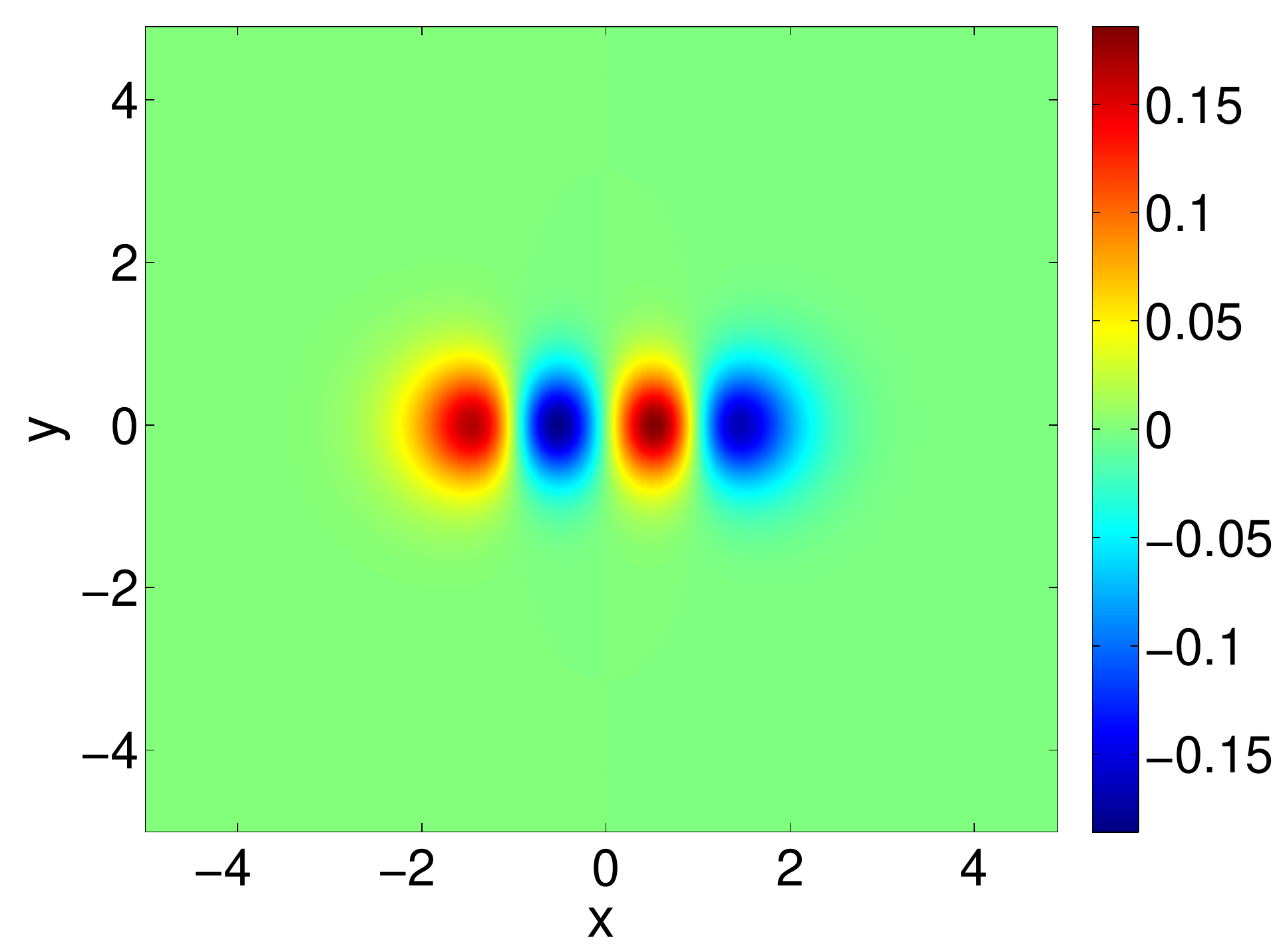} \\
\includegraphics[width=.45\textwidth]{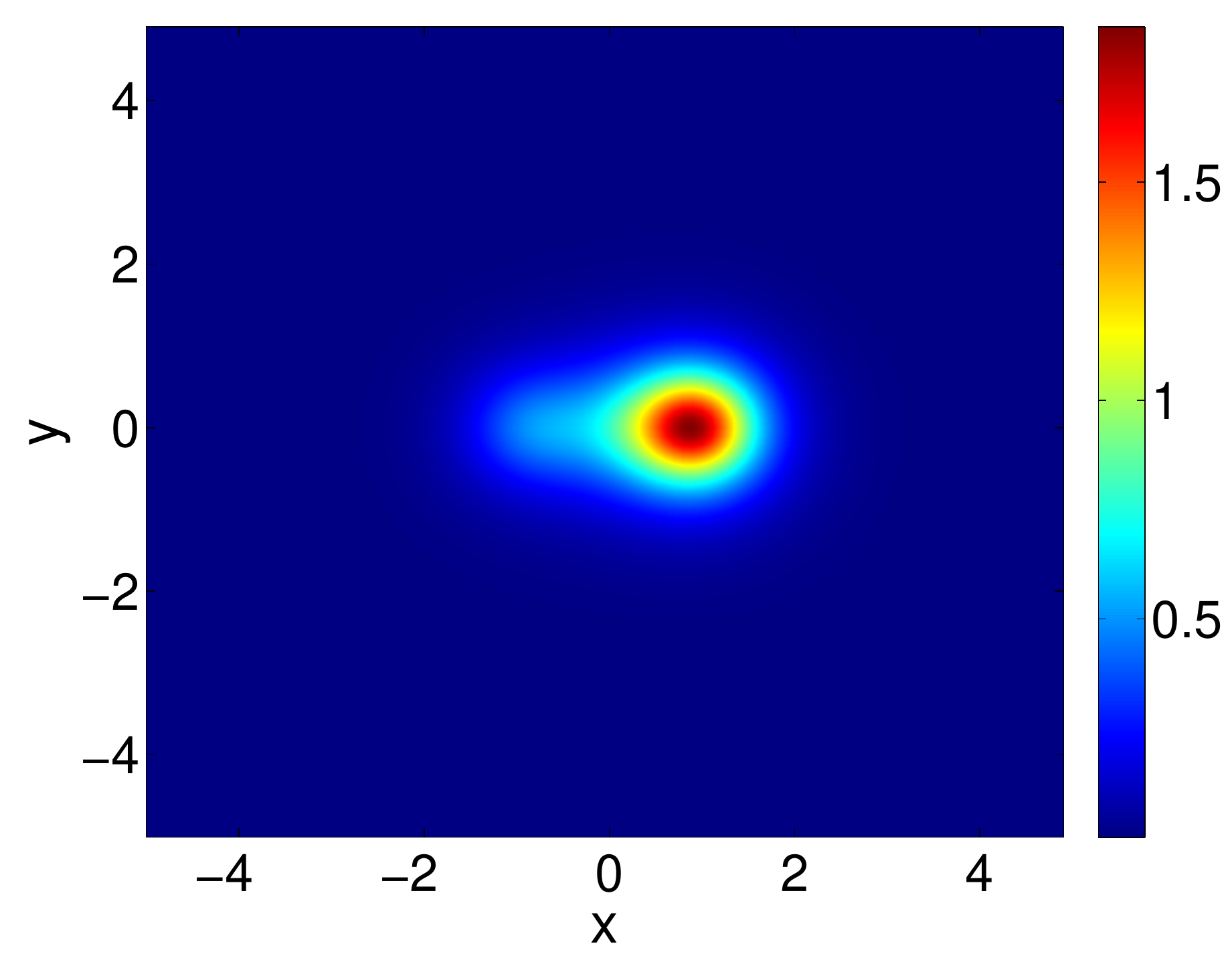} &
\includegraphics[width=.45\textwidth]{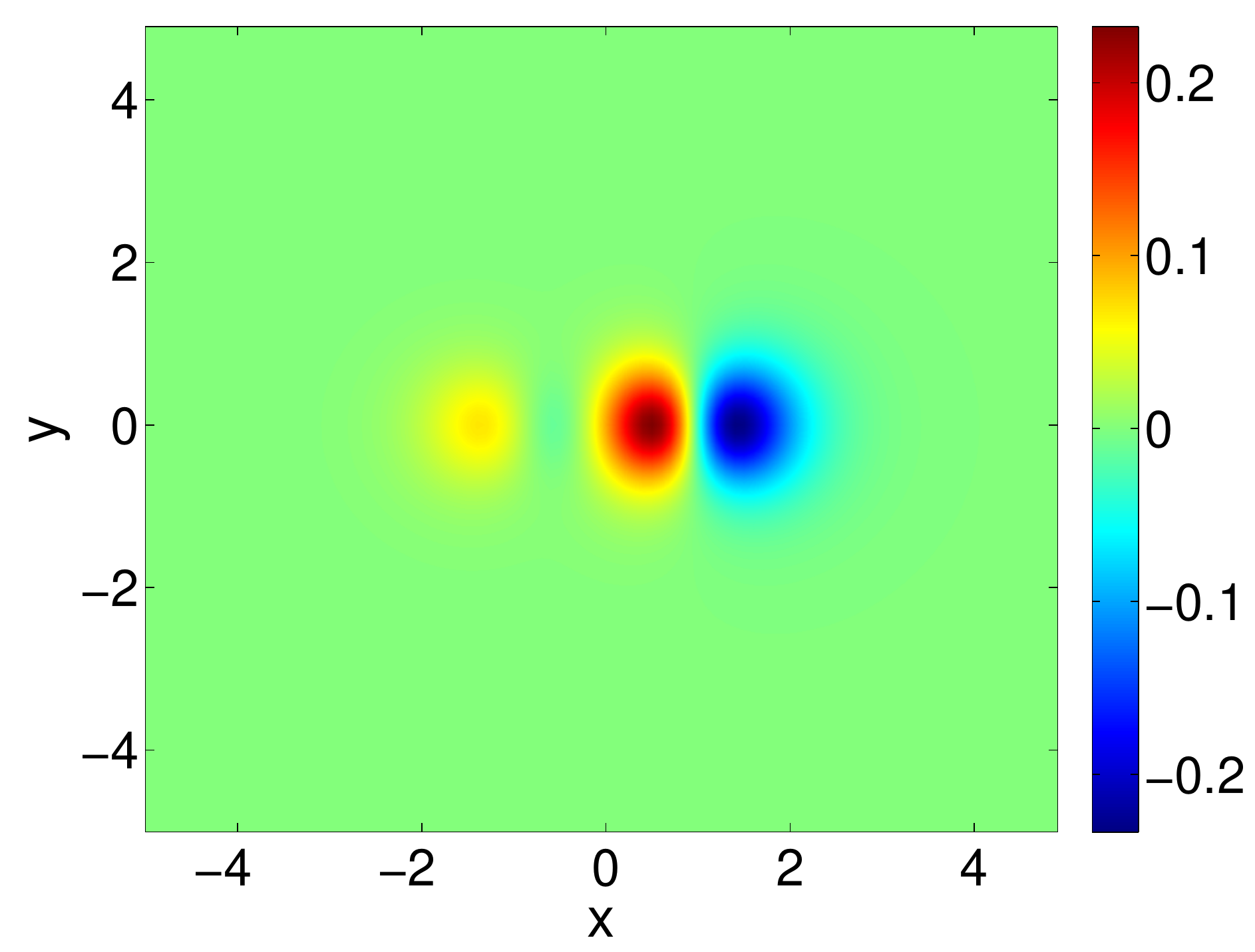} \\
\end{tabular}
\end{center}
\caption{Top row: dependence of the squared $L^2$-norm $P$ of $\mathcal{PT}$-symmetric solitons (blue full line) and asymmetric solitons / optimized beams (red dashed line) with respect to $\mu$ at the 2D potential $\tilde{V}_3(x,y)$ for $A_3=3$ and $k_3=1$.
Middle row: real (left panel) and imaginary (right panel) of the $\mathcal{PT}$-symmetric soliton with $\mu=7$. Bottom row: same as middle panel but for the asymmetric soliton with the same value of $\mu$.}
\label{fig:soliton2D}
\end{figure}

Fig.~\ref{fig:soliton2D} represents $P\equiv N(t=0)$ versus $\mu$ for the symmetric and asymmetric soliton branches; notice that $N(t)$ is now defined as

\begin{equation}
    N(t)=\int |\psi(x,y,t)|^2 \mathrm{d}x\mathrm{d}y.
\end{equation}

One can observe that the asymmetric branches exist for $\mu\geq6.3837$.
%$, but it seems not to exist a point where these branches link to the symmetric solitons branch; for this reason, we cannot say that the asymmetric branches bifurcate from the $\mathcal{PT}$-symmetric branch.
The figure also shows the profile of solitons at $\mu=7$, the same value that was taken in \cite{ch16}. Notice that all the soliton profiles are symmetric with respect to the $y$-axis; the symmetric solitons present a couple of humps at $(x=\pm x_1,y=0)$ whereas the asymmetric solitons only possess a single hump at $(x=x_2,y=0)$. We have only shown solitons with $x_2>0$ as the solutions with $x_2<0$ are attained simply by making the transform $u(x,y)\rightarrow u(-x,y)$.

\begin{figure}
\begin{center}
\begin{tabular}{cc}
\includegraphics[width=.45\textwidth]{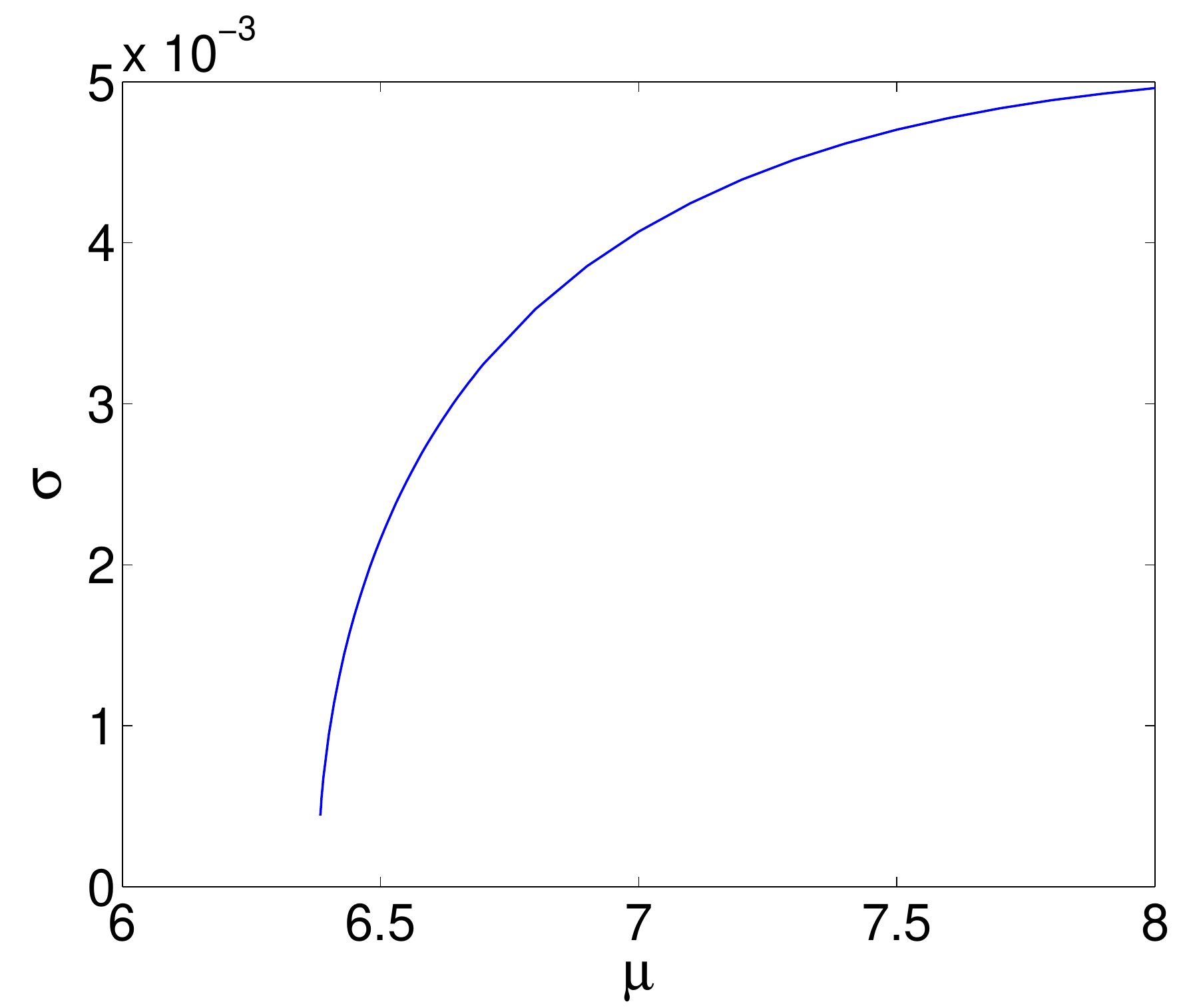} &
\includegraphics[width=.45\textwidth]{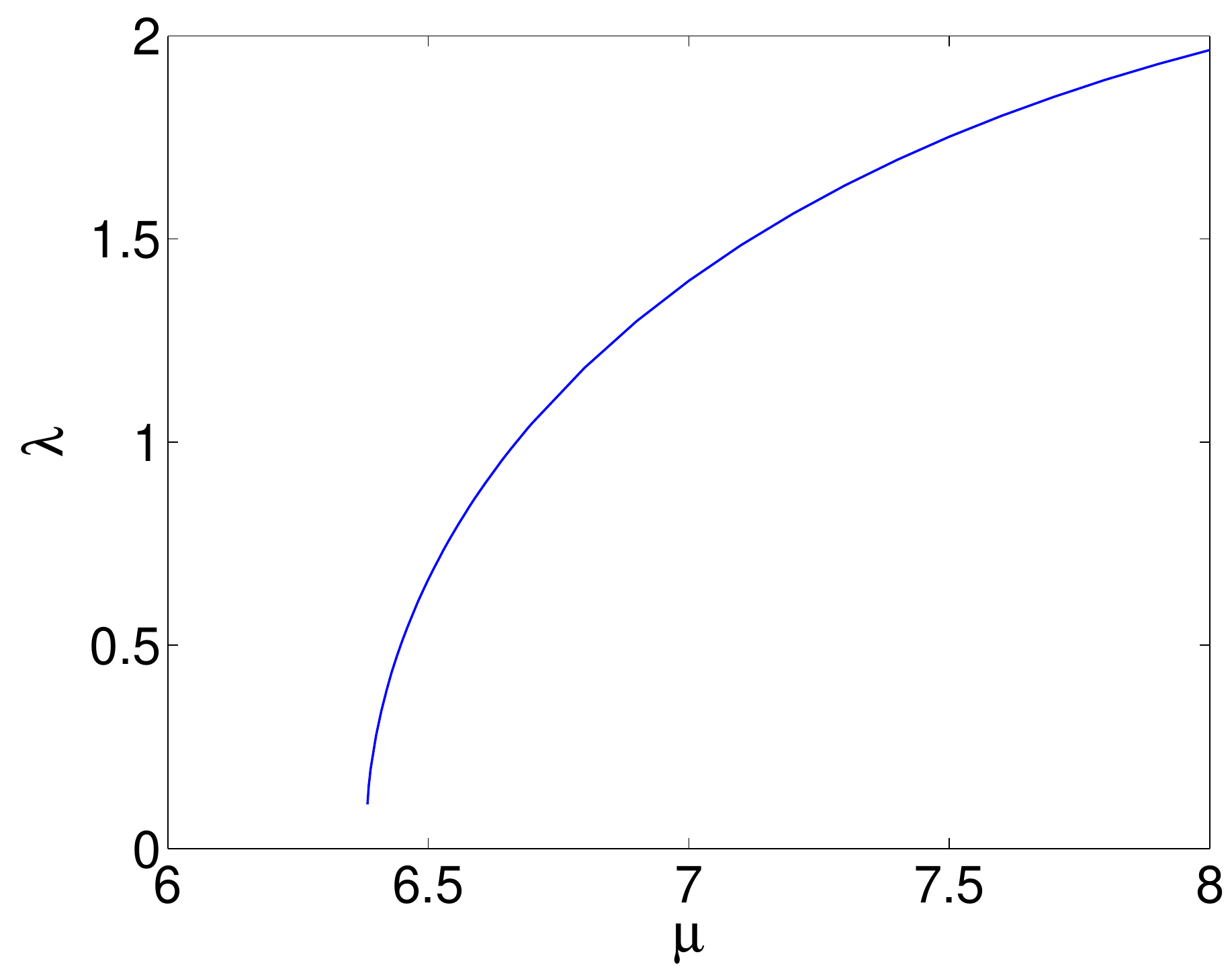} \\
\end{tabular}
\end{center}
\caption{Dependence of diagnostic quantities $\sigma$ (left) and $\lambda$ (right) as a function of $\mu$ for optimized beams at the 2D potential $\tilde{V}_3(x,y)$ with $A_3=3$ and $k_3=1$.}
\label{fig:plane3}
\end{figure}

We have also computed the diagnostic quantities $\lambda$ and $\sigma$ [see (\ref{eq:lambda}) and Eqs. (\ref{eq:sigma}){, with $S$ adapted to 2D domains}] for the asymmetric soliton and depicted them in Fig.~\ref{fig:plane3}. Again, we have considered asymmetric solitons branches centred at $x=x_2>0$. In that case, the norm grows with time, as corresponds to $\sigma>0$ and $\lambda>0$ whereas the opposite takes place if $x_2<0$. We can observe, as in the 1D case, a clear correlation between both quantities.

\begin{figure}
\begin{center}
\begin{tabular}{cc}
\includegraphics[width=.45\textwidth]{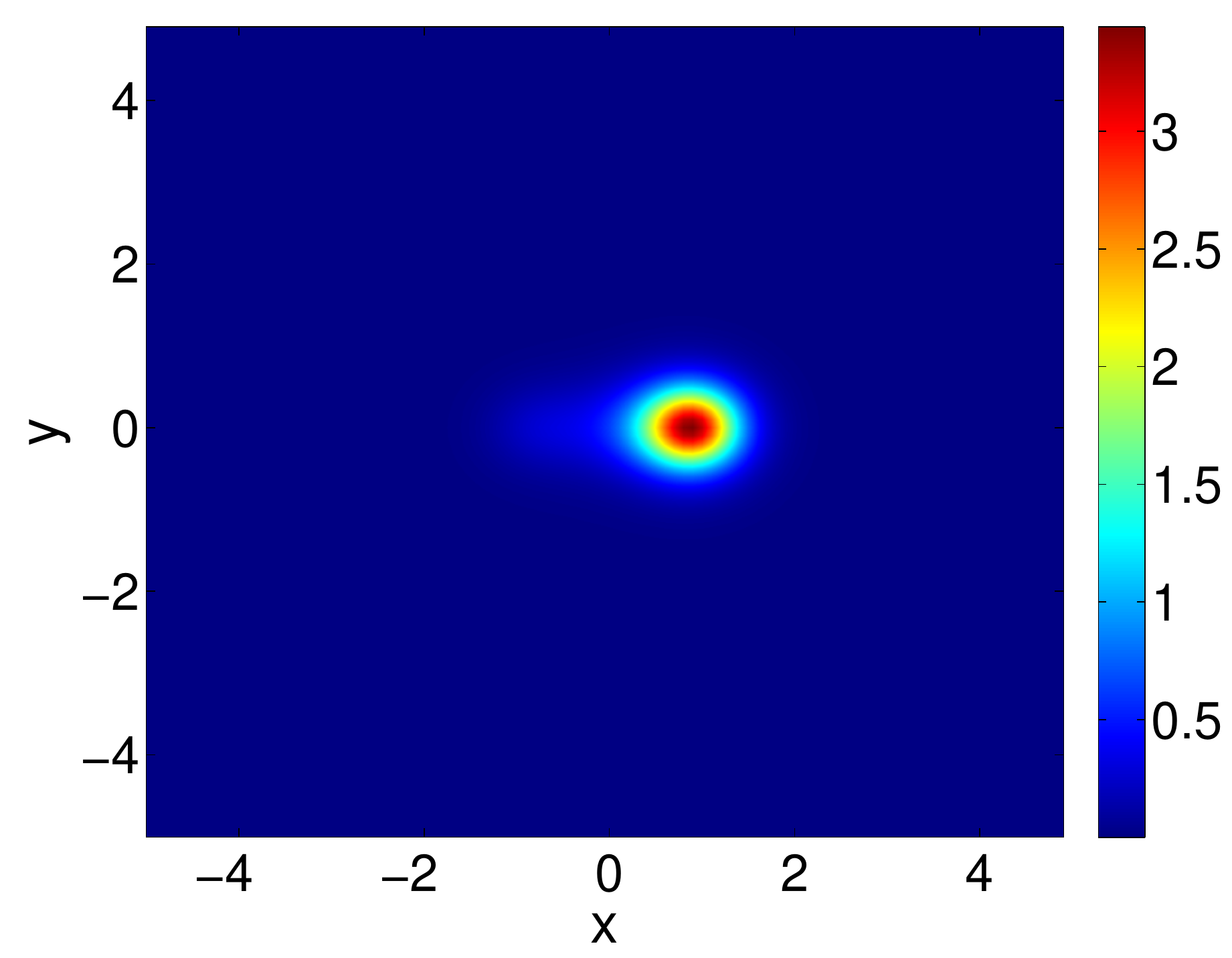} &
\includegraphics[width=.45\textwidth]{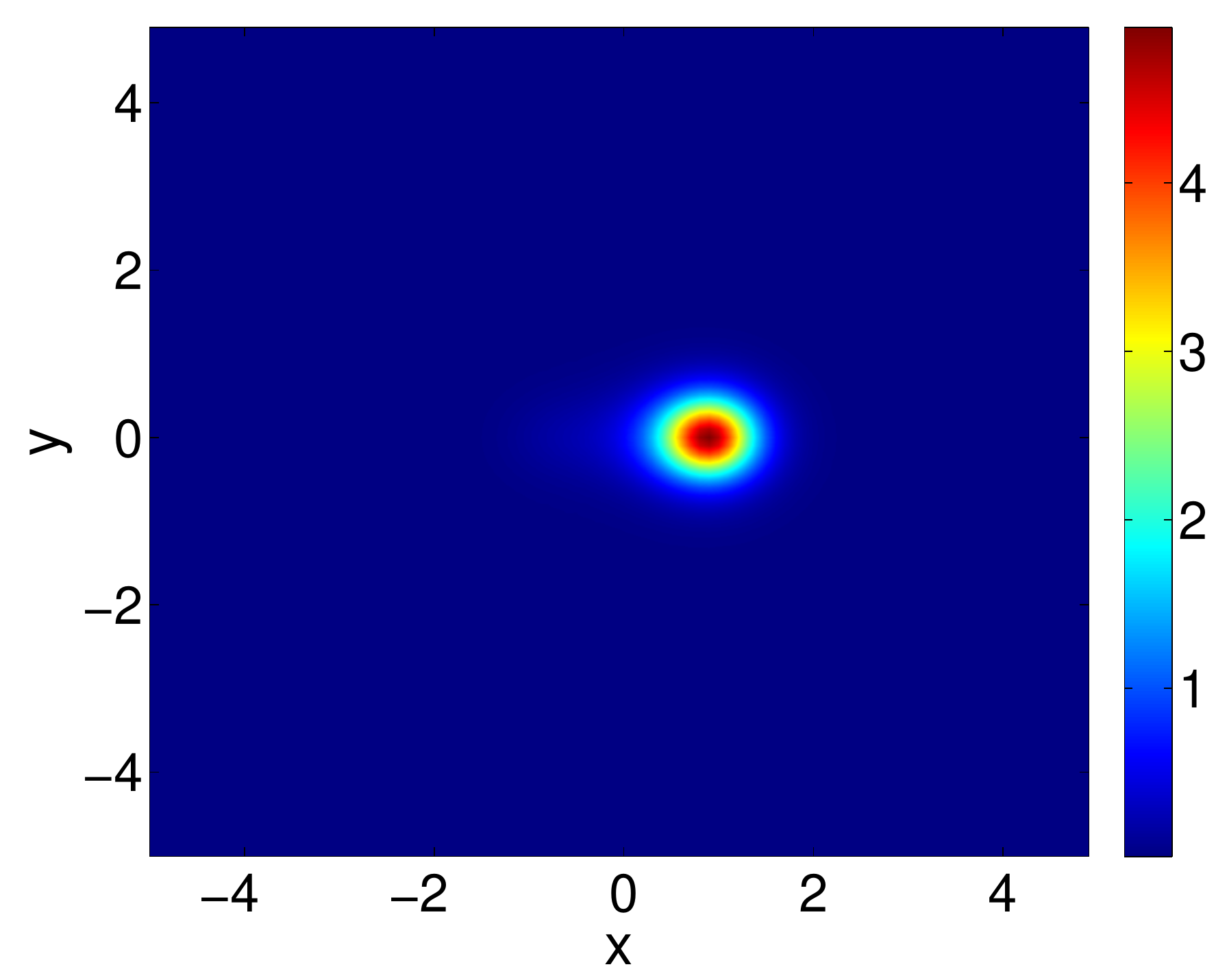} \\
\includegraphics[width=.45\textwidth]{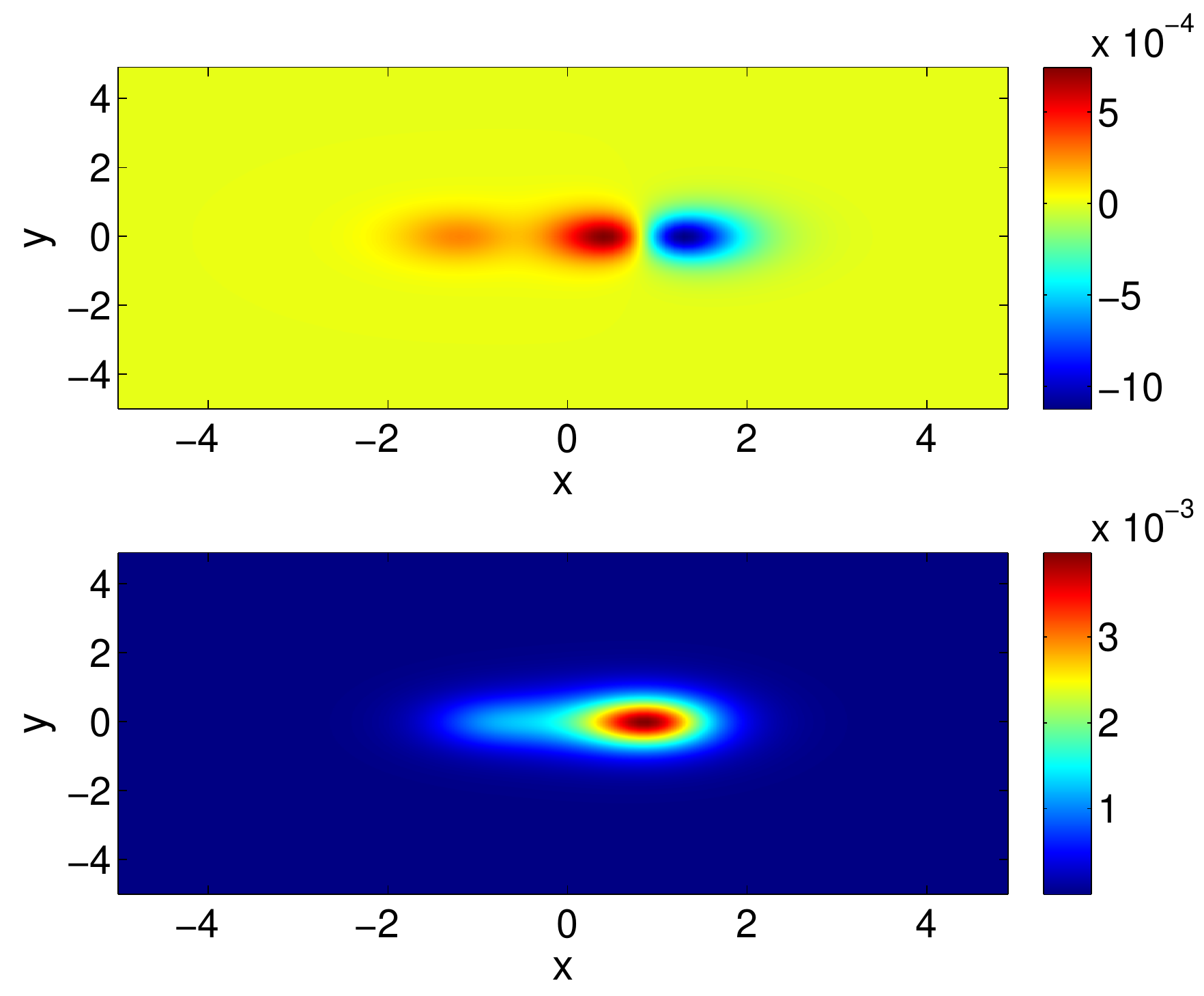} &
\includegraphics[width=.45\textwidth]{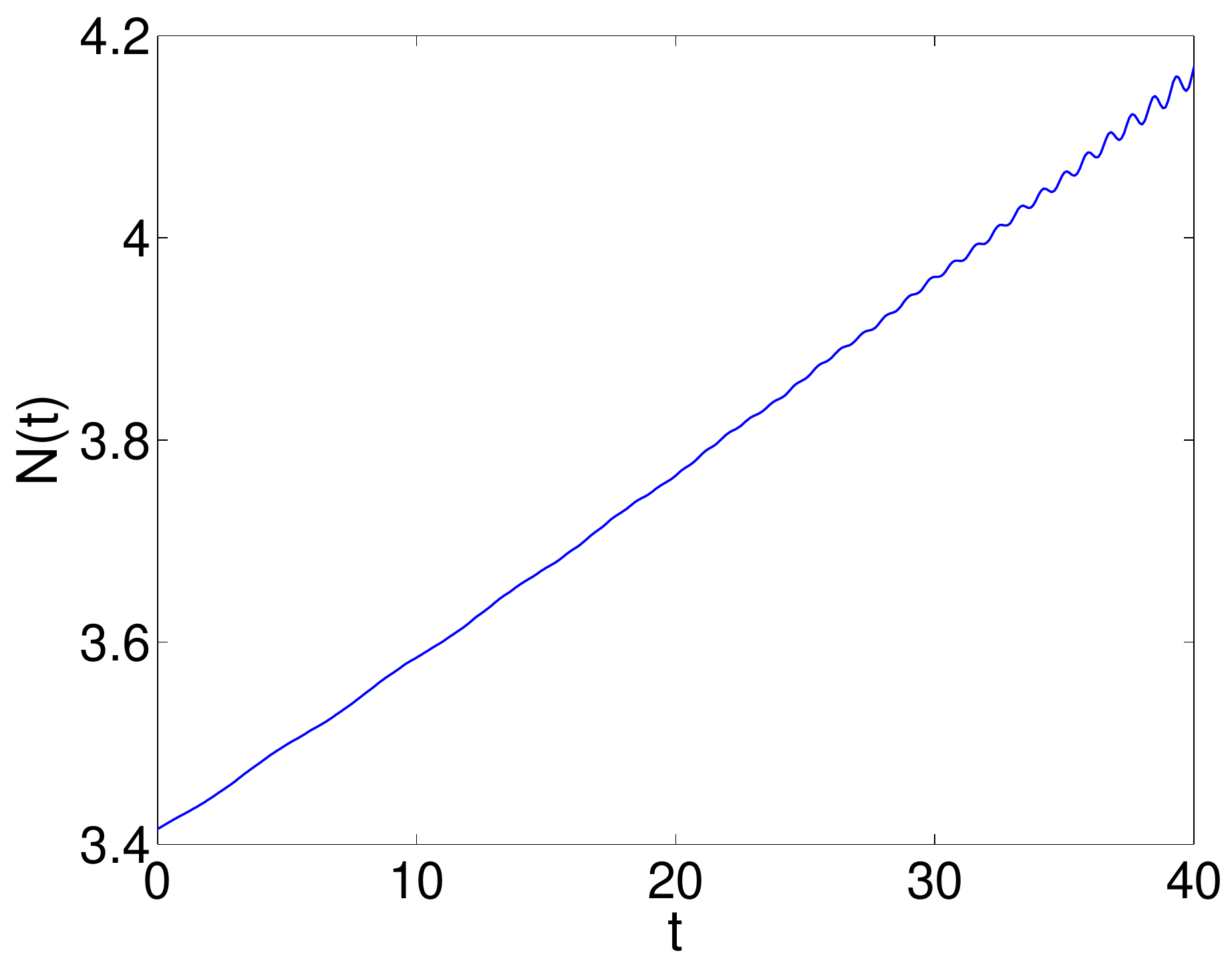} \\
\end{tabular}
\end{center}
\caption{
Optimized beam dynamics in the 2D potential $\tilde{V}_3(x,y)$ for $A_3=3$, $k_3=1$ and $\mu=7$. The top panels show the density profile at $t=0$ (left) and $t=40$ (right). The bottom left panel shows the real and imaginary part of $F[u]$ and the bottom right panel shows the evolution of the norm $N(t)$.
The values of diagnostic quantities are $\lambda=1.39$ and $\sigma=4.07\times10^{-3}$.}
\label{fig:dynamics6}
\end{figure}

\begin{figure}
\begin{center}
\begin{tabular}{c}
\includegraphics[width=.9\textwidth]{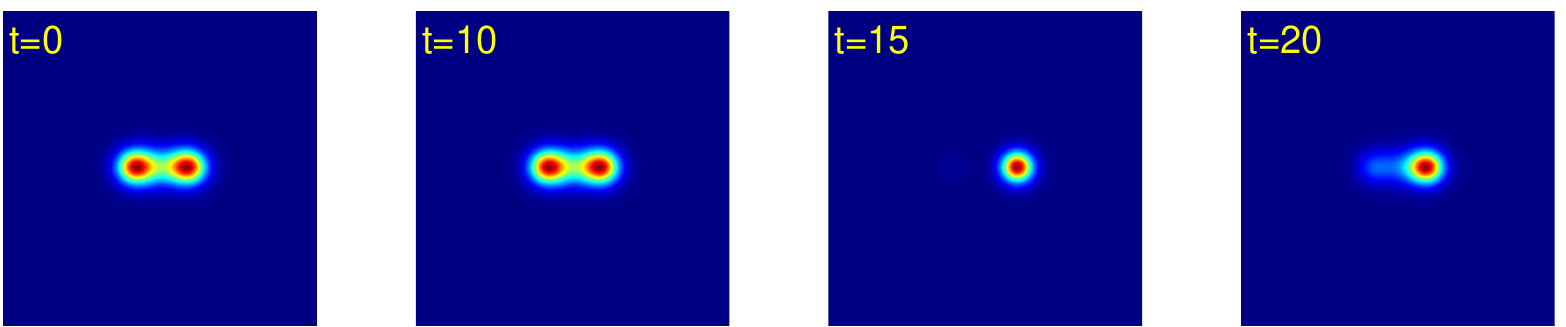} \\
\includegraphics[width=.9\textwidth]{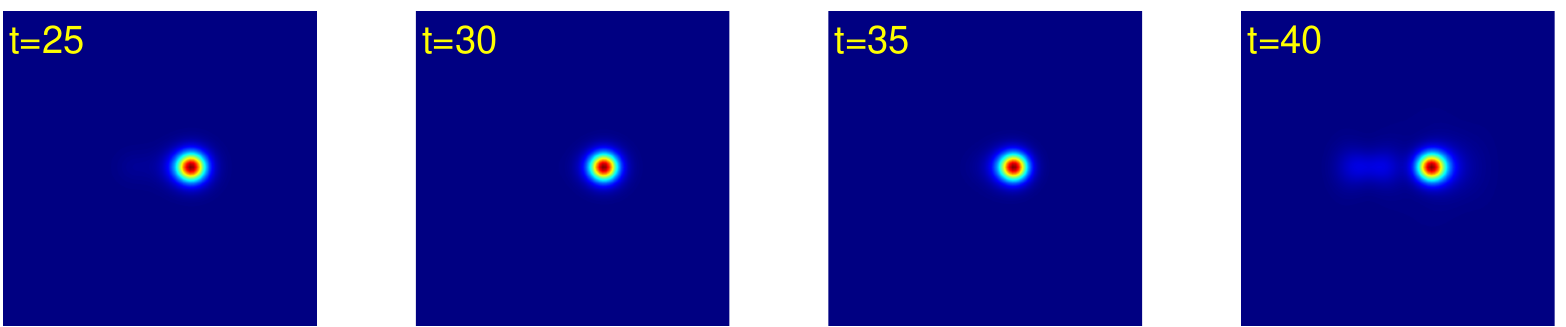} \\
\includegraphics[width=.45\textwidth]{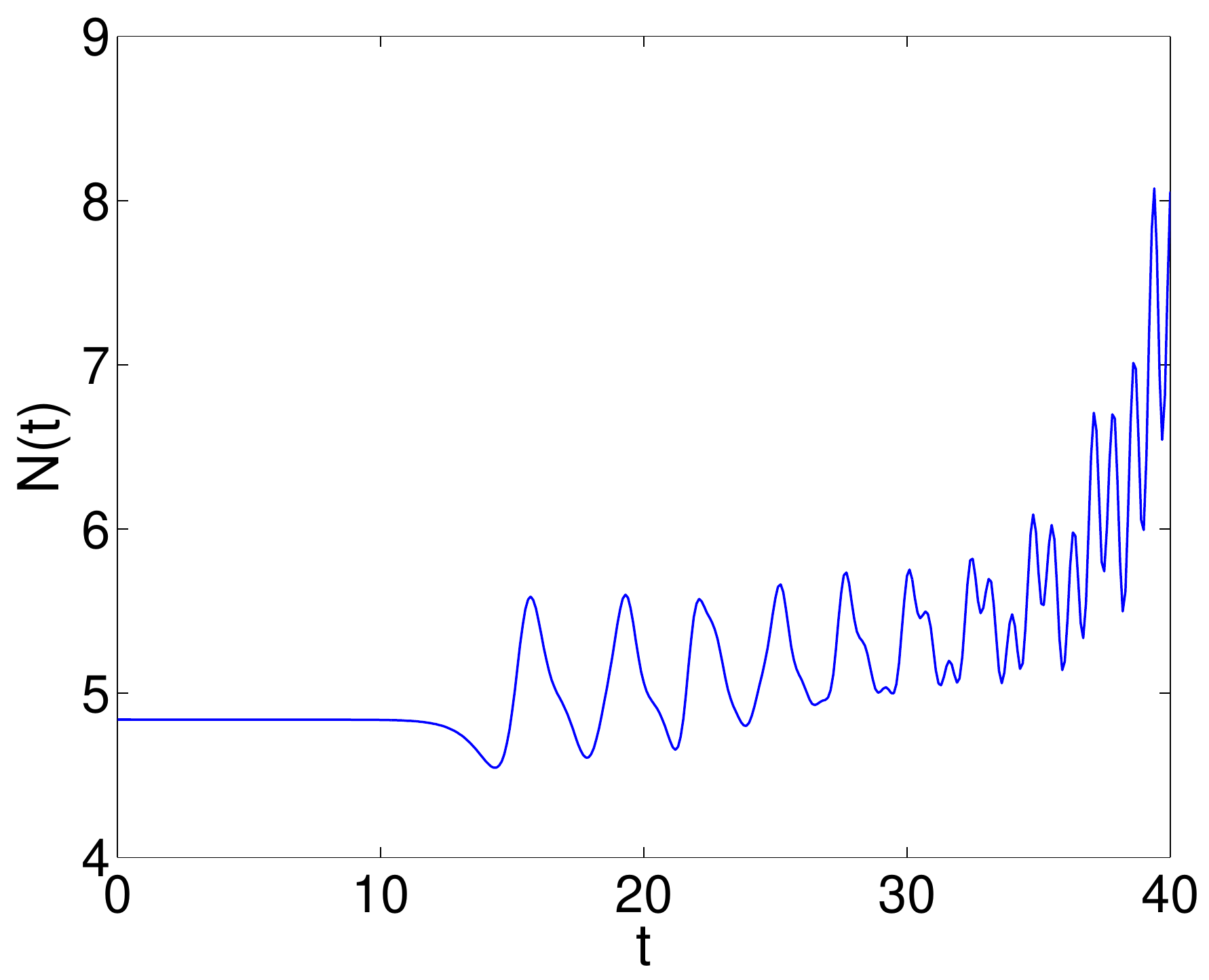}
\end{tabular}
\end{center}
\caption{
Unstable $\mathcal{PT}$-symmetric solitons dynamics in the 2D potential $\tilde{V}_3(x,y)$ for $A_3=3$, $k_3=1$ and $\mu=7$. The top panels show snapshots of the density profile evolution. The bottom shows the evolution of the norm $N(t)$.}
\label{fig:dynamics7}
\end{figure}

Finally, we show in Fig.~\ref{fig:dynamics6} and \ref{fig:dynamics7} the dynamics of the asymmetric and $\mathcal{PT}$-symmetric solitons with $\mu=7$. As it was pointed out in \cite{ch16}, the $\mathcal{PT}$-symmetric solitons are unstable past the ``bifurcation'' point, i.e. when they coexist with the asymmetric branch; as we have shown in Fig. \ref{fig:dynamics7}, they tend to a state similar to the asymmetric soliton, although displaying some density oscillations. However, it was claimed in the same reference that the asymmetric solitons were stable. %This statement is wrong, as it is clearly demonstrated in Fig.~\ref{fig:dynamics6}. But, in any case,
For the optimized beam profiles that we have obtained, as shown
in Fig.~\ref{fig:dynamics6}, the dynamical evolution does not dramatically
alter
the shape of the beam, yet it leads to slow growth of $N(t)$.

We also considered the stability of the $\mathcal{PT}$-symmetric branch
past the relevant bifurcation point. A spectral stability analysis shows that for $\mu\gtrsim6.40$, the solitons become exponentially unstable as an eigenvalue pair becomes real.
%Consequently, although the asymmetric solitons formally are
%optimized beams, the $\mathcal{PT}$-symmetric branch must bifurcate with another branch, and the only candidate seems to be the corresponding to the asymmetric family.
Interestingly, although the asymmetric solitons are actually optimized
beams (i.e. solutions with minimal $||F(u)||$ but not exact solutions),
they might be
%are typically
more robust than the exact solutions of the NLS equation corresponding to the $\mathcal{PT}$-symmetric branch, past the corresponding destabilization
point; compare the associated dynamics of Fig.~\ref{fig:dynamics7} with those
of Fig.~\ref{fig:dynamics6}.

\section{Conclusions \& Future Work}

In the present work, we have revisited a variant of $\mathcal{PT}$-symmetric
systems. In particular, we have examined multi-parametric potentials
whose parameters control, on the one hand, the potential departure
from the $\mathcal{PT}$-symmetric case (such as $x_d$ herein), and on
the other hand, the potential degeneracy of the conditions
(\ref{eqn0}) for stationary solutions --motivated by the recent
works of~\cite{Kom15,kom15b}. We have confirmed the
results of the important recent contribution of~\cite{ny16},
suggesting that in the absence of a special form of the complex
potential, no true stationary solutions are found to exist.
On the other hand, that being said, we have identified beams
that come very close to satisfying the stationary equations.
The dynamics of these beams indicate a {\it slow} departure
from such a configuration. In fact, diagnostics identifying
the rate of this growth and connecting it to the
proximity of the profiles to a stationary solution (via
$||F[u]||$) were developed and numerically evaluated, both in
1D and in 2D.

%Perhaps contrary what one might expect,
%we have found that parametric families of solutions exist
%for arbitrary values of both parameters. The two parameters
%were individually identified as triggering different types
%of instabilities, one with oscillatory growth associated with the
%(former) translational mode, and the other with exponential
%growth associated with the so-called growth mode. Variants
%of these instabilities which typically change the soliton
%mass and induce its mobility were also found when both
%parameters acted concurrently. Nevertheless, their interplay
%was found to be such that stable parameter intervals could
%be identified for such ``dissipative solitons''.
%The stabilizing effect in the vicinity of $\delta=0$
%(i.e., where the condition $V' \propto W$ is
%satisfied) with respect to soliton mass variation may be quite
%important for photonics applications.
%Indeed, a
%solitary wave with continuously increasing mass (power) is undesirable,
%while for a wave of continuously decreasing mass not only the power is lost,
%but moreover there is no trapping in the photonic structure,
%since the soliton width increases and the soliton finally ``feels''
%no trapping potential.

Naturally, this work poses a number of questions for the future.
One of the most notable such concerns the most general conditions
(on, say, a complex potential) under which one may expect to find
(or not) families of stationary solutions. $\mathcal{PT}$-symmetry
is a sufficient but not a necessary condition for such existence
and extending beyond it seems of particular interest.
The conjecture of~\cite{ny16} is that the potentials
$V+i W=- (g^2 + i g'(x)) + c$ represent the generic scenario
is plausible, but it would be particularly interesting to produce
a proof, perhaps revisiting more systematically the relevant shooting
argument.
It is also important to highlight that such shooting arguments
are {\it only} valid in one spatial dimension.
Hence, examining generalizations of the present setting
to higher dimensions is of particular interest
in their own right. We have briefly touched upon
this aspect here, based on the earlier works of~\cite{yang2d,ch16},
but clearly further efforts are necessary to provide
a definitive reply in this direction.
%In that case, identifying the proper
%generalization of the above potentials and their
%potential sufficiency for stationary solutions are topics
%worthy of further exploration in their own right.

%\blindtext

\section*{Acknowledgments}

J.C.-M. thanks financial support from MAT2016-79866-R project (AEI/FEDER, UE). P.G.K. gratefully acknowledges the support of NSF-PHY-1602994, the Alexander von Humboldt Foundation, the Stavros Niarchos Foundation via the Greek Diaspora Fellowship Program, and the ERC under FP7, Marie Curie
Actions, People, International Research Staff Exchange Scheme (IRSES-605096).
The authors gratefully acknowledge numerous valuable discussions
with and input from Professor Jianke Yang during the course of this
work.

\appendix
\section*{Appendix: The Levenberg--Marquardt algorithm}
\label{subsec:LMA}

Classical fixed-point methods like Newton--Raphson cannot be used for solving the  problem $F[u(x)]=0$ in the setting considered in the context
of this Chapter, essentially because there might not exist a $u(x)$ that
fulfils this relation (to arbitrarily prescribed accuracy). However, it is possible to find a function $u(x)$ that can minimize $F[u(x)]$. To this aim, an efficient method is the Levenberg--Marquardt algorithm (LMA, for short), which is also known as the damped least-square method. This method is also used to solve nonlinear least squares curve fitting \cite{leven,marq}. LMA is implemented as a black box in the Optimization Toolbox of Matlab~\textsuperscript{TM} and in MINPACK library for Fortran, and can be considered as an interpolation between the Gauss-Newton algorithm and the steepest-descent method or viewed as a damped Gauss-Newton method using a trust region approach. Notice that LMA can find exact solutions, in case that they exist, as it is the case of the results presented, e.g., in Ref.~\cite{dcd16}.

Prior to applying LMA, we need to discretize our equation (\ref{eq:stat}). Thus, we take a grid $x_n=-L/2+nh$ with $n=0,1,2\ldots M$ and $L$ being the domain length, and denote $u_n\equiv u(x_n)$ and $F_n\equiv F[u(x_n)]$. With this definition $u_{xx}$ can be cast as $(u_{n+1}+u_{n-1}-2u_n)/h^2$. In order to simplify the notation in what follows, let us call $\mathbf{u}\equiv \{u_n\}_{n=1}^M$ and $\mathbf{F}(\mathbf{u})\equiv \{F_n\}_{n=1}^M$. We will also need to define the Jacobian matrix $\mathbf{J}(\mathbf{u})\equiv \{J_{n,m}\}_{n,m=1}^M$ with $J_{n,m}=\partial_{u_m}F_n$. In the presently considered
optimization framework, $\mathbf{F}(\mathbf{u})$ is also knows as the
residue vector.

Let us recall that fixed point methods typically seek a solution by performing the iteration $\mathbf{u}_{j+1}=\mathbf{u}_j+\mathbf{\delta}_j$ from the seed $\mathbf{u}_0$ until the residue norm $||\mathbf{F}(\mathbf{u})||$ is below the prescribed tolerance; here $\mathbf{\delta}_j$ is dubbed as the search  direction. In the Newton--Raphson method, the search direction is the solution of the equation system $\mathbf{J}(\mathbf{u}_j)\mathbf{\delta}_j=-\mathbf{F}(\mathbf{u}_j)$. If the Jacobian is non-singular, the equation system can be easily solved
(as a linear system); however, if this is not the case, one must look for alternatives like the linear least square algorithm. It was successfully used for some of the authors for solving the complex Gross--Pitaevskii equation that describes the dynamics of exciton-polariton condensates \cite{pola1,pola2,pola3}. This technique also allowed us to find optimized beams in the present problem, but presented poor convergence rates, as we were unable to decrease the residue norm
controllably below the order of unity.

As fixed point methods are unable to give a reasonably small residue norm, we decided to use a trust-region reflective optimization method. Such methods consist of finding the search direction that minimizes the so called merit function

\begin{equation}\label{eq:merit}
     m(\mathbf{\delta})=\frac{1}{2}\mathbf{F}(\mathbf{u})^\mathrm{T}\mathbf{F}(\mathbf{u})+
     \mathbf{\delta}^\mathrm{T}\mathbf{J}(\mathbf{u})^\mathrm{T}\mathbf{F}(\mathbf{u})+
     \mathbf{\delta}^\mathrm{T}\mathbf{J}(\mathbf{u})^\mathrm{T}\mathbf{J}(\mathbf{u})\mathbf{\delta}.
\end{equation}

In addition, $\mathbf{\delta}$ must fulfill the relation

\begin{equation}\label{eq:trust}
    ||\mathbf{D}\cdot\mathbf{\delta}||<\Delta,
\end{equation}
where $\mathbf{D}$ is a scaling matrix and $\Delta$ is the radius of the trust region where the problem is constrained to ensure convergence. There are several trust-region reflective methods, with the LMA being
the one that has given us the best results for the problem at hand. This is a relatively simple method for finding the search direction $\mathbf{\delta}$ by means of a Gauss-Newton algorithm (which is mainly used for nonlinear least squares fitting) with a scalar damping parameter $\lambda>0$ according to:

\begin{equation}\label{eq:GN}
    (\mathbf{J}(\mathbf{u}_j)^\mathrm{T}\mathbf{J}(\mathbf{u}_j)+\lambda_j\mathbf{D})\mathbf{\delta}_j=
    -\mathbf{J}(\mathbf{u}_j)^\mathrm{T}\mathbf{F}(\mathbf{u}_j)
\end{equation}
with $\mathbf{D}$ being the scaling matrix introduced in Eq. (\ref{eq:trust}). There are several possibilities for choosing such matrix. In the present work, we have taken the simplest option, that is $\mathbf{D}=\mathbf{I}$ (the identity matrix), so (\ref{eq:trust}) simplifies to $||\mathbf{\delta}_j||<\Delta$. Notice that for $\lambda_j=0$, (\ref{eq:GN}) transforms into the Gauss-Newton equation, while for $\lambda_j\rightarrow\infty$ the equation turns into the steepest descent method. Consequently, the LMA interpolates between the two methods.
Notice also the subscript in $\lambda_j$: this is because the damping parameter must be changed in each iteration, with the choice of a suitable $\lambda_j$
constituting the main difficulty of the algorithm.

The scheme of the LMA is described in a quite easy way in Numerical Recipes book \cite[Chapter 15.5.2]{numrec} and is summarized below:

\begin{enumerate}

\item

Take a seed $\mathbf{u}_0$ and compute $||F(\mathbf{u}_0)||$

\item

Choose a value for $\lambda_0$. In our particular problem, we have taken $\lambda_0=0.1$.

\item

Solve the equation system (\ref{eq:GN}) in order to get $\mathbf{\delta}_0$ and compute $||\mathbf{F}(\mathbf{u}_0+ \mathbf{\delta}_0)||$

\item

\begin{itemize}

\item

If $||F(\mathbf{u}_0+\mathbf{\delta}_0)||\geq||\mathbf{F}(\mathbf{u}_0)||$, then take $\lambda_1=10\lambda_0$ and $\mathbf{u}_1=\mathbf{u}_0$, as with this choice of $\lambda_0$ the residue norm has not decreased.

\item

If $||F(\mathbf{u}_0+\mathbf{\delta}_0)||<||\mathbf{F}(\mathbf{u}_0)||$, then take $\lambda_1=\lambda_0/10$ and $\mathbf{u}_1=\mathbf{u}_0+\mathbf{\delta}_0$, as with this choice of $\lambda_0$ has succeeded in decreasing the residue norm.

\end{itemize}

\item

Go back to step 3 doing $\lambda_0=\lambda_1$ and $\mathbf{u}_0=\mathbf{u}_1$

\end{enumerate}

This algorithm is repeated while $||\mathbf{F}(\mathbf{u})||$ is above the prescribed tolerance.


\begin{thebibliography}{100}

  \bibitem{Bender1} Bender, C. M.; Boettcher, S.  Real Spectra in Non-Hermitian Hamiltonians Having $\mathcal{PT}$ Symmetry, {Phys. Rev. Lett.} {\bf 80}, 5243-5246 (1998).

\bibitem{Bender2} Bender, C. M.; Brody, D. C.; Jones, H. F.  Complex Extension of Quantum Mechanics,
    {Phys. Rev. Lett.} {\bf 89}, 270401 (2002).

\bibitem{Ruter} Ruter, C. E.; Makris, K. G.; El-Ganainy, R.;  Christodoulides, D. N.;  Segev, M.; Kip, D. Observation of parity-time symmetry in optics, {Nat. Phys.}, {\bf 6}, 192-195 (2010).

\bibitem{Peng2014} Peng, B.; Ozdemir, S. K.; Lei, F.;  Monifi, F.;  Gianfreda, M.; Long, G. L.;  Fan, S.; Nori, F.;  Bender, C. M.;  Yang, L.,   Parity-time-symmetric whispering-gallery microcavities,   { Nat. Phys.} {\bf 10}, 394-398
  (2014).

\bibitem{peng2014b} Peng, B.;  Ozdemir, S. K.; Rotter, S.; Yilmaz, H.;  Liertzer, M.; Monifi, F.; Bender, C. M.; Nori, F.; Yang, L.,
  Loss-induced suppression and revival of lasing,  {Science}  {\bf 346}, 328-332
  (2014).


\bibitem{ncomms2015} Wimmer, M.; Regensburger A.; Miri, M.-A.;
  Bersch, C.; Christodoulides, D.N.; Peschel, U.;
  Observation of optical solitons in PT-symmetric lattices,
  {Nature Comms.} {\bf 6}, 7782 (2015).

\bibitem{RevPT} Suchkov, S. V.; Sukhorukov, A. A.; Huang, J.; Dmitriev, S. V.; Lee, C.;  Kivshar, Yu. S.,  Nonlinear switching and solitons in PT-symmetric photonic systems. {Laser Photonics Rev.} {\bf 10}, 177-213 (2016).

\bibitem{Konotop}  Konotop, V. V.; Yang, J.; Zezyulin, D. A. Nonlinear waves in $\mathcal{PT}$-symmetric systems,  {
  Rev. Mod. Phys.} {\bf 88}, 035002 (2016).

\bibitem{Schindler1}  Schindler, J.;  Li, A.;  Zheng, M. C.;  Ellis, F. M.; Kottos,  T.  Experimental study of active LRC circuits with $\mathcal{PT}$ symmetries. {Phys. Rev. A} {\bf 84}, 040101 (2011).

\bibitem{Schindler2} Schindler, J.;  Lin, Z.;  Lee, J. M.; Ramezani, H.;  Ellis, F. M.; Kottos, T., $\mathcal{PT}$-symmetric electronics. {J. Phys. A: Math. Theor.} {\bf 45}, 444029 (2012).

\bibitem{Factor}  Bender, N.; Factor, S.;  Bodyfelt, J. D.; Ramezani, H.;  Christodoulides, D. N.;  Ellis, F. M.; Kottos, T.  Observation of Asymmetric Transport in Structures with Active Nonlinearities, {Phys. Rev.  Lett.} {\bf 110}, 234101 (2013).

\bibitem{Bender3}  Bender, C. M.;  Berntson, B.; Parker, D.; Samuel, E.  Observation of $\mathcal{PT}$ Phase Transition in a Simple Mechanical System. {\em Am. J. Phys.}
{\bf 81}, 173-179 (2013).

\bibitem{ptnlde} Cuevas-Maraver, J; Kevrekidis, P. G.; Saxena, A.; Cooper, F.;
  Khare, A.; Comech, A.; Bender, C. M.
  Solitary Waves of a $\mathcal{PT}$-Symmetric Nonlinear Dirac Equation,
  IEEE J. Select. Top. Quant. Electron. {\bf 22}, 5000109 (2016).

\bibitem{ptkg} Demirkaya, A; Frantzeskakis, D. J.; Kevrekidis, P. G.; Saxena, A.;
Stefanov, A. Effects of parity-time symmetry in nonlinear Klein-Gordon models and their stationary kinks,  Phys. Rev. E {\bf 88}, 023203 (2013); see also:
Demirkaya, A; Kapitula, T.; Kevrekidis, P. G.; Stanislavova M.; Stefanov, A.
  On the Spectral Stability of Kinks in Some PT-Symmetric Variants of the
  Classical Klein–Gordon Field Theories,
  Stud. Appl. Math. {\bf 133}, 298--317 (2014).

\bibitem{abdu}  Tsoy, E. N.; Allayarov, I.M.; Abdullaev, F.Kh.
  Stable localized modes in asymmetric waveguides with gain and loss,
  Opt. Lett. {\bf 39}, 4215--4218 (2014).

\bibitem{kono} Konotop, V.V.; Zezyulin, D. A.
  Families of stationary modes in complex potentials,
  Opt. Lett. {\bf 39}, 5535--5538 (2014).

\bibitem{jian} Yang, J.
  Partially $\mathcal{PT}$-symmetric optical potentials with
  all-real spectra and soliton families in multi-dimensions",
  Opt. Lett. {\bf 39}, 1133--1136 (2014).

\bibitem{jenn} D'Ambroise, J.; Kevrekidis, P. G.
  Existence, stability and dynamics of nonlinear modes
  in a 2D partially $\mathcal{PT}$-symmetric potential,
  Appl. Sci. {\bf 7}, 223 1--10 (2017).


\bibitem{Kom15}
Kominis, Y. Dynamic power balance for nonlinear waves in unbalanced gain and loss landscapes.
\newblock Phys. Rev. A \textbf{92}, 063849 (2015)

\bibitem{kom15b} Kominis, Y.
  Soliton dynamics in symmetric and non-symmetric complex potentials,
  Opt. Commun. {\bf 334}, 265--272 (2015).


\bibitem{mja} Ablowitz M. J.; Segur, H. {\it Solitons and the Inverse Scattering
Transform} (SIAM, Philadelphia, 1981).

\bibitem{ny16} Nixon S. D.; Yang, J. Bifurcation of Soliton Families from Linear Modes in Non-$\mathcal{PT}$-Symmetric Complex Potentials.
\newblock Stud. App. Maths. \textbf{136} (2016) 459.

\bibitem{dcd16} Dutykh, D.; Clamond, D.; Dur\'an, \'A. Efficient computation of capillary-gravity generalised solitary waves. Wave Motion \textbf{65} (2016) 1.

\bibitem{aubry} Johansson, M.; Aubry, S.
  Growth and decay of discrete nonlinear Schr\"odinger breathers interacting with internal modes or standing-wave phonons,
  Phys. Rev. E {\bf 61}, 5864--5879 (2000).

\bibitem{yang2d} Yang, J. Symmetry breaking of solitons in two-dimensional complex potentials. Phys. Rev. E \textbf{91} (2015) 023201.

\bibitem{ch16} Chen, H.; Hu, S. The asymmetric solitons in two-dimensional parity-time symmetric potentials. Phys. Lett. A \textbf{380} (2016) 162.


\bibitem{leven} Levenberg, K. A Method for the Solution of Certain Non-Linear Problems in Least Squares. Quarterly of Applied Mathematics \textbf{2} (1944) 164.

\bibitem{marq} Marquardt, D. An Algorithm for Least-Squares Estimation of Nonlinear Parameters. SIAM Journal on Applied Mathematics \textbf{11} (1963) 431

\bibitem{pola1} Cuevas, J.; Rodrigues, A.S., Carretero-Gonz\'alez, R.; Kevrekidis, P.G.; Frantzeskakis, D.J. Nonlinear excitations, stability inversions, and dissipative dynamics in quasi-one-dimensional polariton condensates. Physical Review B \textbf{83} (2011) 245140.

\bibitem{pola2} Rodrigues, A.S.; Kevrekidis, P.G.; Cuevas, J.; Carretero-Gonz\'alez, R.; Frantzeskakis, D. J. Symmetry-breaking effects for polariton condensates in double-well potentials. In B. A. Malomed (Ed.), Spontaneous Symmetry Breaking, Self-Trapping, and Josephson Oscillations. Springer Verlag. 2013, (pp. 509–-529)

\bibitem{pola3} Rodrigues, A.S.; Kevrekidis, P.G.; Carretero-Gonz\'alez, R.; Cuevas-Maraver, J.; Frantzeskakis, D. J.; Palmero, F. From nodeless clouds and vortices to gray ring solitons and symmetry-broken states in two-dimensional polariton condensates. Journal of Physics: Condensed Matter \textbf{26} (2014) 155801.

\bibitem{numrec} Press, W.H.; Teukolsky, S.A.; Vetterling, W.T.; Flannery, B.P. Numerical recipes. The art of scientific computing. Third Edition. Cambridge University Press (2007).

\end{thebibliography}
\end{document}